\documentclass[times, twoside]{style}
\usepackage{blindtext}

\usepackage{amsmath, amssymb}
\usepackage{graphicx} %
\usepackage{url}
\usepackage{booktabs}
\usepackage{xcolor}
\usepackage{multirow}

\usepackage[normalem]{ulem}

\definecolor{kscolor}{RGB}{44, 162, 95}

\definecolor{ehcolor}{RGB}{129, 33, 181}

\definecolor{titlecolor}{RGB}{44,162,95}

\usepackage[scaled]{helvet}

\usepackage[T1]{fontenc} %

\makeatletter
\let\@linenumberpar\par
\def\par{\@linenumberpar\@ifstar{\@linenumberpar}{\@linenumberpar\@linenumberpar}}
\makeatother

\leadauthor{Zentner} 

\begin{document}

\title{\color{titlecolor}Combinatorial decision-making driven by multicomponent surface condensates}

\author[a]{Aidan Zentner}
\author[b,c,g]{Ethan V. Halingstad}
\author[d]{Cameron Chalk}
\author[a,e]{Michael P. Brenner}
\author[f,g]{Arvind Murugan}
\author[d]{Erik Winfree}
\author[b,c,g,1]{Krishna Shrinivas}
\affil[a]{School of Engineering and Applied Sciences, Harvard University, 29 Oxford St, Cambridge, MA 02138}
\affil[b]{Department of Chemical and Biological Engineering, Northwestern University, 2145 Sheridan Road, Evanston, IL 60208}
\affil[c]{Center for Synthetic Biology, Northwestern University, 2145 Sheridan Road, Evanston, IL 60208}
\affil[d]{Computation and Neural Systems, California Institute of Technology, 1200 E California Blvd, Pasadena, CA 91125}
\affil[e]{Department of Physics, Harvard University, 17 Oxford St, Cambridge, MA 02138}
\affil[f]{Department of Physics, University of Chicago, 929 E 57th St, Chicago, IL 60637}
\affil[g]{NSF-Simons National Institute for Theory and Mathematics in Biology, Chicago, IL 60611}

\maketitle

\begin{abstract}
Living organisms rely on molecular networks, such as gene circuits and signaling pathways, for information processing and robust decision-making in crowded, noisy environments. Recent advances show that interacting biomolecules self-organize by phase transitions into coexisting spatial compartments called condensates, often on cellular surfaces such as chromatin and membranes. In this paper, we demonstrate that multicomponent fluids can be designed to recruit distinct condensates to surfaces with differing compositions, performing a form of surface classification by condensation. We draw an analogy to multidimensional classification in machine learning and explore how hidden species, analogous to hidden nodes, expand the expressivity and capacity of these interacting ensembles to facilitate complex decision boundaries. By simply changing levels of individual species, we find that the same molecular repertoire can be reprogrammed to solve new tasks. Together, our findings suggest that the physical processes underlying biomolecular condensates can encode and drive adaptive information processing beyond compartmentalization.
\end{abstract}

\section*{Introduction}

Living organisms process information through networks of interacting constituents spanning molecular to ecological scales. 
In cells, classic examples include gene regulatory circuits and signal transduction pathways where molecular features such as binding and copy number combine to drive biological decisions such as discrimination, feedback control, adaptation, and bistability \citep{alon_introduction_2019,hartwell_molecular_1999,lim_design_2013,parres-gold_contextual_2025}.
Although biological pathways are often described as modular \citep{hartwell_molecular_1999}, where a dedicated decision-making module drives distinct downstream events, some computational capability is embedded in processes that appear to serve different cellular tasks.
For instance, the very act of building a macromolecular assembly can encode and interpret high-dimensional inputs to trigger context-specific outcomes \citep{conrad_self-assembly_1989,winfree1998algorithmic,murugan_multifarious_2015,zhong_associative_2017,woods_diverse_2019,evans_pattern_2024}.
As another example, while genetic control circuits can be engineered to reduce fluctuations in molecular concentrations \citep{hasty_engineered_2002,frei_genetic_2022}, the same control naturally emerges from the thermodynamics that underlies single-species phase separation \citep{klosin_phase_2020}.
More generally, this kind of embedded and distributed computational power is often quite robust due to the underlying collective physics that drives it.

Recently, biomolecular condensation has emerged as a conserved mechanism for spatially organizing the cellular milieu across the tree of life \citep{Banani2017,shin_liquid_2017,lyon_framework_2020}.
Rather than being well-mixed, molecules in cells often self-organize to form dozens of coexisting compartments called condensates.
These compartments condense multiple biomolecules through phase transitions \citep{Banani2017,choi_physical_2020,julicher_droplet_2024}, typically around intracellular surfaces.
Prominent examples span gene regulatory condensates that form at specific DNA \citep{hnisz_phase_2017,shrinivas_enhancer_2019,hur_cdk-regulated_2020,morin_sequence-dependent_2022,pancholi_rna_2021,shrinivas2020dewdrops} or RNA scaffolds \citep{bond_paraspeckles_2009,snead_immiscible_2025}, and signaling condensates that are membrane-localized \citep{case_regulation_2019,zhao_phase_2020,snead_condensate-membrane_2025}.
At many surfaces, a particular combination of surface-resident molecules (i.e. “inputs”) like DNA-bound transcription factors or membrane-localized receptors facilitates assembly of specific multicomponent condensates.
These condensates, in turn, selectively recruit biomolecules (i.e. “outputs”) like polymerases or signaling messengers from the cellular milieu to drive surface-specific downstream functions---like, for instance, activating certain genes but not others.
In multicomponent fluids such as biomolecular condensates, the mapping from molecular parameters to
emergent high-dimensional phase behavior is typically nonlinear \citep{sear_instabilities_2003,sollich_predicting_2001,jacobs_predicting_2013,jacobs_self-assembly_2021,chen_programmable_2023,zwicker_evolved_2022,shrinivas_multiphase_2022,shrinivas_phase_2021,qiang_scaling_2024,rouches_surface_2021,mao_phase_2019}.
Leveraging this, recent theoretical \citep{braz_teixeira_liquid_2024,chalk_learning_2024} and experimental  \citep{abraham_nucleic_2024, do_engineering_2022,stewart_modular_2024,fabrini_co-transcriptional_2024,gong_computational_2022,takinoue_dna_2023,udono_dna_2023} work highlights the potential of condensates to perform functions beyond compartmentalization, including computations.
 A key question is what capabilities and constraints govern the ability of multicomponent biomolecular fluids to perform computations through condensation.

In this paper, we explore how biomolecular fluids can perform classification by selectively assembling distinct condensates on different surfaces. First, we model the exchange of molecules between a surface---characterized by its composition of surface-resident input species---with the broader cellular milieu, or “reservoir”.  
By exploiting differentiable methods, we tune molecular parameters such as intermolecular interactions and reservoir makeup to imbue fluids with desired phase behavior. With this framework, we demonstrate that designed fluids can deploy distinct condensates on surfaces that only subtly differ in their input compositions. This surface classification is driven by the formation of condensates that recruit to certain surfaces, but not others, high concentrations of an output molecule necessary for executing specific downstream functions. The addition of extra hidden species that can interact with all other species but cannot functionally substitute output molecules enhances the capacity to sculpt complex decision boundaries. We show that this expanded expressivity is driven by encoding novel phases that are distinct in hidden species composition but recruit the same outputs. Once designed, we show that simply adjusting hidden species levels in the reservoir enables the same molecular repertoire to classify new tasks. Together, our study suggests that the physics underlying multicomponent condensates offers flexible and versatile mechanisms for information processing in living and synthetic systems.

\begin{figure*}[t!]
    \centering
    \includegraphics[width=0.85\linewidth]{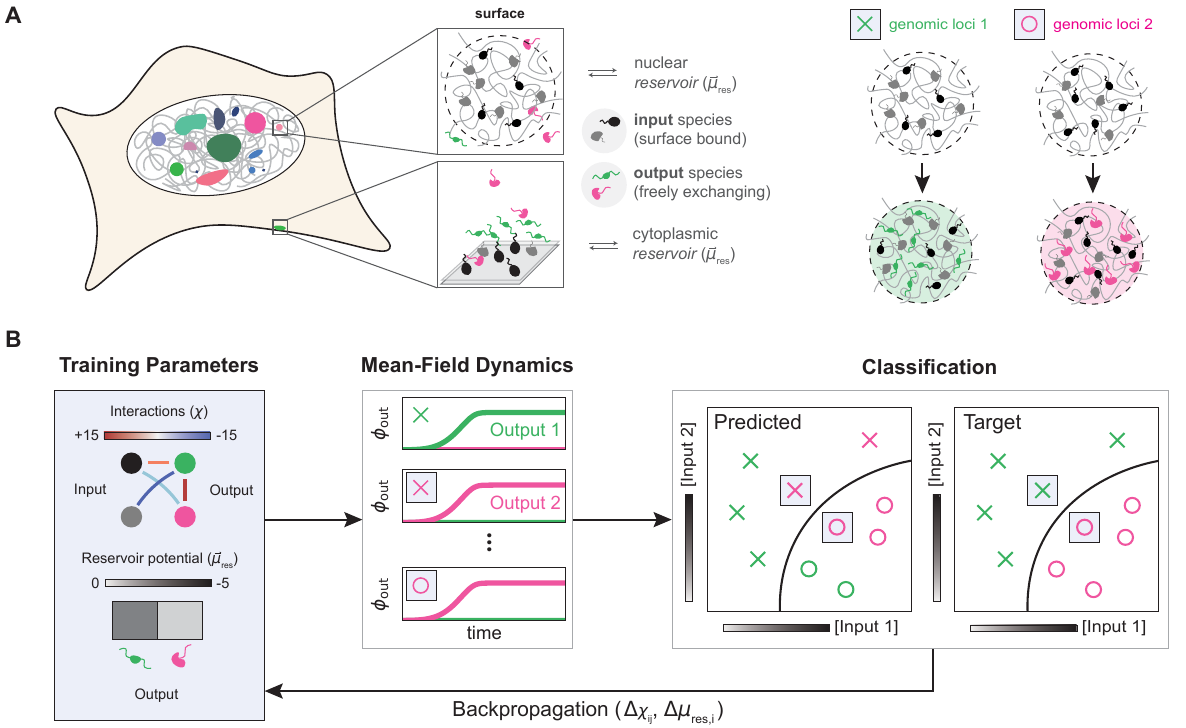}
    \caption{\textbf{(A)} The model is motivated by cellular condensates that form on surfaces such as DNA and bilayers. Surface-localized species (e.g., transcription factors on DNA) are treated as input species (black, gray), while others freely exchange with a reservoir. Output species (green, pink) are exchanging molecules that drive downstream functions - for example, polymerases (DNA) or signaling activators (membranes). Right panel highlights this where polymerases are recruited to active genes (green) and repressors to silenced genes (pink).  \textbf{(B)} Motivated by (A), we consider a simplified model in which  surfaces characterized by the presence of different combinations of input species recruit distinct output species from an infinite reservoir. (Left) The key parameters of the model are the interactions $\chi$ between the species and the reservoir chemical potentials $\vec{\mu}_\text{res}$. (Middle) We consider the evolution of surfaces in the well-mixed, mean-field limit. (Right) The recruitment of distinct outputs is accomplished through forming surface-specific condensates. The phase boundary across which the composition abruptly changes functions as a decision boundary to classify surfaces. (Loop) We use JAX to iteratively tune $\chi$ and $\vec{\mu}_\text{res}$ with the goal of recruiting the desired output species (based on the phase label, x vs o) for each training data point.}
    \label{fig:model}
\end{figure*}

\section*{Model Framework}

\subsection*{Motivation}

Surface condensation plays a key role in regulating intracellular processes, such as the formation of activating or silencing condensates on distinct genetic loci or varying signaling condensates on the plasma membrane (Fig. \ref{fig:model}A) \citep{case_regulation_2019,shrinivas_enhancer_2019}. Typically, the combination of loci-specific DNA-bound transcription factors (or surface-localized “input” species) facilitates the assembly of particular condensates. These loci-specific condensates, in turn, selectively recruit either gene-activating polymerases (an example of an “output” species, Fig. \ref{fig:model}A, green) or gene-silencing repressors (an example of another “output” species, Fig. \ref{fig:model}A, pink) that drive distinct downstream functions. Beyond input and output species, transcriptional cofactors and chromatin remodelers (“hidden” species) often regulate phase behavior and molecular recruitment but ultimately do not directly drive output response. Surface-localized receptor combinations (inputs), downstream messengers or transcription factors (outputs), and adaptors/kinases (hidden species) play analogous roles in membrane-localized condensation. Although these represent different biological pathways, they share similarities in that surface-specific properties enable the assembly of function-specific condensates, a form of classification by condensation.

This motivates a minimal model for the surface condensation of molecules from a complex cellular milieu. The cellular milieu is modeled as an infinite molecular reservoir that exchanges molecules with a surface of volume $V$. 
Here, $V$ describes an \textit{effective} 3D region that contains both the biological scaffold (such as a membrane or a DNA locus) and the neighborhood with which it interacts. 
In the model, species are partitioned into three types. First, each surface is characterized by the composition of input species that are localized to it; these species do not exchange with the reservoir. The reservoir consists of output and hidden species that, unlike input species, exchange freely with the surface.

We note two assumptions underlying our minimal formulation of the surface and reservoir (expanded upon in the next section and SI Note 7). First, the physical structure of the scaffold is not explicitly considered. Instead, the effective volume $V$ is simply defined as a region of scaffold with a well-mixed composition of input species. Second, we posit that the reservoir is maintained at a constant chemical potential, without describing its bulk behavior/composition explicitly. As a consequence of these assumptions, we do not model how a condensate adheres to or reshapes a lower-dimensional scaffold and cannot directly distinguish whether surface condensates undergo pre-wetting or wetting \cite{morin_sequence-dependent_2022}. We focus only on the composition of the condensate that forms at a well-mixed surface equilibrating with a reservoir.

With this model framework, our goal is to design a molecular network such that one specific output molecule is recruited to surfaces with specific combinations of input molecules, and a different output molecule is recruited to surfaces with other input combinations. The recruitment of distinct outputs to surfaces with specific combinations of input molecules is possible when the molecular network encodes multiple types of condensates, i.e., multiple phases where each is enriched in only one output species. The formation of one condensate over another in response to subtle differences in input combinations represents a sharp phase transition that can, in principle, be exploited to engineer for ultra-sensitive switches in the recruitment of different output molecules by designing phase boundaries in the space of input composition \citep{chalk_learning_2024,murugan_could_2025} (Fig. \ref{fig:model}B).

\subsection*{Model formulation}

Towards this goal, we model a multicomponent fluid with $N$ solute species and an additional solvent species. These $N$ solutes consist of $N_\text{in}$ input, $N_\text{out}$ output, and $N_\text{h}$ hidden species ($N=N_\text{in}+N_\text{out}+N_\text{h}$). For simplicity, the sizes of all species are assumed to be equal to the volume $\nu$ of the solvent molecule, and the mean volume fraction of species $i$ is therefore related to the absolute number of particles $n_i$ within the surface by $\phi_i=n_i\nu/V$. We work in the mean-field limit and assume that the surface remains well-mixed. The surface is therefore described by its mean composition vector, labeled as $\vec{\phi}\equiv\vec{\phi}_\text{in}\circ\vec{\phi}_\text{out}\circ\vec{\phi}_\text{h}$, where $\circ$ indicates vector concatenation and
\begin{align}
    \phi_\text{in}&=(\phi_{\text{in},1}, ..., \phi_{\text{in},N_\text{in}})\\
    \phi_\text{out}&=(\phi_{\text{out},1}, ..., \phi_{\text{out},N_\text{out}})\\
    \phi_\text{h}&=(\phi_{\text{h},1}, ..., \phi_{\text{h},N_\text{h}})\ .
\end{align}
The subvectors denote the input, output, and hidden composition vectors, respectively, and the total volume fraction of (non-solvent) species is $\phi_T=\sum_{i=1}^N\phi_i$. The surface only exchanges hidden and output species with the infinite reservoir. Within our framework, we do not prescribe any specific model of the reservoir (see SI Note 7) and assume that it can maintain output and hidden species at a constant chemical potential described by
\begin{equation}
    \vec{\mu}_\text{res}=\left(\mu_{\text{out},1}^\text{res},\dots,\mu_{\text{out},N_\text{out}}^\text{res}, \mu_{\text{h},1}^\text{res},\dots,\mu_{\text{h},N_\text{h}}^\text{res}\right)\ .
\end{equation}

The non-dimensionalized free energy density of such a surface is
\begin{equation}
    \Omega_\text{surface} = \beta\nu f(\vec{\phi},\chi)-\beta\vec{\mu}_\text{res}\cdot\vec{\phi}_\text{oh}
    \label{eq:grand_potential}
\end{equation}
where $\vec{\phi}_\text{oh}=\vec{\phi}_\text{out}\circ\vec{\phi}_\text{h}$ and $\beta=1/k_\text{B}T$ is the inverse temperature. The quantity $\beta\vec{\mu}_\text{res}\cdot\vec{\phi}_\text{oh}$ therefore describes the external coupling of the output and hidden species to the reservoir. $f$ is the \textit{internal} free energy density of the surface, approximated in Flory-Huggins theory as
\begin{equation}
    \begin{aligned}
        \beta\nu f(\vec{\phi},\chi) &= \sum_{i=1}^N\phi_i\log{\phi_i}+(1-\phi_T)\log{(1-\phi_T)}\\
        &\qquad\qquad\qquad+\frac{1}{2}\sum_{i=1}^N\sum_{j=1}^N\phi_i\chi_{ij}\phi_j
    \end{aligned}
    \label{eq:flory-huggins}
\end{equation}
where $\chi$ is the effective interaction matrix,
\begin{equation} \label{eq:flory-energy}
    \chi_{ij}=z\beta\left(\epsilon_{ij}-\frac{\epsilon_{ii}+\epsilon_{jj}}{2}\right)
\end{equation}
and $z$ is the number of nearest interacting neighbors, with $\epsilon_{ij}$ being the microscopic nearest-neighbor contact energy between species $i$ and species $j$. Note that since $\chi_{ii}=0$ by definition, the $i=j$ terms do not contribute to eq. \ref{eq:flory-huggins}. 
Further, we assume negligible effective solute-solvent interactions such that $\chi_{i0}=0$ for all $i= 1,2,..N $, where $0$ indexes the solvent (see SI Note 1 for an extended discussion of this assumption and implications for experimental realization).

We next write a dynamical model to probe the steady-state composition of a surface characterized by a fixed input species composition $\vec{\phi}_\text{in}$. The volume fractions of all non-input species evolve over time due to the exchange with the reservoir until a steady-state is reached. We treat these compositional dynamics as near-equilibrium relaxation that, to a first approximation, is driven by linear gradients of the free energy with respect to the surface's composition \citep{hohenberg_theory_1977}. Here, we assume that solvent molecules have much faster dynamics than solutes, which improves numerical stability of the optimization but does not affect the steady-state (Fig. S1). Thus, the temporal evolution of the surface composition $\vec{\phi}_{\text{oh}}$ of the exchanging output and hidden species is written as (SI Note 1)
\begin{equation}
    \frac{d \vec{\phi}_\text{oh}}{dt}\approx -
    \textbf{D}\frac{d\Omega_\text{surface}}{d\vec{\phi}_\text{oh}} = -\beta\textbf{D} (\vec{\mu} - \vec{\mu}_\text{res})
    \label{eq:model-A}
\end{equation}
where $\beta \mu_i=\partial(\beta\nu f)/\partial\phi_{\text{oh},i}$ is the intrinsic (non-dimensionalized) chemical potential of exchanging species $i$.  $\textbf{D}$ is the $(N_\text{out}+N_\text{h})\times(N_\text{out}+N_\text{h})$ mobility matrix that sets the rate of exchange between the surface and reservoir and is chosen, for simplicity, to be diagonal, identical for solutes, and consistent with Fick’s law at dilute equilibrium conditions \citep{shrinivas_phase_2021} (SI Note 1). At steady state, the surface and reservoir must have identical chemical potentials in the non-input species but can have distinct compositions---a feature of multiphase systems that we aim to exploit.

\subsection*{Designing multiphase classifiers}

With this forward model, our goal is to identify an effective interaction matrix $\chi$ and reservoir chemical potential $\vec{\mu}_\text{res}$  (at $\beta=1$) such that, for a surface defined by a given input vector $\vec{\phi}_\text{in}$, the steady state is enriched in the desired output species and depleted in all other outputs (Fig. \ref{fig:model}B). This output convention is akin to ``one-hot'' representations common in machine learning. Tuning these parameters allows for the construction of free energy landscapes for which the input concentration $\vec{\phi}_\text{in}$ of a surface determines the free energy minimum to which it relaxes. In this case, surfaces enriched in different output species are separated in the input space by a discontinuous phase boundary that behaves functionally as a decision boundary.

To train this model, we employ a differentiable implementation of the above dynamical description amenable to gradient-based optimization methods that minimize a loss function \citep{bradbury_jax_2018}. We require that our loss function captures the following physical criteria at any surface. First, we require the desired output species to be enriched at concentrations exceeding  $\phi_\text{max}=A/N$. Second, we require all  undesired outputs to be depleted at concentrations below $\phi_\text{min}=B/N$. These enrichment and depletion criteria, when satisfied, enforces that the desired output is recruited at high concentrations and exceeds undesired outputs by a factor of at least $A/B$. We choose $A=1.1$ (mild enrichment), and $B=0.25$ (significant depletion).

Training the system entails considering a set of surfaces (or data points), each defined by an input concentration vector and a specification of the desired output species to be recruited. We find empirically that the following loss function gives the best performance in optimizing for the enrichment and depletion criteria:
\begin{equation}
    \mathcal{L}(\chi,\vec{\mu}_\text{res})=\frac{1}{n_\text{batch}}\sum_{a=1}^{n_\text{batch}}l_{j(a)}\left(\chi,\vec{\mu}_\text{res};\vec{\phi}_a\right)
    \label{eq:loss-function}
\end{equation}
where the sum is over $n_\text{batch}$ data points in the training set, data point $a$ corresponds to a surface that reaches steady-state concentrations $\vec{\phi}_a$, and $j(a)$ is the index of the desired output species for data point $a$. We define a function that is monotonic in the extent that $\vec{\phi}_a$ fails to satisfy the enrichment/depletion criteria for a surface:
\begin{align}
    l_j(\chi,\vec{\mu}_\text{res})&=\log{(1+Np_j)}+\sum_{\substack{k=1\\(k\neq j)}}^{N_\text{out}}\log{(1+Nq_k)} \ge 0\\ 
    p_j&=\max\left(0, \phi_\text{max}-\phi_{\text{out},j}\right)\\
    q_k&=\max\left(0,\phi_{\text{out},k}-\phi_\text{min}\right).
\end{align}
The term $l_j$ is therefore at a global minimum when $p_j=q_k=0$, and $\mathcal{L}$ is at a global minimum of $0$ when this condition is satisfied for all data points $a$. 
Because the loss saturates once the thresholds are met, optimization only requires modest enrichment and depletion. However, two surfaces that are highly similar but lie on opposite sides of the decision boundary requires a sharp flip in the enriched output. The optimization therefore leverages the collective physics of phase separation, which produces discontinuous switches in composition across a phase boundary, and yields much larger enrichment ratios in practice.

We minimize $\mathcal{L}$ with respect to $\chi$ and $\vec{\mu}_\text{res}$ over several thousand training epochs using an RMSProp algorithm from the Optax library \citep{bradbury_jax_2018,deepmind_deepmind_2020}. Once trained, we evaluate the performance of the classifier using the classification success $S_c$, defined as the fraction of surfaces whose steady states fully satisfy the enrichment/depletion criteria:
\begin{equation}
    S_c=\frac{1}{n_\text{set}}\sum_{a=1}^{n_\text{set}}\left[1-\Theta(l_{j(a)})\right]
    \label{eq:success-criterion}
\end{equation}
where $n_\text{set}=500$ is the number of points in the validation/test set, and $\Theta(x)=0$ if $x=0$ and is 1 otherwise.

\begin{figure*}[t!]
    \centering
    \includegraphics[width=0.85\linewidth]{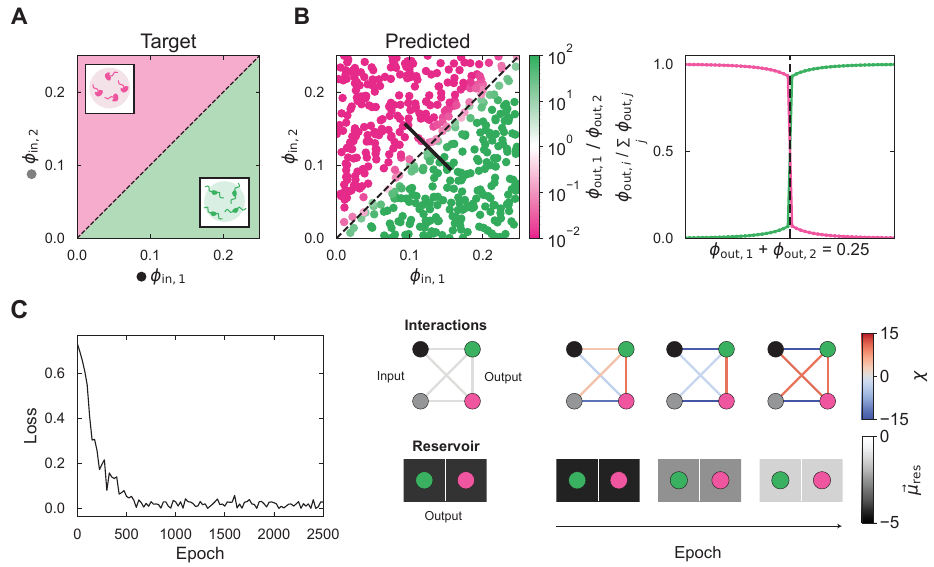}
    \caption{\textbf{(A)} The target linear decision boundary is shown, with each axis being the concentration of one of the input species. Green and pink denote regions where we desire condensates enriched in the green and pink component, respectively. \textbf{(B)} Predictions from the trained model for different input compositions in the test set. The axes depict the input concentrations while each dot is a test input condition, colored by the ratio of the two output species at steady-state (displayed on a log-scale). Along the solid black line, the system undergoes a discontinuous transition in mean-field composition across the boundary, as shown in the right-most panel. \textbf{(C)} Evolution of training loss and parameters over the optimization. The training parameters converge to a solution that is analytically consistent with the formation of a linear decision boundary (SI Note 3).
}
    \label{fig:linear}
\end{figure*}
In training the system over $\chi$ and $\vec{\mu}_\text{res}$, we impose several constraints. 
First, since we are modeling liquid phases, we require that energies be of order $k_\text{B}T$ and therefore enforce that each entry of the chi matrix has $|\chi_{ij}|<15$, which is $\mathcal{O}(z=18)$. Hence, from eq. \ref{eq:flory-energy}, the typical strength of an interaction energy is $|\epsilon_{ij}| \leq k_\text{B}T$ and comparable to thermal scales, as expected in liquid-like mixtures. Second, since we are designing surfaces to only enrich one particular output species, we require that output-output interactions be repulsive, with $\chi_{ij}>10$ for distinct output species $i$ and $j$. Third, we enforce that all output species have the same reservoir chemical potential as a design criterion, which is meant to mimic the surface choosing from outputs that are at ``identical'' potentials in the reservoir. Finally, since input-input interactions and input chemical potentials don't affect the steady-state in the mean-field limit, they are omitted from the model and not treated as free design parameters.

\begin{figure*}[t!]
    \centering
    \includegraphics[width=0.85\linewidth]{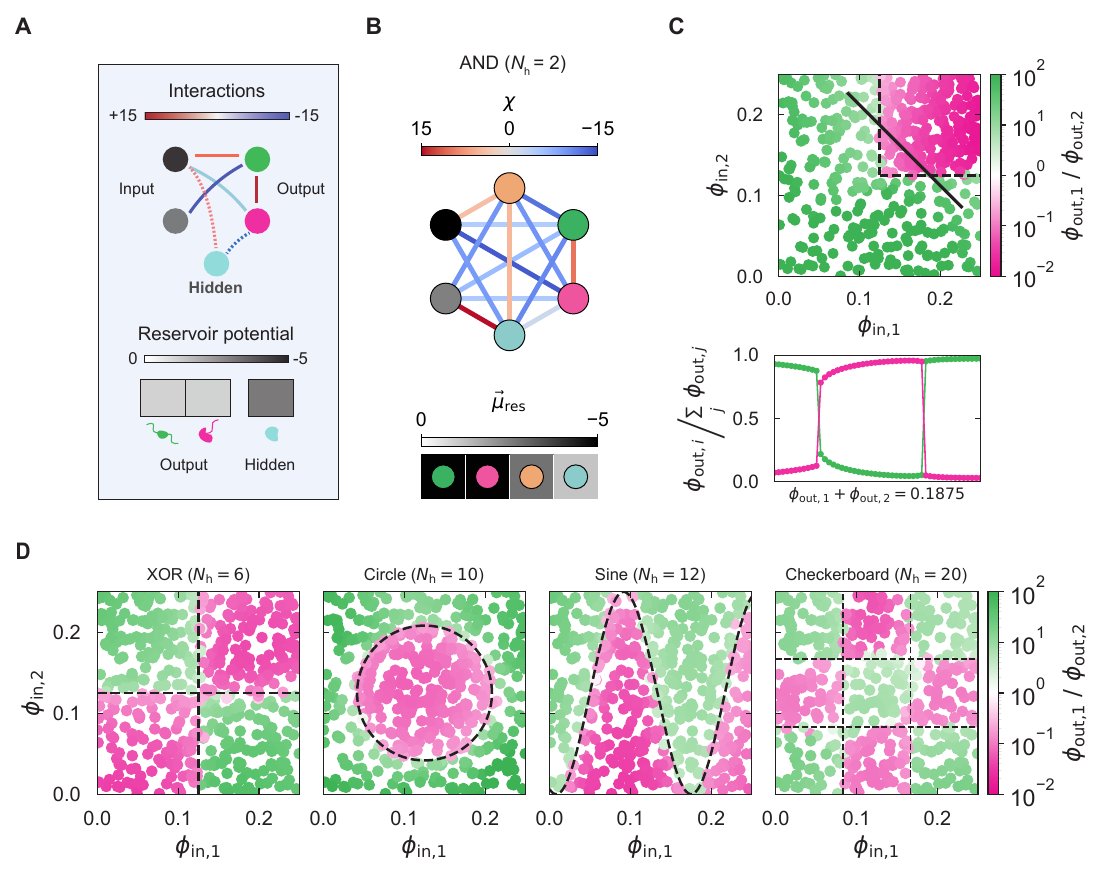}
    \caption{\textbf{(A)} Hidden species, depicted in cyan in the interaction matrix and analogous to hidden nodes in Boltzmann machines, shape emergent overall phase behavior by interacting with input and output species but cannot directly drive output function. \textbf{(B)} Using only 2 hidden species (gold and cyan), we train for parameters to form an AND-like upper quadrant decision boundary in the mean-field limit. \textbf{(C)} Predictions from the trained classifier for different input compositions in the test set, which resembles the desired decision boundary. Along the solid black line, the system undergoes a discontinuous transition in mean-field composition across the boundary. \textbf{(D)} Test predictions for models with other nonlinear decision boundaries in the mean-field limit. (left to right) A XOR boundary trained with 6 hidden species, a circular boundary trained with 10 hidden species, a sine curve boundary trained with 12 hidden species, and a checkerboard boundary trained with 20 species.
}
    \label{fig:complex_boundaries}
\end{figure*}

\section*{Results}

\subsection*{Tuned molecular networks drive linear classification}

Armed with this model, we first aim to create linear classifiers. In a simple mixture comprising only 2 input and 2 output species, our target is to design surface condensates that recruit a specific output molecule (green or pink) depending on which input species is at higher concentration (Fig. \ref{fig:linear}A), i.e., an ultrasensitive ratiometric sensor. With this objective in mind, we initialize a non-interacting liquid mixture and train the molecular interactions and reservoir potential over multiple epochs (Fig. \ref{fig:linear}B). The learned interaction matrix broadly matches physical intuition, with each input preferring to mix with the cognate output and demix from the non-cognate output. Upon testing, our model exhibits a sharp switch in composition across the boundary (Fig. \ref{fig:linear}C, Fig. S2A). This switch is consistent with a first-order phase transition, exhibiting a discontinuous jump in the output composition across the phase boundary. As expected, this discontinuity disappears when interactions are weakened (Fig. S2B).

To understand how the decision boundary emerges from molecular parameters, we develop a simple analytical approach (SI Note 3). We first define the decision boundary as the manifold where all output species are recruited at identical levels. We find that the expressivity (or repertoire of encodable manifolds) of mixtures with 2 inputs and 2 outputs is limited to linear boundaries, and this theoretical prediction is supported by simulation (Fig. S3A, SI Note 3). More generally, our theory predicts that liquids with only input and output species can only typically encode linear decision boundaries in input space (SI Note 3). Consistent with this prediction, we find that our model still sharply classifies higher-dimensional linear manifolds (Fig. S3B). 

To test our model's prediction that purely input-output mixtures cannot classify nonlinear boundaries, we train a 2 input and 2 output mixture to separate an elementary nonlinear manifold: an upper quadrant AND-like distribution, in which one output is recruited only when both inputs are present at high concentrations; otherwise, the other output is recruited. After training, we find that input-output mixtures fail to encode this nonlinear decision boundary, instead showing a best-fit linear approximation (Fig. S4A).

\subsection*{Hidden species expand capacity for nonlinear complex decisions}

\begin{figure*}[t!]
    \centering
    \includegraphics[width=0.85\linewidth]{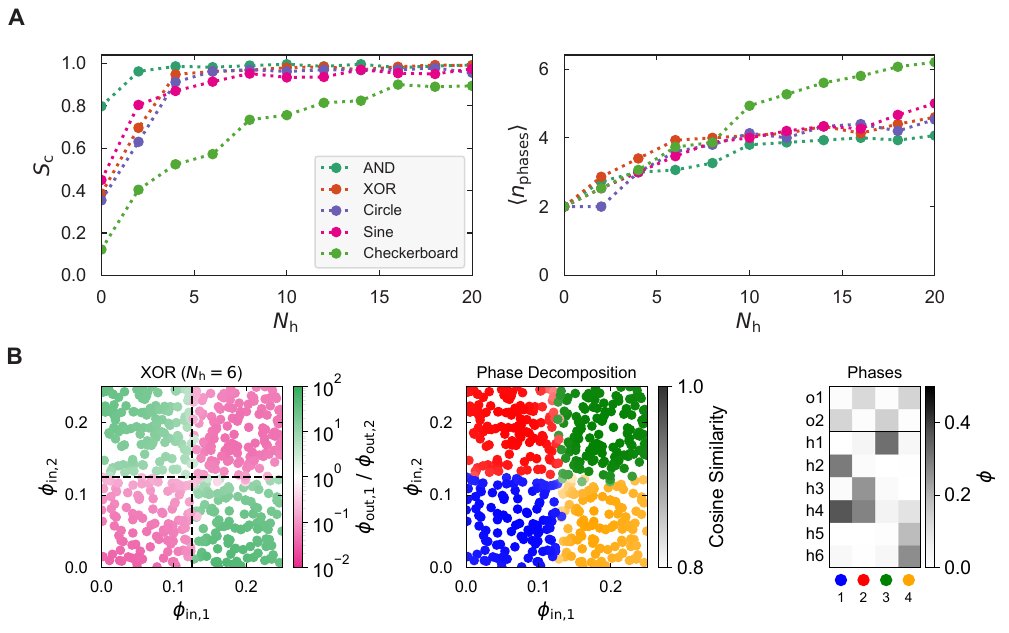}
    \caption{\textbf{(A)} The scaling of the test set classification success $S_\text{c}$ (left) and the number of phases, averaged across molecular networks corresponding to different minima of the loss function (right), as a function of the number of hidden species in the model. \textbf{(B)} When defined in terms of output species composition alone, the XOR liquid (left) shows two distinct phases, each repeated twice in the four quadrants of input space. However, each of these quadrants corresponds to a distinct phase (middle) if phases are distinguished by the composition of hidden species. The color of a point indicates the phase that the surface exhibits, while the color intensity of the point indicates the cosine similarity between the surface's phase vector and the mean concentration vector of all surfaces within the phase (right). The compositions of the output and hidden species presented in the right-most panel represents the mean composition of surfaces that cluster into a phase (see SI Note 5).
}
    \label{fig:scaling}
\end{figure*}

The inability to form nonlinear decision boundaries with simple input-output networks motivates the introduction of hidden species. In our model, hidden species are similar to output species in that they can interact with all molecules and be recruited to surfaces from the reservoir, thereby influencing the steady-state condensate that forms. However, their steady-state concentrations are taken to be irrelevant in performing the classification of the surface; they therefore play a role analogous to that of hidden nodes in a neural network \citep{ackley_learning_1985,chalk_learning_2024}.

Extending our analytical approach, we find that the addition of a single hidden species offers enough flexibility to encode decision boundaries of varying curvatures (Fig. S3C, SI Note 3). We thus explore classification of complex, high-dimensional decision boundaries by including multiple hidden species (Fig. \ref{fig:complex_boundaries}A).

First, we demonstrate the effectiveness of hidden species by programming an AND-like upper quadrant decision boundary with two additional hidden species (gold and cyan in Fig. \ref{fig:complex_boundaries}B). 
Analyzing the trained molecular network reveals a complex interplay of interactions that leads to essentially binary responses in the output species recruited to the surface (Fig. \ref{fig:complex_boundaries}C, Fig. S4). Like the linear classifier, and consistent with a phase transition, our trained AND system exhibits a sharp switch in composition across the decision boundary (Fig. \ref{fig:complex_boundaries}C). 
With the addition of more hidden species, the model can encode increasingly nonlinear decision boundaries such as XOR, circle, sinusoidal, and checkerboard patterns (Fig. \ref{fig:complex_boundaries}D). 
Similar to the AND boundary, each of these systems exhibit sharp, discontinuous switches in the recruited species across the boundary consistent with first-order phase transitions (Fig. S5). The trained parameters for each decision boundary are shown in Fig. S6.%

\subsection*{Hidden species expand capacity by encoding multiple modular, encrypted phases}

To understand how hidden species enhance expressivity, we trained mixtures with varying numbers of hidden species to solve a range of decision boundaries, and evaluated the classification success (as defined in 
eq. \ref{eq:success-criterion}). We find that the addition of hidden species improves classification but saturates beyond a decision-boundary specific threshold (Fig. \ref{fig:scaling}A, left). While surface condensates correctly enrich the pertinent output, we find that surfaces with the same output molecules often recruit varying concentrations of hidden species. To better understand this, we estimated how many distinct phases were formed as defined by the \textit{overall} composition of hidden and output species on surfaces.
Collecting the compositions across multiple surfaces ($n_\text{set}$ test points into a matrix of size $n_\text{set}\times(N_\text{out}+N_\text{h})$), we perform principal component analysis and use a Marchenko-Pastur \citep{marcenko_distribution_1967} based threshold to estimate the number of distinct phases from the significant eigenmodes. We then perform hierarchical clustering to identify the average composition of each phase (see SI Note 5).
We find that for each decision boundary, the number of steady-state phases, averaged across a collection of molecular networks corresponding to different minima of the loss function, grows with  hidden species (Fig. \ref{fig:scaling}A, right). This suggests that encoding multiple phases plays an important role in improving expressivity of multicomponent condensates.

To explore this deeper, we consider the trained XOR liquid with 6 hidden species (Fig. \ref{fig:scaling}B). Compositional analysis reveals that the XOR decision boundary is achieved through 4 \textit{distinct phases}. For example, areas with high output 1 (green output) are encoded by 2 distinct phases (e.g., red and yellow phases, or phases 2 and 4) that recruit different hidden species but the same output species. The concentration vectors within a phase show very little variance, as evidenced by their high cosine similarity. Identifying each point with an independent surface, our model shows that multiple surfaces that condense the same output recruit distinct hidden species, and thus vary in phase composition. Biologically, such a solution might look like condensates that drive gene activation at different DNA loci by recruiting high concentrations of the functional polymerase but varying concentrations of coactivator molecules. Thus, the encoding of multiple \textit{encrypted} phases, which differ in hidden species but recruit similar output molecules, is the primary mechanism by which hidden species improve expressivity.

\begin{figure*}[t!]
    \centering
    \includegraphics[width=0.95\linewidth]{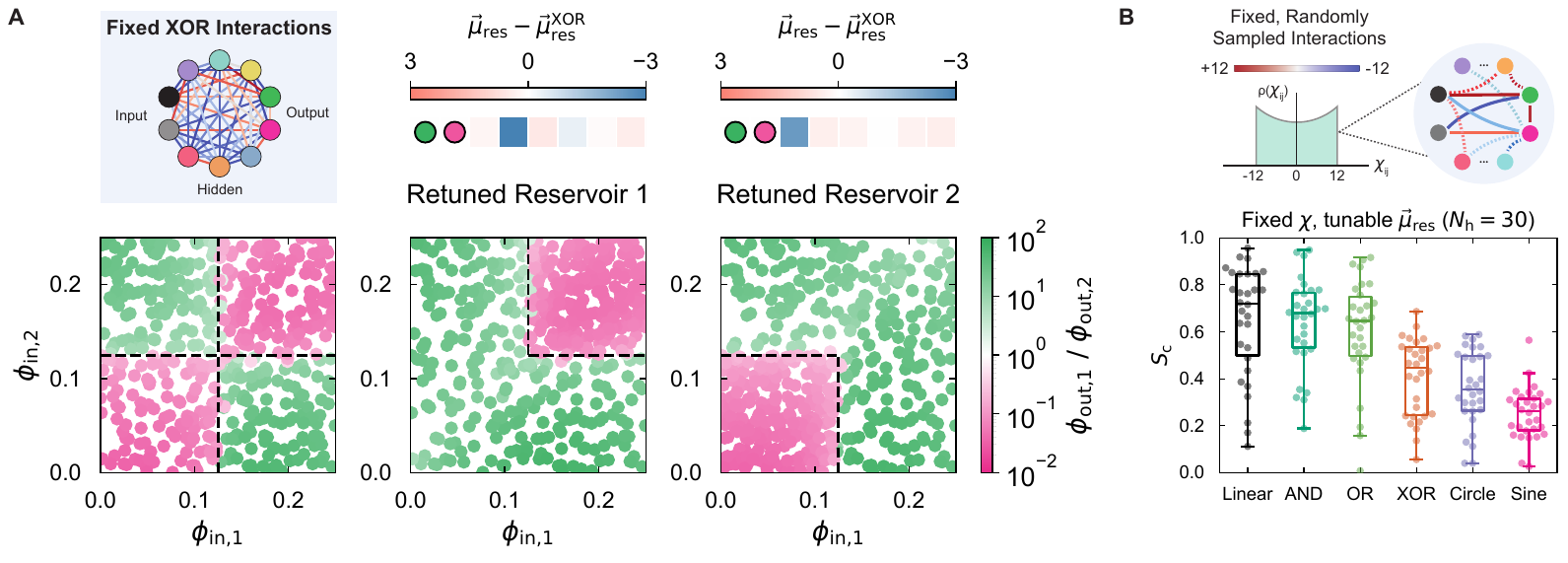}
    \caption{\textbf{(A)} With the $N_\text{h}=6$ mixture previously trained to solve a XOR boundary (left), the same interaction network can also solve an AND decision boundary (middle) and OR decision boundary (right) by selectively tuning the reservoir potentials. \textbf{(B)} The capacity for classification by liquids with random interaction networks. (left) A random interaction matrix with $N_\text{h}=30$ is sampled from a probability distribution (see SI Note 5) and remains immutable; subsequently, reservoir potentials are tuned to solve a classification task. (right) The success rate $S_\text{c}$ across multiple trajectories (30 per task, where each has a different random interaction network) for various classification tasks.}
    \label{fig:reversoir-tuning}
\end{figure*}

In the XOR liquid, we find that the 4 distinct encrypted phases modularly partition the input space into quadrants, such that groups of related inputs drive condensation of a particular phase. When we extend this analysis to other nonlinear decision boundaries, we find that hidden species generally learn modular representations of related input surfaces (Figs S7-S11). We next explore whether we can repurpose this modular multiphase representation learned by hidden species for other tasks.

\begin{figure*}[t!]
    \centering
    \includegraphics[width=0.85\linewidth]{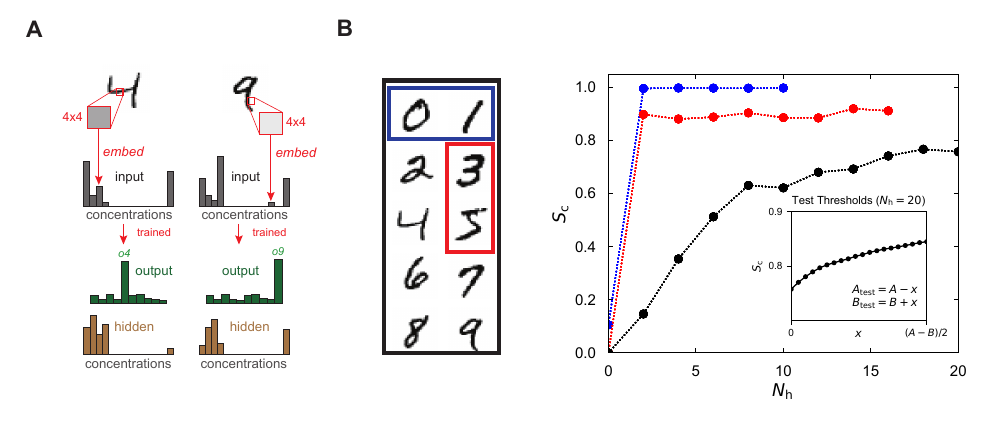}
    \caption{\textbf{(A)} Solving classification of the MNIST dataset involves embedding pixel grayscale values into many input species concentrations, then training interaction parameters and reservoir concentrations for the formation of condensates enriched in 1 of 10 output species. \textbf{(B)} Classification success $S_\text{c}$ for 0 vs 1 (blue), 3 vs 5 (red), and 10-digit MNIST (black) tasks as a function of number of hidden species. For the 10 digit MNIST dataset, the classification success reaches a plateau below 1 at approximately 15-20 hidden species. The inset shows the plateau value when the success criterion is made more lenient by decreasing $A_\text{test}=A-x$ and increasing $B_\text{test}=B+x$, where $A=1.1$ and $B=0.25$ are the threshold values defined in the loss function used to train the classifier. In the extreme case where $A_\text{test}=B_\text{test}$, the success is approximately $\sim85\%$.}
    \label{fig:mnist}
\end{figure*}

\subsection*{Changing reservoir composition of hidden-species drives solution of new classification tasks}

Motivated by the modularity of encoded phases, we hypothesized that once trained with sufficient hidden species, the same molecular ensembles could be adapted to solve new decision tasks by simply tuning the reservoir of hidden species without changing interactions.  This idea is analogous to machine-learning architectures comprising modules where an upstream (typically randomly-wired) network remains fixed and solutions to new tasks are achieved by training only the parameters of a small downstream network \citep{rahimi_random_2007,tanaka_recent_2019}. To demonstrate this idea, we revisit the trained XOR liquid and ask whether it can be repurposed to solve AND or OR decision boundaries only by changing reservoir composition. We find that changing the potential of a few key hidden species is sufficient to finetune the same molecular mixture to perform distinct tasks (Fig. \ref{fig:reversoir-tuning}A).

Given this finding, we next explored whether liquids \textit{without} designed interactions, e.g., with randomly chosen molecular interactions $\chi_{ij}$, could nevertheless be trained to classify surfaces through fine-tuning the reservoir alone. To test this, we generated liquids with 2 inputs, 2 outputs, and a large number of hidden species ($N_\text{h}=30$, SI Note 6). The interactions between species were sampled from a near-uniform distribution such that $|\chi_{ij}|\lesssim 12$; for a fixed decision boundary, we report the distribution of model performance over $n=30$ different interaction networks (SI Note 6). Through training only the reservoir makeup, we show that liquids with randomly chosen and fixed interactions $\chi$ contain the ability to model both linear and nonlinear decision boundaries, albeit with decreasing performance as we increase the complexity of the decision boundaries that we seek to approximate (Fig. \ref{fig:reversoir-tuning}B).

Our results suggest that rather than constantly redesigning or evolving new interactions, the physics of surface condensation provides a flexible mechanism to redeploy the same molecular repertoire to solve new tasks by adjusting compositions of the reservoir. An analogous idea has been explored previously by Elowitz and coauthors in the context of BMP signaling and dimerization networks \citep{parres-gold_contextual_2025,su_ligand-receptor_2022} where they show that tuning stoichiometries but not binding affinities in dilute molecular ensembles can facilitate solving distinct tasks. Together, this highlights that the physics embedded in collective molecular networks permits flexible computations at distinct hierarchies.

\subsection*{Surface condensates classify high-dimensional datasets}

Our motivation for physically embedded computation in phase separation is to understand how cells might process chemical stimuli through concentration-dependent condensation, not to build a general-purpose classifier for arbitrary domains (e.g., distinguishing cat vs. dog images). In the same spirit, related work has evaluated physical systems as classifiers of physical stimuli in many domains \citep{stern_learning_2023}, ranging from molecular concentrations to mechanical forces \citep{stern_supervised_2021,stern_supervised_2020}.  Molecular examples include winner-take-all reaction networks \citep{cherry_scaling_2018}, self-assembly with Hebbian-like interactions \citep{zhong_associative_2017,evans_pattern_2024}, and multicomponent liquids \citep{braz_teixeira_liquid_2024,chalk_learning_2024}. Nevertheless, to evaluate expressivity of these physical systems on high-dimensional inputs in a standardized way, we follow this literature and use symbolic ML datasets as benchmarks, not as an end in themselves: each feature is reinterpreted as a molecular concentration and presented to the system as a physical stimulus.

We start by classifying the near-linear Seaborn Iris dataset, which comprises 4 analog flower features (petal and sepal length and width) and 3 output labels (flower species). The value of the $j$’th feature $x_{aj}$ of the $a$’th data point $x_a$ is encoded as an input concentration according to the linear, scaled mapping $\phi_{aj}=\phi^0\left(\frac{x_{aj}-\min_a x_{aj}}{\max_a x_{aj}-\min_a x_{aj}}\right)$, where $\min_a x_{aj}$ and $\max_a x _{aj}$ denote the minimum and maximum $x_{aj}$ across all data points $a$, respectively, and $\phi^0=0.5/4=0.125$, such that the input species can occupy a maximum of half of the volume. We incorporate 1 output species per label to mimic a species-specific molecule. Once trained, we demonstrate that this 7-component mixture can directly classify the IRIS dataset without the use of hidden nodes (Fig. S12).

Next, we turn to the higher-dimensional MNIST dataset, a collection of labeled hand-drawn images of digits, to study how our model generalizes to larger interaction networks. 
We first coarse-grain each grayscale image from $28\times28$ to $7\times7$ by averaging pixel values in a $4\times4$ block and assign each pixel in the reduced image to an input species. Then, we map the volume fraction $\phi_{\text{in},i}$ of an input species to its corresponding pixel value ($x_i$) by $\phi_{\text{in},i}=\phi^0(x_i/255)$, where $\phi^0=0.5/49\approx0.01$. We train the mixture to initially discriminate between two digits, achieving strong performance with just 2 hidden nodes. However, digits that are traditionally harder to distinguish (Fig. \ref{fig:mnist}B, red, 3 vs 5) reached lower performance levels compared to easier ones (Fig. \ref{fig:mnist}B, blue, 0 vs 1). Extending the model to simultaneously classify all ten digits requires more hidden nodes ($\sim15$) and saturating performance is lower (Fig. \ref{fig:mnist}B, black). As we relax the classification stringency by requiring lower and lower excess of the desired output species over the undesired ones without retraining the system (Fig. \ref{fig:mnist}B inset), the success in classifying MNIST increases from $\sim75\%$ to a saturating test success of $\sim85\%$. Generally, the ability to design condensates with large numbers of species for high-dimensional classification is improved with hidden nodes but typically saturates. 

These results are generally consistent with recent findings from \citep{chalk_learning_2024}, where the authors develop a 3D lattice condensate model and train it with a probabilistic learning algorithm derived from classical Boltzmann machines to classify MNIST digits with $\sim 75\%$ accuracy. Their lattice liquid with a “semipermeable membrane” is conceptually equivalent to our approximation of “surface-localized inputs”. In \citep{evans_pattern_2024}, which explores crystalline self-limited assembly, MNIST digit classification is similarly demonstrated in a theoretical model with $\sim85-90\%$ accuracy depending on the design constraints. 
Together, these results highlight the potential of multicomponent mixtures to classify high-dimensional decision boundaries despite differences in microscopic physics, training algorithms, design constraints, and problem encodings.

\subsection*{Mean-field solutions translate to successful classifiers in 3D lattice liquids}

\begin{figure*}[t!]
    \centering
    \includegraphics[width=0.85\linewidth]{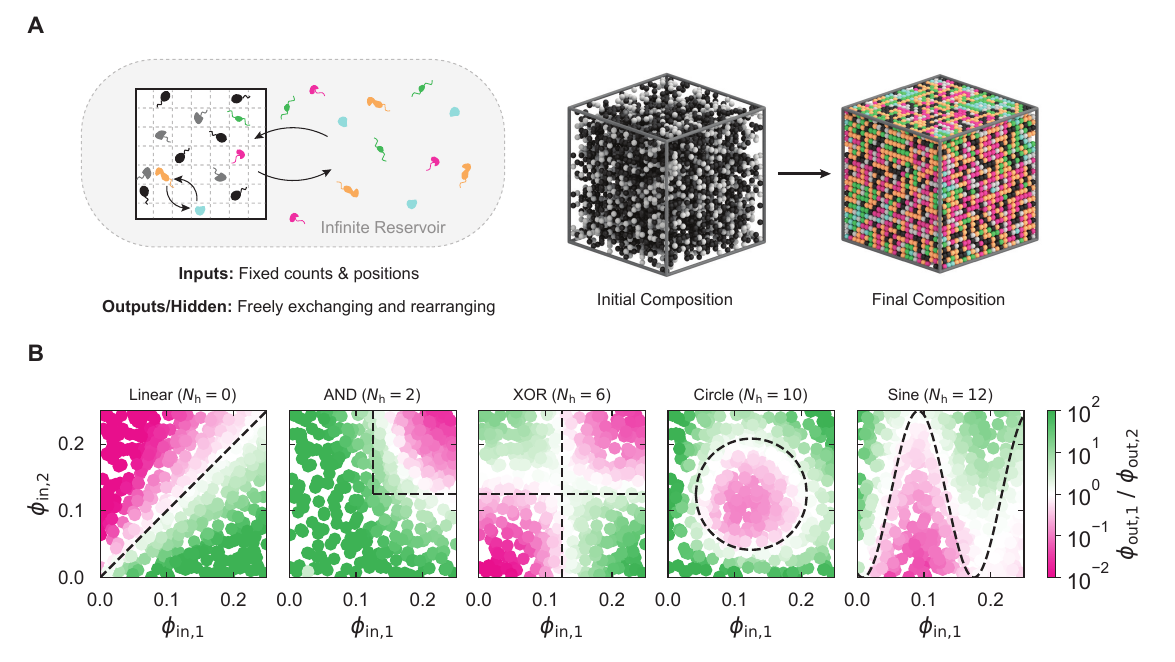}
    \caption{\textbf{(A)} Schematic of 3D lattice liquid. Inputs are clamped to lattice positions, and output and hidden species can exchange freely. On the right is the evolution of a lattice ($L=24$) from initial to final configurations. \textbf{(B)} 3D lattice liquid ($L=24$) classification using mapped mean-field parameters: (left to right) Linear boundary with 0 hidden species, AND boundary with 2 hidden species, XOR boundary with 6 hidden species, Circle boundary with 10 hidden species, and Sine boundary with 12 hidden species.}
    \label{fig:lattice}
\end{figure*}

We next aim to understand whether the mean-field design of liquids transfers to a more detailed 3D model that explicitly captures spatial correlations. Following earlier work \citep{jacobs_predicting_2013,jacobs2017phase,chalk_learning_2024}, we adopt a lattice liquid formulation in which we treat a surface as a lattice of length $L\times L\times L$ with 1 molecule per site. Interactions between 18 nearest neighbors, i.e., those within a $\sqrt{2}$ lattice distance, contribute to the overall energy of the system, which thus depends on the spatial configuration of molecules (SI Note 4). To mimic our mean-field treatment of surface-localized and well-mixed input species, we fixed their counts and positions on the lattice, thereby treating them as immobile and non-exchanging in the canonical ensemble (Fig. \ref{fig:lattice}A). Output and hidden species are allowed to exchange with the reservoir at a fixed chemical potential, i.e., in the grand-canonical ensemble. We sample this mixed-ensemble model through parallelized Monte-Carlo simulations to ensure sufficiently equilibrated thermodynamic properties and compositions (SI Note 4, Fig. S13B).
We 
simulate the lattice liquid with trained parameters from the mean-field model and evaluate it's ability to classify surfaces. The designed mean-field interactions are rescaled to account for the number of nearest-neighbors to parameterize this lattice liquid. We find empirically that decreasing temperature (or increasing $\beta$) sharpens the decision boundary in the lattice model (Fig. S13A), and all test data shown is at $\beta=2$ in Fig. \ref{fig:lattice}.

Using liquids trained on a range of decision boundaries reported in Figs \ref{fig:linear}-\ref{fig:complex_boundaries}, we parameterize and sample the equilibrium configurations of the 3D lattices. Overall, we find that lattice liquids broadly encode  similar classification boundaries as their mean-field counterparts (Fig. \ref{fig:lattice}B), with a few key differences. Near the decision manifolds, we find that lattice liquids exhibit more continuous variation than the abrupt jumps observed in mean-field liquids. 
By analyzing the phase structure \textit{within} each surface, we find that this blurring likely arises from coexisting, spatially isolated pockets of phases that each comprise different output species (SI Note 5, Fig. S14).
Away from the boundary, output species ratios still reach $10-100\times$ ratios of correct over incorrect species (see Fig. S15). Finally, we find that as the decision boundary increases in complexity, and thus requires more hidden species, the asymptotic classification success in the 3D liquid typically decreases (Fig. \ref{fig:lattice}B, Fig S13A).
The broad agreement between mean-field and 3D lattice liquids supports the generality of our results and motivates experimental testing.

\section*{Discussion}

Across the tree of life, biomolecules in cells can self-organize into membraneless organelles called condensates that regulate biological pathways. Motivated by this fact, we explore the computational capabilities that are embedded in and arise from the physical processes shaping condensation in multicomponent mixtures. We find that multicomponent liquids can recruit distinct molecules (and thus condensates) to surfaces that differ only subtly in their composition of surface-resident ``input'' molecules. This high-dimensional surface classification is offered as a model of how cells might assemble transcriptionally active condensates at certain genetic loci (with a particular combination of DNA-bound transcription factors that recruits RNA Polymerases) but repressive ones at other DNA surfaces (with a different combination of transcription factors that recruits repressors). The function of a condensate is therefore driven by the recruitment of a single surface-specific ``output'' molecule. We note that our framework can be easily adjusted to model other pathways where condensates may need to recruit more complex compositions of functional output species, as illustrated in Fig. S16. Together, our work suggests that emergent condensation in multicomponent liquids like the cellular milieu can drive  information processing that may be necessary for regulating complex biological functions.

We show that inclusion of hidden species---molecules that shape condensation but do not drive downstream function---expands \textit{expressivity} \citep{vazirani_approximation_2013}, i.e., the ability to encode increasingly complex classification boundaries. We find that hidden species improve expressivity through encoding novel phases that differ in composition of hidden molecules but still recruit the same \textit{functional} output species. The role of such species could be played by different coactivators that recruit the same polymerases to drive gene activity \citep{pei_transcription_2025}, varying co-receptors and adaptor proteins that recruit the same downstream kinase to membranes to propagate signaling cascades \citep{case_regulation_2019}, and more generally by regulatory molecular cascades. In addition, hidden species simultaneously facilitate \textit{adaptability} by allowing reuse of the same molecular interaction networks, including even purely random ones, to perform distinct tasks (Fig. \ref{fig:reversoir-tuning}A-B) simply by changing makeup of the cellular milieu. This adaptability loosely mimics cell-type specific expression, in which cellular compositions can use the same genetically-encoded molecular ensemble to drive different gene programs with the same functional molecular output species---a feature that emerges in other multicomponent biomolecular networks \citep{antebi_combinatorial_2017,klumpe_context-dependent_2022,parres-gold_contextual_2025}. More generally, the features of multicomponent phase separation naturally provide cells with regulatory knobs such as changing composition (by expression) or interactions (by post-translational modifications) to leverage condensate-mediated computations.
Finally, since condensates \textit{in vivo} are typically characterized by visualizing only a subset of hidden and/or output species, it is possible that (a) condensates that appear similar (by hidden species) could carry out distinct functions (by recruiting distinct output species that are not visualized), and vice versa (b) condensates that appear distinct (by hidden species) could still perform similar functions (by recruiting similar output species that are not visualized). Together, our work advocates for an expanded view, including through the ``hidden-output'' axis, of how and where function can be encoded by condensates in biological systems. 

We show that increasing hidden species generally improves the precision of classification. To better contextualize this computational power, we outline in SI Note 3 an idealized construction, subject to key assumptions, that maps the essential physics underlying our surface condensation model to well-understood computational structures, namely, piecewise linear machines combined with winner-take-all dynamics and their corresponding Voronoi cells~\citep{duda_pattern_1966,aurenhammer_voronoi_1991,maass2000computational}. Within this construction, we argue that surface condensates can universally approximate arbitrary decision boundaries through scaling the number of hidden nodes (Fig. S17). Systematically dissecting and exploring the assumptions that underlie this construction is an important area for future work. In our paper, we also find that classification power eventually saturates with more hidden nodes (Fig. \ref{fig:scaling}). This could arise from a limitation in our optimization formulation, including in our choice of loss function or parameter constraints, that may make it difficult to find global minima of the loss landscape. 
Second, the saturation could point to a more fundamental limit that arises from two competing physical constraints in our model: (a) with more species, there is an overall dilution that increases the entropic cost of condensation, and (b) the requirement of liquid-like condensates, i.e., energy scales of order $k_\text{B}T$, limits the enthalpic stabilization that is possible to encode in our simple model of pairwise interactions. While not captured in our simple thermodynamic model, biology points to the need for more complex models that may expand the scope of computations possible through condensation---including through leveraging higher-order interactions such as discrete sticker-spacers or excluded-volume interactions that expand capacity of the underlying free-energy landscape \citep{choi_physical_2020,bhandari_structurefunction_2021,cappa_phase-field_2025}, multimerization domains that reduce entropic costs of demixing \citep{rana_asymmetric_2024}, and more generally, out-of-equilibrium reaction cascades that provide additional axes for tunable multiphase behavior.

The balance of entropy-energy trade-offs direct surfaces with differing input compositions to recruit distinct condensates and behave as a classifier. 
Our model has partial parallels to well-known architectures in inference---for example, the free energy governing phase behavior in our model resembles that of a Hopfield network \citep{hopfield_neural_1982,braz_teixeira_liquid_2024}.  Our model more closely resembles Boltzmann machines \citep{ackley_learning_1985,chalk_learning_2024} in that we exploit hidden species to encode more complex stimuli-response behaviors, i.e., higher expressivity \citep{raghu_expressive_2017,murugan_could_2025}. A key distinction, more generally, is that condensation is typically collectively encoded and does not directly enable sequential information processing characteristic of modern deep learning networks. However, unlike these networks, condensation, and many other forms of equilibrium molecular computation, typically expend very little energy and are compatible with massively parallel computation---the same milieu of molecules can classify multiple spatially separated surfaces.

While we focus on classification, recent studies argue for broader computational capabilities embedded in multicomponent liquids. 
For example, \citep{braz_teixeira_liquid_2024} explores the capacity of condensates to store and retrieve compositions as stable phases (or memories) analogous to Hopfield models trained with the Hebbian rule, identifying key roles for three-body interactions for effective memorization. 
More recently, \citep{chalk_learning_2024}  train molecular parameters of 3D lattice liquids for multiple tasks, showing that equilibrium multicomponent liquids can form complex spatial architectures, drive Hopfield-like retrieval of memories upon contacting a patterned surface, and 
perform MNIST digit classification.
In the MNIST classification task, our model performance saturates ($\sim85\%$), potentially hinting at limitations in the physics of condensation and/or in the choice of data encoding. More generally, it would be valuable to delineate and contrast the principles and limits of computations performed by different physical systems with and in addition to condensation---for example,
dimerization networks \citep{parres-gold_contextual_2025}, self-assembly \citep{foley2025membraneassociatedselfassemblycellulardecision, murugan_multifarious_2015,zhong_associative_2017,woods_diverse_2019,evans_pattern_2024,winfree1998algorithmic}, mechanical systems \citep{fletcher2010cytoskeleton,yasuda_mechanical_2021}, and stochastic biomolecular reactions \citep{winfree2019chemical,soloveichik2008computation,floyd2025limits}.

We characterize the computational capabilities of programmed multiphase fluids that are trained through gradient-descent based global optimization routines. While we focus on the system's capacity for classification and not the biological plausibility of the learning rule, the ways by which molecular networks can learn, potentially autonomously, or be trained represents an important area for investigation. For instance, \citep{chalk_learning_2024} train lattice liquids using sleep-wake training rules, a set of spatially local rules based on competition between Hebbian learning and anti-Hebbian unlearning as in classical Boltzmann machines \citep{ackley_learning_1985}.
In conjunction with other recent studies, these suggest that molecular networks can be trained \textit{in situ} through physical learning rules that directly modify parameters like interactions or composition \citep{chalk_learning_2024,poole_chemical_2017,stern_learning_2023,poole_detailed_2022}. In particular, we show that only changing levels of hidden and output species in the reservoirs of trained fluids---a variable amenable to modification in living systems---enables adaptation to new tasks (Fig. \ref{fig:reversoir-tuning}A). If the levels of reservoir species could be directly regulated by condensate formation---for example through engineered genetic feedback circuits where condensation of output species alters gene expression of reservoir species \citep{zhang2025phase}---this would permit learning over longer time-scales. Together, these hint at biologically plausible mechanisms for autonomous and continual learning in biomolecular fluids without any electrical computers in the loop. 

Finally, we demonstrate concordance of our mean-field designs with function in a 3D lattice model that explicitly captures spatial correlations that are relevant \textit{in vitro}. In our framework, the surface is treated implicitly, characterized instead by the host of surface-resident inputs. Recent theories have begun to relate the explicit physical properties of biological scaffolds, both DNA sequence aspects \cite{shrinivas_enhancer_2019,morin_sequence-dependent_2022} and reversible membrane binding  \citep{rouches_surface_2021,zhao_2021_thermodynamics}, to the resulting phase behavior.
These models describe how condensates can wet (condense on a lower-dimensional surface as well as in the bulk), pre-wet (condense only on a lower-dimensional surface), or drive coupled phase transitions between the scaffold along with the soluble proteins. Designing over such models, in turn, could enable more expressive computations by directly optimizing surface features and using complex phase behaviors.  
Towards experimental realization, promising avenues include (a) designed DNA \citep{abraham_nucleic_2024,evans_pattern_2024,jacobs_assembly_2025} molecules, along with surface-functionalized or immobilized DNA strands, and (b) emerging synthetic biology approaches \citep{dai_engineering_2023,qian_synthetic_2022,wei_nucleated_2020,lyons_functional_2023} that combine genetic reporter systems with coexpression of phase-separation proteins. More generally, the confluence of machine-learning, physics-based models, and multiplexed experimental techniques will inform future opportunities to dissect as well as design biological computation through condensation.

\subsection*{Limitations of the study}

In this paper, we explore how the emergent physics underlying condensation in multicomponent liquids can classify surfaces with distinct compositions. Towards this, we introduce a simple mean-field description of liquids comprising molecules of identical size with pairwise interactions. As a consequence, we are unable to explore the computational capacity and constraints that are afforded through explicit consideration of complex molecules---including effects of polydispersity, higher-order interactions, and anisotropic molecular architectures that all typify biomolecules. We focus on mean-field surface condensation from a large (infinite) cellular reservoir that we posit maintains any learned chemical potential. Thus, a limitation of the model is that molecules are maintained at well-mixed compositions inside the surface and intra-surface demixing is not explicitly studied. Thus, further studies are required to explicitly study the effects of finite copy numbers, surface-surface competition, and dynamics of nucleation. Since our model does not explicitly specify the mechanisms by which the reservoir can be maintained, either in biological or physical systems,  new models that explicitly consider specific reservoirs will provide insights and offer pathways to design molecular networks more amenable to experimental instantiation. 

\section*{Acknowledgments}

We thank Francesco Mottes, Ryan Krueger, Mason Rouches, William Jacobs, Suriyanarayanan Vaikunthanathan, David Zwicker, Serena Carra, and members of the Murugan, Brenner, Winfree, and Shrinivas labs for helpful discussions on this manuscript.
E.V.H. and K.S. acknowledge support from NSF NRT 2021900 - Synthesizing Biology Across Scales and Northwestern University.
A.Z. and M.P.B. acknowledge support from NSF AI Institute of Dynamic Systems 2112085.
E.W. and C.C. acknowledge support from NSF CCF/FET 2008589 and 2212546.
A.M acknowledges support from the National Science Foundation
through the Center for Living Systems (grant no. 2317138) and DMR-2239801.
K.S. acknowledges helpful discussions related to this paper at the Kavli Institute for Theoretical Physics (KITP) workshop: Physical Principles Shaping Biomolecular Condensates supported in part by NSF PHY-2309135 and the Gordon and Betty Moore Foundation Grant No. 2919.02.
This research was supported in part by grants from the NSF (DMS-2235451) and Simons Foundation (MPS-NITMB-00005320) to the NSF-Simons National Institute for Theory and Mathematics in Biology (NITMB). The computations in this paper were, in part, run on the FASRC Cannon cluster supported by the FAS Division of Science Research Computing Group at Harvard University, and in part, through the computational resources provided by the Quest high performance computing facility at Northwestern University.

\section*{Code availability}

We make code and figure data available via the
following GitHub repository: \url{https://github.com/shrinivaslab/2025_zentner_multiphase_classification}

\section*{Bibliography}

\bibliography{main}

\onecolumn
\newpage

\renewcommand{\thefigure}{S\arabic{figure}}
\setcounter{figure}{0}  %

\renewcommand{\thetable}{S\arabic{table}}
\setcounter{table}{0}  %

\renewcommand{\theequation}{S\arabic{equation}}
\setcounter{equation}{0}  %

\captionsetup*{format=largeformat}
\newcommand{\secondtitle}{%
    \begin{flushleft}
        {\fontsize{14}{16}\bfseries \color{titlecolor} Supplementary Information: Combinatorial decision-making driven by multicomponent surface condensates}\\[1em]
        \vspace*{0.2cm}
    \end{flushleft}
}

\secondtitle

\section{Model A Dynamics}\label{sec:SI-Dynamics}
\subsection*{Deriving Mean-Field Dynamics}

In this section, we derive the mean-field dynamics for an effective interaction matrix $\chi$ and reservoir potential $\vec{\mu}_\text{res}$ used in the manuscript. For simplicity in deriving these dynamics, we make no distinction between input, output, and hidden species, and we assume that inputs can also exchange with the reservoir; we relax this assumption at the end of the derivation by setting their mobilities to $0$. Accordingly, the surface exchanges with an infinite reservoir held at a chemical potential vector $\vec{\mu}''_\text{res}$. To accommodate indexing the solvent when required, we extend objects ($\vec{\phi}$, $\chi$, $\vec{\mu}''_\text{res}$) to have an additional $0$ index to denote the solvent, such that when the solvent is included as an explicit variable, we index from $0$ to $N$ rather than from $1$ to $N$. We therefore define the length $N+1$ chemical potential vector $\vec{\mu}'_\text{res}\equiv0\circ\vec{\mu}^{(\text{in})}_\text{res}\circ\vec{\mu}_\text{res}$, where $\mu'_{\text{res},i}=\mu''_{\text{res},i}-\mu''_{\text{res},0}$ is the reservoir potential of species $i$ relative to the solvent. Note that the input chemical potential $\vec{\mu}^{(\text{in})}_\text{res}$ is irrelevant to the dynamics since these species do not exchange with the reservoir. Written in this way, the vector $\vec{\mu}_\text{res}$ is the same as in the manuscript.

When the concentration vector $\vec{\phi}$ is treated as a function of space, the Landau-Ginzburg Hamiltonian describes the effective free energy of the surface as
\begin{equation} \label{eq:LGH}
    \beta\mathcal{H}=\int dV\left[\Omega_\text{G}\left(\vec{\phi},\chi\right)+\frac{\kappa}{2}\left(\nabla\vec{\phi}\right)^2\right]
\end{equation}
where $\left(\nabla\vec{\phi}\right)^2=\sum_{i=0}^{N}\sum_{n=1}^{d}\left(\partial_{x_n}\phi_i\right)^2$. 
The $\kappa$
term penalizes spatial gradients in the homogeneous system. The grand-potential $\Omega_\text{G}$  describes an effective free-energy  for the surface exchanging with an infinitely large reservoir and includes a Lagrange-multiplier term that enforces overall conservation of mass. For an open system at some initial composition, the relaxation to steady-state is driven by an exchange of species without conserving counts. Near equilibrium, Model A dynamics \cite{hohenberg_theory_1977} characterizes these relaxation dynamics as purely downhill: the decrease in the overall free energy of the system is, to a first approximation, driven by linear gradients of the free energy with respect to the system's composition. Since we assume our surface remains well-mixed, we neglect the contributions from spatial gradients. We also
neglect the interfacial energy between the surface volume and the reservoir, assuming that the free-energy is dominated by bulk contributions. Thus, the temporal evolution of the average volume fraction $\phi_i$ of species $i$ within the system can be written as
\begin{align}
        \frac{\partial\phi_{i}(t)}{\partial t}&= - \sum_{j=0}^{N}D'_{ij}\frac{\delta(\beta \mathcal{H})}{\delta \phi_{j}}+\eta_i(t)\\
    &= -\sum_{j=0}^{N}D'_{ij} \frac{\partial \Omega_\text{G}}{\partial\phi_j}+\eta_i(t)
\end{align}
where $\textbf{D}'$ is the mobility matrix, again with index 0 corresponding to the solvent, that sets the rate of exchange between the system and reservoir. The above equation reflects the fact that, rather than purely decreasing energies, Model A dynamics also explicitly permits modeling the effect of temporally uncorrelated thermal fluctuations, described by $\eta_i$, such that
\begin{equation}
    \langle \eta_i(t)\eta_j(t')\rangle=2\beta D'_{ij}\delta_{ij}\delta^{(d)}(t-t').
\end{equation}
We mention this term for completeness but focus on the purely deterministic limit in this paper. Thus, the effective dynamics of the system's composition as it exchanges with the reservoir are given by
\begin{equation}
    \frac{d \phi_i}{dt} \approx -\sum_{j=0}^{N}D'_{ij} \frac{\partial\Omega_\text{G}}{\partial\phi_j}
    \label{eq:model-A-explicit}
\end{equation}

In writing our solute dynamics in the main manuscript, we treat the solvent implicitly. We first show below that this is tacit to assuming that the solvent molecules rearrange and equilibrate quickly to any small changes in solutes. The free energy is
\begin{equation}
    \Omega_\text{surface}=\sum_{i=0}^{N}\phi_i\log{\phi_i}+\frac{1}{2}\sum_{i=0}^N\sum_{j=0}^N\phi_i\chi_{ij}\phi_j-\vec{\mu}''_\text{res}\cdot\vec{\phi}
\end{equation}

There is still a constraint $\sum_{i=0}^{N} \phi_i=1$. We derive the dynamics of the system assuming Model A dynamics, where the mobility matrix $\textbf{D}'$ is determined by first assuming that the mobility follows Fick's law of diffusion in the dilute limit, such that $D'_{ij}=d_i\phi_i\delta_{ij}$ \cite{shrinivas_multiphase_2022}, and imposing the constraint via a Lagrange multiplier, so that $\Omega_\text{G} = \Omega_\text{surface} - \lambda \left(\sum_{i=0}^{N} \phi_i - 1\right)$. The dynamics from eq. \ref{eq:model-A-explicit} are therefore

\begin{equation}
    \frac{\partial \phi_i}{\partial t} = -d_{i}\phi_i\left(\frac{\partial\Omega_\text{surface}}{\partial \phi_i}  - \lambda\right)=-d_{i}\phi_i\left[\beta\left(\mu''_i-\mu''_{\text{res},i}\right)-\lambda\right]
\end{equation}
where
\begin{equation}
    \beta\mu''_i = 1 + \log{\phi_i}+\sum_{j=0}^{N}\chi_{ij}\phi_j
\end{equation}
The constraint is given by $\frac{d\Omega_\text{G}}{d\lambda}=0$, which once differentiated is
\begin{equation}
    \sum_{i=0}^N\frac{\partial \phi_i}{\partial t}=-\sum_{i=0}^N d_{i}\phi_i\left[\beta\left(\mu''_i-\mu''_{\text{res},i}\right)-\lambda\right]=0\qquad\Longrightarrow\qquad \lambda=\frac{\sum_{j=0}^Nd_{j}\phi_j\beta\left(\mu''_j-\mu''_{\text{res},j}\right)}{\sum_{k=0}^Nd_{k}\phi_k}
\end{equation}
Substituting $\lambda$ and grouping the terms gives the dynamics
\begin{equation}
    \frac{\partial\vec{\phi}}{\partial t}=-\textbf{D}'\beta\left(\vec{\mu}''-\vec{\mu}''_{\text{res}}\right),\qquad\qquad D'_{ij} = d_i\phi_i\left( \delta_{ij} - \frac{d_j\phi_j}{\sum_{k=0}^N d_k\phi_k}\right)
\end{equation}

We now apply model assumptions. We assume first that $d_{0}\gg d_{i}$ for $i>0$ (implying that the solvent relaxes much faster than the solutes). Second, we assume that there are negligible effective solute-solvent interactions (taking $\chi_{0j}=0$ for all $j$). This assumption implies that each solute is ``well-solvated'' and thus cannot phase separate on its own, i.e., when present in a binary mixture of only solute and solvent.  In a fluid with $N$ solutes and a solvent, this assumption maintains the same scaling of free parameters with increasing species ($\sim N^2$) but restricts the absolute number of such parameters with  $N$ fewer $\chi$ parameters to train. In this paper, we focus on this restricted limit where individual solutes are well-solvated, and in fact, are treated as effectively inert. Correspondingly, as explained in the lattice description (see SI Note \ref{sec:SI-Lattice}), we enforce $\chi_{0j}=0$ by setting both $\epsilon_{jj}=\epsilon_{0j}=0$. Thus, the solutes we design exhibit multiphase behavior only through interaction with other types of solutes. This bears direct parallels with the restriction of individual neurons or cells forming associative connections with other cells but no self-connections as originally formulated in Hopfield models \citep{hopfield_neural_1982}. 

We note that this approximation poses challenges for experimental realization since it requires free-tuning of inter-species interactions while keeping self-interactions negligible. This poses issues when considering systems such as DNA nanostars that operate primarily through attractive, sequence-based interactions, where positive (or repulsive) $\chi_{ij}$ cannot be generically invoked without also allowing self interactions (i.e. pairs of species that have strong self-interaction but weak cross-interactions). In general, relaxing this assumption, allows $N$ free parameters ($\chi_{0j}$) to be freely tuned and could, in principle, increase expressivity of the solution. One such instantiation, in terms of the microscopic contact energies introduced in eq. 7, would stipulate that microscopic solvent interactions are zero ($ \epsilon_{0j}=0\ \forall j $) and solute self-interactions $\epsilon_{jj}$ are thus freely tuneable. In this paper, for simplicity, we focus on the theoretical limit of only inter-species interactions.

We can express the system dynamics in terms of the solute concentrations,
\begin{align}
    \frac{\partial\phi_0}{\partial t}&\approx\sum_{j=1}^{N}\beta d_j\phi_j\left(\mu'_j-\mu'_{\text{res},j}\right)\\
    \frac{\partial\phi_i}{\partial t}&\approx-\beta d_i\phi_i\left(\mu'_i-\mu'_{\text{res},i}\right),\qquad i>0
    \label{eq:solvent-dynamics}
\end{align}
or
\begin{equation}
    \frac{\partial\vec{\phi}}{\partial t}\approx-\textbf{D}_\text{f}'\beta\left(\vec{\mu}'-\vec{\mu}'_{\text{res}}\right),\qquad\qquad \left(\textbf{D}'_\text{f}\right)_{ij} = \begin{cases}
        d_j\phi_j(\delta_{0j}-1),& i=0\\
        d_i\phi_i\delta_{ij},&i>0
    \end{cases}
    \label{eq:model-A_fast}
\end{equation}
where
\begin{equation}
    \beta\mu'_i = \beta\left(\mu''_i-\mu''_{0}\right)=\log{\phi_i}-\log{(1-\phi_T)}+\sum_{j=1}^{N}\chi_{ij}\phi_j
    \label{eq:mu_analytic}
\end{equation}
is the intrinsic (non-dimensionalized) chemical potential vector and $\phi_T = \sum_{i=1}^{N} \phi_i$ is the total solute volume fraction (i.e. omitting the solvent). The equations for $i>0$ therefore form a matrix equation that is approximately diagonal in the limit of fast solvent dynamics.

Furthermore, this equation provides the solute dynamics used throughout this paper when we set $d_i=0$ for $1\leq i\leq N_\text{in}$ and $d_i=d$ for $i>N_\text{in}$, where $d$ is a constant whose value does not affect the steady state. In this case, the input species are confined to the box, and the solute dynamics can be further simplified to be written only in terms of the $(N_\text{out}+N_\text{h})\times(N_\text{out}+N_\text{h})$ lower block of the full mobility matrix $\textbf{D}'_\text{f}$; this truncated matrix, which we label $\textbf{D}$, is the mobility matrix used in eq. 8. Likewise, because the solvent is being treated implicitly and the inputs cannot exchange with the reservoir, the reservoir can be described by the length-$(N_\text{out}+N_\text{h})$ chemical potential vector $\vec{\mu}_\text{res}$ that includes only the output and hidden species---the convention used throughout the paper.

The solvent equation ($i=0$) follows from---and implies---the constraint $\phi_0=1-\phi_T$, since the solvent balances the flux of the solutes.  Empirically, we also find that relaxing this assumption (by taking the solvent to have finite mobility compared to the solutes) essentially does not alter steady-state (or ``performance'') of trained multiphase fluids (Fig. \ref{fig:sup-dynamics}).

\subsection*{Parameterizing the Dynamics}

This section offers a parametrization for the case where we have fast solvent dynamics. Again, for the simplicity of the derivation, we consider all solutes (including input species) as mobile and therefore also take the reservoir potential vector to again be $\vec{\mu}'_\text{res}$, which is measured with respect to the solvent chemical potential.

\subsubsection*{Proposed Parametrization}

In practice, the logarithmic terms in eq. \ref{eq:mu_analytic} become unstable for $\phi_i\to0$ or $\phi_T\to1$, making even the simplified model in eq. \ref{eq:model-A_fast} difficult to integrate. Eq. \ref{eq:mu_analytic} therefore suggests the following parametrization:
\begin{equation}
    x_i = \log{\left(\frac{\phi_i}{1-\phi_T}\right)}\qquad\Longleftrightarrow\qquad\phi_i = \frac{\exp{(x_i)}}{1+\sum_{j=1}^{N}\exp{(x_j)}} %
    \label{eq:x_param}
\end{equation}
In this parametrization, the chemical potential vector simplifies to
\begin{equation}
    \beta\vec{\mu}'(\vec{x}, \chi)=\vec{x}+\chi\vec{\phi}(\vec{x})
    \label{eq:mu_x_analytic}
\end{equation}
This parametrization has the benefit that it spans all real numbers, thereby transforming the problem from a system of ODEs with constraints $\phi_i>0$ and $\phi_T<1$ to one that is unconstrained.

The inverse Jacobian of this transformation is
\begin{equation}
    (\textbf{J}^{-1})_{ij}=\frac{\partial x_i}{\partial\phi_j} = \frac{1}{\phi_j}\delta_{ij}+\frac{1}{1-\phi_T}
\end{equation}
As a result, the time-evolution of $\vec{x}$ is governed by
\begin{equation}
    \frac{d\vec{x}}{dt}
    =\sum_{j=1}^{N} \frac{d\vec{x}}{d\phi_{j}} \frac{d\phi_{j}}{dt}=
    - \sum_{j,k=1}^{N}\frac{d\vec{x}}{d\phi_{j}} \left(\textbf{D}_\text{f}'\right)_{jk}\frac{\partial\Omega_\text{G}}{\partial \phi_{k}} = -\beta\textbf{J}^{-1}\textbf{D}_\text{f}'\left[\vec{\mu}'(\vec{x},\chi)-\vec{\mu}'_\text{res}\right]
    \label{eq:model_A-in-x}
\end{equation}
where $\vec{\mu}'(\vec{x},\chi)$ is as defined in eq. \ref{eq:mu_x_analytic}.
The transformed mobility matrix has components
\begin{equation}
    \left(\textbf{J}^{-1}\textbf{D}_\text{f}'\right)_{ij}=\delta_{ij} + d_j\exp{(x_j)}
\end{equation}

In the case of fixed input concentrations, we take $d_{i}=0$ for $1 \leq i \leq N_\text{in}$, such that $\phi_{i}'(t)=0$ for the input species. Note that in the new parametrization, the input parameters $\vec{x}_\text{in}=\vec{x}_\text{in}(\vec{\phi})$ are no longer constant. However, the total input concentration of the surface is fixed at $\sum_{1 \leq j\leq N_\text{in}} \phi_{j} \equiv \phi_{\text{in},T}$, and thus the non-input $x_i$ can be evolved without needing to simultaneously evolve the input $x_i$, since
\begin{equation}
    1+\sum_{j=1}^{N}\exp{(x_j)}=\frac{1+\sum_{j>N_\text{in}}\exp{(x_j)}}{1-\phi_{\text{in},T}}
\end{equation}
This relation allows for the non-input components of $\vec{\phi}(\vec{x})$ in eq. \ref{eq:mu_x_analytic} to be written independent of the input $x_i$ coordinates. In turn, eq. \ref{eq:model_A-in-x} depends only on the (fixed) values of $\vec{\phi}_\text{in}$ and the non-input parameters $x_i$.

\newpage
\section{Training the model\label{sec:SI-Training}}
\subsection*{Learning Rules}

We optimize over both the $\chi$ matrix and the reservoir chemical potential $\vec{\mu}_\text{res}$, with the target being an enrichment in the desired output species for a given input concentration vector $\vec{\phi}_\text{in}$. Unlike in the case of an artificial neural network, where there are no physical constraints on the weights assigned to the hidden layers, a surface is constrained to have total volume fraction 1. Therefore, unlike the unconstrained problem, the cross-entropy of the output vector is not a favorable loss function, because imposing that the desired output concentration be as close as possible to 1 depletes the volume fraction available to the hidden species, thereby limiting their effectiveness. We require instead that the following criteria be captured by our loss function:
\begin{enumerate}
    \item The final concentration of the desired output species should be above some threshold value $\phi_\text{max}=A/N$, where $N$ is the total number of particle species and $A$ is a value to be specified.

    \item The final concentrations of the undesired output species should be below some threshold $\phi_\text{min}=B/N$, where $B$ is a value to be specified.
\end{enumerate}
These two criteria in turn enforce that the ratio of desired to undesired outputs should be above a set threshold $A/B$, and that this ratio is attained with a sufficiently enriched output species. In principle, for a steady-state concentration vector $\vec{\phi}$ where the $j$'th output is desired to be enriched, the above criteria could have been satisfied by using, instead of $l_j$ in eq. 9, the function
\begin{equation}
    \tilde{l}_j(\chi, \vec{\mu}_\text{res})=-\sum_{k\neq j}\log\left[\frac{\min\left(1,\phi_{\text{out},j}/\phi_\text{max}\right)}{\max\left(1,\phi_{\text{out},k}/\phi_\text{min}\right)}\right]
    \label{eq:loss_first_try}
\end{equation}
which enforces that the ratio in the argument of the log be as close as possible to 1, and therefore that $\phi_{\text{out},j}/\phi_{\text{out},k}>\phi_\text{max}/\phi_\text{min}$, while the numerator and denominator are independent of $\vec{\phi}_\text{out}$ when their values are above and below (respectively) their corresponding threshold values. This function indeed allows for successful decision boundaries to be sculpted. In practice, we further adjust this loss empirically to improve training. In particular, we use the fact that
\begin{align}
    -\log{\left(\frac{\min{(1,x)}}{\max{(1,y)}}\right)}&=-\log{(\min(1,x))}+\log{(\max{(1,y)})}\\
    &=-\log{\left(1-\max{(0,1-x)}\right)}+\log{\left(1+\max{(0,y-1)}\right)}\\
    &=\log{\left(1+\max{(0,1-x)}\right)}+\log{\left(1+\max{(0, y-1)}\right)}+\mathcal{O}(\Delta x^2)
\end{align}
where $\Delta x = 1-x$ is small. Motivated by this expansion, and combined with empirical tests, we use the loss function
\begin{equation}
    l_j(\chi, \vec{\mu}_\text{res})=\log\left(1+A\max\left(0,1-\frac{\phi_{\text{out},j}}{\phi_\text{max}}\right)\right)+\sum_{k\neq j}\log\left(1+B\max\left(0, \frac{\phi_{\text{out},k}}{\phi_\text{min}}-1\right)\right)
\end{equation}
We've deviated from the expansion of eq. \ref{eq:loss_first_try}  by dropping the factor of $(N-1)$ that would otherwise be on the $i$-dependent logarithm and also by introducing the hyperparameters $A=\phi_\text{max} N$ and $B=\phi_\text{min}N$ as prefactors in the logarithms. We find that this choice for the loss function gives strong results near decision boundaries, and we therefore favor $l_j$ over $\tilde{l}_j$ for training our molecular networks. Substituting the expressions for $A$ and $B$ results in the form of the loss function in the main text. As discussed in the main text, we further constrain the magnitude of $\chi$ such that $|\chi_{ij}|<15$. Following eq. 7 and the lattice definition of 18 nearest neighbors (SI Note \ref{sec:SI-Lattice}), this constraint corresponds to magnitudes of $\epsilon_{ij}\sim k_\text{B}T$ for individual molecular bonds and is consistent with liquid-like mixtures where molecules can rearrange with thermal fluctuations.

\subsection*{Hyperparameter choices} We minimize $\mathcal{L}$ with respect to $\chi$ and $\vec{\mu}_\text{res}$ over several thousand training epochs using an RMSProp algorithm from the Optax library \cite{bradbury_jax_2018,deepmind_deepmind_2020} with an initial learning rate of 0.01, followed by several thousand more epochs with a learning rate of 0.001 to improve convergence. We use 5000 training points and a mini-batching scheme where $n_\text{batch}=128$ randomly selected training points are evaluated at each epoch. Once trained, we construct a validation set of 500 data points to validate the classifier. For a given number of hidden species, we perform this optimization procedure over 15 initial guesses in the loss landscape. Using the definition of success $S_\text{c}$ in eq. 13, the trained model performance is evaluated on the validation set and the best performing model is subsequently applied to an independent test set (of same size as the validation set) and depicted in figures.

\newpage
\section{Decision boundary\label{sec:SI-Decision}}
To understand the constraints on the shapes the decision boundary can encode in our model, we first provide insights with a simplified model only 2 inputs and 2 output species, and expand in later sections to explore the effect of adding more species. Without loss of generality, and consistent with our design articulated in the main text, we assume $\beta=1$ below.

\subsection*{2 input + 2 output + 0 hidden species}
In the case of mixtures with 2 input species (with species labels $i=1,2$) and 2 output species (with species labels $i=3,4$), the concentration vector is given by $\vec{\phi} = (\phi_{\text{in},1},\phi_{\text{in},2},\phi_{\text{out},1},\phi_{\text{out},2})$. The decision boundary is defined as the manifold where output species are equally recruited, with $\phi_{\text{out},1}=\phi_{\text{out},2}\equiv\phi_o$. For a trained mixture with parameters $(\chi,\vec{\mu}_\text{res})$, the steady-state conditions of eq. \ref{eq:model-A_fast} along this manifold are
\begin{align}
    \mu^\text{res}_{\text{out},1} &= \log(\phi_{o}) - \log(1-\phi_T) + \sum_{j=1}^4 \chi_{3j} \phi_j\\
    \mu^\text{res}_{\text{out},2} &= \log(\phi_{o}) - \log(1-\phi_T) + \sum_{j=1}^4 \chi_{4j} \phi_j
\end{align}
Defining $\Delta\mu_{\text{res},\text{out}} = \mu^\text{res}_{\text{out},1}-\mu^\text{res}_{\text{out},2}$, the difference of the above two equations is independent of the specific value of the output concentrations,
\begin{equation} \label{eq:2-2-linear}
    \Delta\mu_{\text{res},\text{out}} = (\chi_{31} - \chi_{41})\phi_{\text{in},1} +  (\chi_{32} - \chi_{42})\phi_{\text{in},2}
\end{equation}
and defines a decision manifold across which the recruited output species changes from species 1 to species 2. Recall that all diagonal elements of $\chi$ are 0 by definition, and the output-output interaction contributions cancel exactly thanks to the symmetry $\chi_{ij}=\chi_{ji}$. The decision boundary described in eq. \ref{eq:2-2-linear} is therefore exactly linear in the inputs. Fig. \ref{fig:sup-theory}A shows two theoretically computed linear boundaries using eq. \ref{eq:2-2-linear}.

\subsection*{$N_\text{in}$ input + $N_\text{out}$ output + 0 hidden species} \label{sub: m-n-0}

Generalizing to $N_\text{in}$ input species (with species labels $i=1,\dots,N_\text{in}$) and $N_\text{out}$ output species (with species labels $i=N_\text{in}+1,\cdots,N_\text{in}+N_\text{out}$, such that $\phi_{N_\text{in}+n} = \phi_{\text{out},n}$ for $n=1,\dots,N_\text{out}$), we see that the decision boundary between any two output species at equal concentrations ($\phi_{\text{out},n} \equiv \phi_i = \phi_o = \phi_j \equiv \phi_{\text{out},m}$) can similarly be written as
\begin{equation} \label{eq:m-n-linear-all}
    \Delta\mu_{\text{res},\text{out}}^{(n,m)} =  \sum_{k=1}^{N_\text{in}}(\chi_{ik} - \chi_{jk})\phi_{\text{in},k} + \sum_{\substack{k=N_\text{in}+1\\ (k\neq i,j)}}^{N_\text{in}+N_\text{out}}(\chi_{ik} - \chi_{jk})  \phi_{k} (\vec{\phi}_\text{in},\vec{\mu}_\text{res},\chi)  
\end{equation}
where $\Delta\mu_{\text{res},\text{out}}^{(n,m)} = \mu^\text{res}_{\text{out},n}-\mu^\text{res}_{\text{out},m}$. The first term is the generalization of eq. \ref{eq:2-2-linear} to sum over all inputs, and second term is a sum over the remaining (non-boundary) output species. Unlike the previous case, here we treat the concentrations of non-boundary output species as nonlinear functions of input species, and as such they could encode more complex boundaries. Strictly speaking, as the energy landscape may have multiple local minima, the final output concentrations may not be uniquely determined by the input concentrations; however, in this work, training appears to avoid this situation for the cases we have tested, in part, because during training, the initial hidden/output concentrations are randomly assigned at different epochs.
When training is successful, target surfaces typically enrich a single output species with the others being depleted, and the resulting output concentrations will have $\phi_k \ll1$ for $k\neq i,j$. Since the $\chi$ matrix has components that are constrained to be $|\chi_{ij}|<\chi_\text{max}$, the second term in eq. \ref{eq:m-n-linear-all} should therefore be negligible for solutions obeying the loss criterion. As a result, trained mixtures of surface condensing species form generalized linear boundaries as a function of input species concentrations:
\begin{equation} \label{eq:m-n-linear}
    \Delta\mu_{\text{res},\text{out}}^{(n,m)} \approx  \sum_{k=1}^{N_\text{in}} (\chi_{ik} - \chi_{jk})\phi_{\text{in},k}
\end{equation}
Fig. \ref{fig:sup-theory}B shows two theoretically computed boundaries using eq. \ref{eq:m-n-linear} for $N_\text{in}=2$, $N_\text{out}=3$, compared with numerical results. This linearity breaks down in the vicinity of points in the input space where multiple classes meet, in which case there are more than two relevant output species, and the second term in eq. \ref{eq:m-n-linear-all} is no longer negligible. The decision boundaries therefore resemble hyperplanes far from regions of multiclass intersection, but can also potentially be nonlinear.

\subsection*{2 input + 2 output + 1 hidden species}

In general, with the inclusion of hidden nodes, the equations become analytically intractable. Here we consider the inclusion of a single hidden species and show that this is sufficient for producing nonlinear decision boundaries. We consider a concentration vector as defined above, where indices 1 and 2 correspond to input species, 3 and 4 correspond to output species, and an additional component (species index 5) corresponds to the hidden species. The concentration vector then reads as $\vec{\phi} = (\phi_{\text{in},1},\phi_{\text{in},2},\phi_{\text{out},1},\phi_{\text{out},2},\phi_\text{h})$. When $\phi_{\text{out},1}=\phi_{\text{out},2}=\phi_o$, the decision boundary follows eq. \ref{eq:2-2-linear} with the the modification
\begin{equation} \label{eq:2-2-1-first}
    \Delta\mu_\text{res,out} = (\chi_{31} - \chi_{41})\phi_{\text{in},1} +  (\chi_{32} - \chi_{42})\phi_{\text{in},2} +  (\chi_{35} - \chi_{45})\phi_\text{h} (\vec{\phi},\vec{\mu}_\text{res},\chi)
\end{equation}
While the concentration of the hidden species is (assumed to be) an implicit function dependent on input concentrations, the shape of this boundary is generically hard to interpret. The chemical potential of the hidden species is given by
\begin{equation*}
    \mu_\text{h} = \log(\phi_\text{h})-\log{(1-\phi_T)} + \sum_{k=1}^{5} \chi_{5k}\phi_{k}
\end{equation*}
Since $\chi_{5,5}=0$, the sum on the right hand side is independent of $\phi_\text{h}$, and we can therefore isolate for $\phi_\text{h}$ as
\begin{equation}
    e^{\mu_\text{h}}=\frac{\phi_\text{h}}{(1-\phi_{\text{in},T}-2\phi_o)-\phi_\text{h}}\exp{\left(\sum_k\chi_{5k}\phi_k\right)}\qquad\Longrightarrow\qquad \phi_\text{h} = \frac{1-\phi_{\text{in},T}-2\phi_o}{1+\exp{\left(-\mu_\text{h}+\sum_k\chi_{5k}\phi_k\right)}}
\end{equation}
and the decision boundary in the input space thus obeys
\begin{equation}
\label{eq:2-2-1-implicit}
    \Delta\mu_\text{res,out} = (\chi_{31} - \chi_{41})\phi_{\text{in},1} +  (\chi_{32} - \chi_{42})\phi_{\text{in},2} + \frac{(\chi_{35} - \chi_{45})\left((1-2\phi_o)-\phi_{\text{in},1}-\phi_{\text{in},2}\right)}{1+\exp{\left(-\mu_\text{h} + (\chi_{35}+\chi_{4,5})\phi_o)\exp(\chi_{51}\phi_{\text{in},1}+\chi_{52}\phi_{\text{in},2}\right)}}
\end{equation}

Fig. \ref{fig:sup-theory}C shows two nonlinear decision boundaries for systems with a single hidden species, computed theoretically from eq. \ref{eq:2-2-1-implicit}, against numerical results. To see how this rather complex equation permits nonlinear boundaries, it is instructive to look at the following limit when (a) input-output interactions are identical across species, (b) output-hidden interactions are non-zero and different, and (c) hidden-input interactions are strong and of opposing signs. Here, the boundary will primarily be defined by variances in the relative interaction of the hidden species with the two inputs. The features of such a decision boundary in the input plane can be computed as an implicit derivative from the decision boundary since output concentrations at the decision boundaries are low ($\phi_o\approx0$) and hence eq. \ref{eq:2-2-1-implicit} is of the form $f(\phi_{\text{in},1},\phi_{\text{in},2})=0$. As such,
\begin{align}
    \frac{d\phi_{\text{in},2}}{d\phi_{\text{in},1}} &= -\frac{df/d\phi_{\text{in},1}}{df/d\phi_{\text{in},2}}
    =-\frac
    {
    1+g(\mu_\text{h},\vec{\phi}_\text{in})(1+\chi_{51}(1-\phi_{\text{in},1}-\phi_{\text{in},2}))
    }
    {
    1+g(\mu_\text{h},\vec{\phi}_\text{in})(1+\chi_{52}(1-\phi_{\text{in},1}-\phi_{\text{in},2}))
    }
\end{align}
where
\begin{equation}
    g(\mu_\text{h},\vec{\phi}_\text{in}) = \exp{\left(-\mu_\text{h})\exp(\chi_{51}\phi_{\text{in},1}+\chi_{52}\phi_{\text{in},2}\right)}
\end{equation}

While the prefactor functions $g$ are always positive, the second factor of the form $\chi_{5j} (1-\phi_{in,1}-\phi_{in,2})$ for $j\in[1,2]$ can change the sign and magnitude of the whole term depending on a particular choice of parameters. In general, this implies that not only is the decision boundary nonlinear (i.e. non-uniform slope for changing magnitude) but also capable of changing curvatures (i.e. changing of slope signs).

\subsection*{$N_\text{in}$ input + $N_\text{out}$ output + $N_\text{h}$ hidden species}

In general, since $\phi_{\text{out},n}$ and $\phi_{\text{h},m}$ are implicit functions of the input volume fractions $\phi_{\text{in},k}$, one cannot assume any particular shape of the decision boundary on the input space.

\clearpage

\subsection*{A potential route for a universal approximation construction}

A hallmark of general-purpose machine learning architectures is that there is a well-defined sense in which they can approximate any target function of any complexity by scaling their size. The universal approximation theorem for multilayer sigmoidal feedforward networks used in early backpropagation algorithms is a canonical example of this kind of argument \cite{hornik_multilayer_1989,duda_pattern_2012}. Motivated by this, and making the assumptions outlined below, we discuss a path towards showing that arbitrary continuous decision boundaries can be achieved by surface condensates by increasing the number of hidden species. 

\subsubsection*{Linear decision boundaries and connection to winner-take-all dynamics}  \label{subsub: WTA-linear}
We first revisit the mixture considered above, consisting of $N_\text{in}$ input species and $N_\text{out}$ output species (and 0 hidden species), where all the output species are strongly mutually repulsive and thus form distinct condensates that are each enriched only in one output species. The decision boundary between a condensate of output species $m$ and another with output species $n$ is given by eq. \ref{eq:m-n-linear} (where $\phi_{\text{out},m}\approx\phi_{\text{out},n}=\phi_o \gg \phi_{\text{out},i \neq m,n})$, rewritten for simplicity as
\begin{align}
 \sum_{k=1}^{N_\text{in}} (\chi_{mk} - \chi_{nk})\phi_{\text{in},k} - (\mu_{\text{out},m}^\text{res}-\mu_{\text{out},n}^\text{res}) =0  
\end{align}
As originally noted, the decision boundaries are linear planes in the input space. Then, the non-dimensionalized energy of a surface (as seen in eq. 5) of a condensate enriched in output species $m$ ($\phi_{\text{out},m}\approx\phi_o$) and only with negligible amounts of other output species can be approximated as:

\begin{equation}
    \begin{aligned} \label{eq:UTA-surface-energy}
        \Omega_\text{surface}^{m} \approx \sum_{i=1}^N\phi_i\log{\phi_i}+(1-\phi_T)\log{(1-\phi_T)} + (\sum_{i=1}^{N_{in}} \chi_{im}\phi_{\text{in},i} - \mu_{m}^\text{res})\phi_{o} 
    \end{aligned}    
\end{equation}

We define the corresponding score function 
\begin{align}\label{eq:linear-weight-bias}
    f_m(\vec{\phi}_\text{in}) = \sum_{k=1}^{N_\text{in}} \chi_{km}\phi_{k} - \mu_{m}^{res}
    &\ = \vec{w}_m \cdot \vec{\phi}_\text{in} + b_m \ \ .
\end{align}
Assuming the output condensate composition doesn't change majorly away from decision boundaries, the free energy difference between surface condensates enriched in output pairs ($m,n$) can be written as:
\begin{equation} \label{eq:linear-score-wta}
    \begin{aligned} 
        \Omega_\text{surface}^{m} - \Omega_\text{surface}^{n} = \phi_o \left( f_m(\vec{\phi}_\text{in})- f_n(\vec{\phi}_\text{in}) \right) 
    \end{aligned}    
\end{equation}

Since the mean-field model drives a steady-state composition that minimizes this free energy, we see that eq. \ref{eq:linear-score-wta} gives rise to a winner-take-all (WTA) form of dynamics.  That is to say, output $n$ dominates other output species in the condensate when
\begin{equation}
    n = \text{argmin}_m \ \ f_m(\vec{\phi}_\text{in}) = \text{argmin}_m \ \ \vec{w}_m \cdot \vec{\phi}_\text{in} + b_m \ \ .
\end{equation}

In this sense, surface condensation behaves as a computational structure with $N_\text{in}$ inputs that are linearly weighted in $N_\text{out}$ ways, such that the winning output is chosen and the others suppressed.  
This exact computational structure has been well studied in the neural computation and machine learning literature, where it is it often described as a single linear layer followed by a winner-take-all (WTA) unit~\cite{duda_pattern_2012,maass2000computational}.  Such WTA circuits have been proposed theoretically and demonstrated experimentally as an architecture for biochemical computation~\cite{kim_neural_2004,genot_computing_2012,genot_scaling_2013,cherry_scaling_2018,chen_synthetic_2024}.  
Notably, the linearly-weighted WTA is a powerful primitive, with a single such unit being able to perform classification that requires a quadratic number of classical linear threshold units~\cite{maass2000computational}.

However, a single linearly-weighted WTA remains computationally limited, in a way that can be visualized geometrically.  Specifically, the region of input space that activates a given output must be a single convex polyhedron, i.e., bounded by planes that are the linear decision boundaries (as in the bottom left of Fig.~\ref{fig:sup-universal-approx}).  This suggests a close connection between linearly-weighted WTA and Voronoi diagrams~\cite{aurenhammer_voronoi_1991} that is worth understanding here.  Given a set of $N_\text{out}$ reference points with centers $\vec{c}_m$, the Voronoi diagram partitions the input space into regions consisting of all points that are closest to a given reference point.  Mathematically, input vector $\vec{\phi_\text{in}}$ is in region $n$ if 
\begin{equation}
    n = \text{argmin}_m \ \ \| \vec{\phi}_\text{in} - \vec{c}_m \| 
    = \text{argmin}_m \ \ (\vec{\phi}_\text{in} - \vec{c}_m ) \cdot (\vec{\phi}_\text{in} - \vec{c}_m )
    = \text{argmin}_m \ \ \vec{w}_m \cdot \vec{\phi}_\text{in} + b_m \ \ .
\end{equation}
for $\vec{w}_m = -\vec{c}_m$ and $b_m=\frac{1}{2} \| \vec{c}_m \|^2$.  So any Voronoi classification can be achieved by a single linearly-weighted WTA. In fact, linearly-weighted WTA can implement a slightly larger class of {\it generalized} Voronoi diagrams in which 
\begin{equation}
     n = \text{argmin}_m \ \ \| \vec{\phi}_\text{in} - \vec{c}_m \|^2 + d_m 
\end{equation}
where each region $m$ is augmented by an ``importance'' factor $d_m$.  In the above construction, the linearly-weighted WTA is obtained simply by increasing $b_m$ by $d_m$.

Thus, we claim that $N_\text{in}$-input, $N_\text{output}$ surface condensates with no hidden species can perform any classification task in which each output class corresponds to a single cell in a generalized Voronoi diagram. However, it is fair to ask whether the above construction results in reasonable physical parameters $\chi_{km}$ and $\mu_m^\text{res}$. For example, the training elsewhere in this paper was restricted to $\| \chi_{ij} \| < 15$ and used $ \| \mu_m^\text{res} \| < 6$.  In the Voronoi approach, we would choose cell centers $\vec{c}$ with $0 < c_i < 1$ since they represent composition fractions, so in fact $\| \chi_{km} \| < 1$ and $\| \mu_m^\text{res} \| < 1$, which ought to be physically feasible.  Note that the decision boundaries are unaffected by scaling $\chi$ and $\mu^\text{res}$, so if phase separation does not result in sharp enough classification, we can increase both by the same factor until either sharp enough performance is achieved or the limits of physicality are hit.

While this argument provides a mostly satisfactory characterization of the computational power of classification by surface condensation with no hidden species, it also illustrates limits to classification without hidden species.
Specifically, non-convex decision boundaries cannot be achieved or even approximated well. Further, the extent to which computational power can be expressed primarily through a linear WTA mechanism, as derived from the approximate surface energy in eq \ref{eq:UTA-surface-energy}, requires further characterization of the key assumption that condensates are essentially ``one-hot'' in output species recruitment far from the decision boundaries.

\subsubsection*{Decision boundary approximation construction}

In the machine learning literature, the limitations of a single linearly-weighted WTA unit can be overcome in several ways, leading to universal approximation theorems for the enhanced computational architectures~\cite{duda_pattern_2012,maass2000computational}.  The most common approach is to allow multi-layer network computation, but perhaps the simplest generalization is known as the {\it piecewise linear machine}~\cite{duda_pattern_1966}: rather than the WTA outputs being the class labels directly, with one class per WTA output, the piecewise linear machine has $N_\text{in}$ input signals routed with linear weights to a single WTA unit with $N_\text{h}$ lines, and winning output $n$ is assigned a class label $l_n \in \{1 \ldots N_\text{out}\}$.  That is, for input vector $\vec{\phi}_\text{in}$ the classification decision is
\begin{equation}   \label{eq:PLM}
    l_n \ \ \ \ \text{if and only if} \ \ \ n = \text{argmin}_m \ \ \vec{w}_m \cdot \vec{\phi}_\text{in} + b_m \ \ \ .
\end{equation}
Essentially, the computation is coloring a (generalized) Voronoi diagram with as many cells as needed to approximate the desired decision boundaries, and then coloring each cell with $N_\text{out}$ colors (as in the bottom right of Fig.~\ref{fig:sup-universal-approx}).

Let us now translate this approach into the language and mechanisms of surface condensation. First, let us informally note that if the surface classification result were indicated by a moiety (``red'', ``green'', ``blue'', ...) of the output species, rather than the full species identity, then the above WTA construct suffices with almost no modification: just ``color'' each output species according to the desired class associated with its Voronoi cell.  However, in this paper we require one of $N_\text{out}$ distinct output species to be recruited.  We will do this by using the above WTA construction to recruit one of $N_\text{h}$ hidden species according to the relevant Voronoi cell, and then have that hidden species recruit a specific output species associated with that cell.  To show that this construction will work, we must show that the hidden unit computation and output computation do not interfere with each other.  Now the construction is given in detail.

We consider the following problem: Given $N_\text{in}$ input species and $N_\text{out}$ output species, we desire a target decision function $g(\vec{\phi}_\text{in}) \in \{1, \ldots, N_\text{out}\}$. When $g(\vec{\phi}_\text{in})=j$, as with the original model definition, we require the output species $j$ to be selectively recruited at much higher concentrations over other output species, i.e., $\phi_{\text{out},j}=\phi_* \gg \phi_{\text{out},k\neq j}$. Note that, in this formulation, we don't require an \textbf{absolute} high value for  $\phi_*$, just that it is much more than other output species. Our goal is to design $\chi$ and $\vec{\mu}_\text{res}$ that can achieve this for arbitrary $g$, to any necessary degree of accuracy. 

To achieve this, we suggest the following construction, outlined in Fig. \ref{fig:sup-universal-approx}. First, we consider a linear partition of  the  Euclidean input plane $\mathbb{R}^{N_\text{in}}$ into $n_\text{p}$ cells. For example, a specific instantiation of this would be the Voronoi tesselation of $n_{p}$ prototypical input concentrations, $ \left( \vec{\phi}_\text{in}^{1},\vec{\phi}_\text{in}^{2}, \ldots , \vec{\phi}_\text{in}^{n_\text{p}} \right) $. Importantly, $n_\text{p} \gg N_\text{out}$ is a free parameter, and as it increases, one can achieve increasingly finer partitioning of the input space. For two-dimensional input, note that with this partitioning, pairs of cells share linear decision boundaries and a finite number of \textit{vertices} where 3 or more decision boundaries meet.

Following our connection to linear partitions in eq. \ref{eq:linear-score-wta}, we propose including 1 species for each of the $n_p$ classes, which we label as \textit{hidden} species ($N_\text{h}=n_\text{p}$). Despite the label name, hidden species are treated similar to outputs, in that  interactions between any two hidden species is unfavorable $\chi_{ij}=(1-\delta_{ij})\chi_\text{pen}$ for hidden species $i$ and $j$, where $\chi_\text{pen}\gg0$.  At strong interactions, this is sufficient to encode for condensates that are each enriched in 1 hidden species and exclude all others. As described by eq. \ref{eq:m-n-linear}, the resulting molecular network encodes linear boundaries between condensates enriched in hidden species $m$ versus hidden species $n$. From the score function eq. \ref{eq:linear-weight-bias}, we see that boundary slopes (weights) and intercepts (biases) are determined by the subset of tunable interactions ($\chi_{km},\mu_{m}^\text{res}$) that are freely chosen for $k$ being any input species and $m$ being any hidden species. For a desired partitioning, these values can be assigned by the system of linear equations or gradient-descent based approaches.

With this construction, we return to the original objective of achieving a decision boundary of type $g(\vec{\phi}_\text{in})$ with $N_\text{out}$ output species. Note that, with $n_p \gg N_\text{out}$, for a specific function $g$, we need to appropriately assign each of the $n_p$ cells to the appropriate output. To accomplish this, we enforce that each hidden species has an attractive interaction with exactly one output species, and no interactions with the others, i.e., for hidden species $n$ and output species $m$, we have $\chi_{nm} = \delta_{l_n,m} \chi_\text{recruit}$ where $l_n$ is the class label as in eqn.~\ref{eq:PLM} and $\chi_\text{recruit}\ll0$. As before, all the $N_\text{out}$ species also have unfavorable interactions between each other. 
Generically, since multiple cells can be colored with the same output, this provides a many-to-one attractive interaction from hidden species to outputs. We stipulate that the output species reservoir potentials are identical and cannot directly interact with inputs. 
This is, in principle, analogous to a ``client'' (output) and ``scaffold'' (hidden) relationship that has been proposed to study biological condensates \citep{Banani2017}. This construction justifies the assumptions that (a) except on decision boundaries, the surface will recruit at most one hidden species $n$ and at most one output species $m$, with the other hidden and output species being negligible, and (b) the amount of solvent within the condensate will be negligible.  To show that the decision boundaries established by the WTA construction remain intact, consider the boundary between some hidden/output pair $n,m$ and a distinct hidden/output pair $n',m'$.  Within a cell, the surface energy is
\begin{equation}
    \begin{aligned}
        \Omega_\text{surface}^{n,m} &= 
        \sum_{i=1}^N\phi_i\log{\phi_i}+(1-\phi_T)\log{(1-\phi_T)} + 
        \frac{1}{2} \sum_{i,j} \phi_i \chi_{ij}\phi_j - \sum_i \mu_i \phi_i \\
        &\approx
        \sum_{i=1}^{N_\text{in}} \phi_i\log{\phi_i}+\phi_n \log \phi_n+\phi_m \log \phi_m + \left(\sum_{i=1}^{N_{in}} \chi_{in}\phi_i - \mu_n^\text{res}\right)\phi_{n} + \left( \phi_n \chi_\text{recruit} \phi_m - \mu_\text{output}^\text{res} \phi_m \right) \\
    \end{aligned}    
\end{equation}
At the classification decision boundary, $\phi_m=\phi_{m'}$ by definition. Following the ``one-hot'' assumption, since  non-input species are negligible compared to relevant hidden species ($n,n'$), $\phi_n \approx \phi_{n'}$, so we assume $\phi_n = \phi_{n'} = \phi_h$.  Thus when considering the difference in energy, most terms cancel out and what remains is
\begin{equation}
     \Omega_\text{surface}^{n,m} - \Omega_\text{surface}^{n',m'} \approx
     \left(\sum_{i=1}^{N_{in}} \chi_{in}\phi_i - \mu_n^\text{res}\right) \phi_h
     - \left(\sum_{i=1}^{N_{in}} \chi_{in'}\phi_i - \mu_{n'}^\text{res}\right) \phi_h
\end{equation}
which gives the same linear decision boundary as before. A different decision boundary can be achieved by simply reassigning the hidden-output map of interactions as above. With sufficiently large $n_\text{p}$, this model should enable for increasingly complex (convex and non-convex) decision boundaries. 

This construction provides sufficient basis for our universal function approximation claim subject to assumptions specified below. 
Specifically, the input plane is  partitioned into linearly separable regions that each exhibit WTA classification, the number of regions can scale with the number of hidden species, and selective mapping of hidden-output species can approximate arbitrary decision functions---analogous to what is known as a piecewise linear machine for pattern recognition \cite{duda_pattern_1966,duda_pattern_2012}. 
It is intriguing to note that although the universality of this construction relies on potentially using many hidden species, any given resulting condensate is dominated by just one hidden species and one output species; the molecular complexity of the system is not reflected in the simplicity of the outcome.
Beyond existence, the surface condensation driven WTA regions are more flexible than the Voronoi-inspired construction above, so they might be able to achieve a given level of approximation with fewer species;
similarly, gradient descent training may be able to find better approximations with fewer species by exploiting non-``one-hot'' hidden representations, direct input-to-output interactions, and other nonlinearities.  

We consider the above to only be a rough sketch for a universal approximation theorem for surface condensation. This framework makes several assumptions (below) that still require rigorous testing and investigation. 

\subsubsection*{Key assumptions}  First, note that this construction is only true away from parts of the input space where multiple decision boundaries meet, i.e., vertices of the decision planes, which we assume only excludes a finite number of points from an infinite plane. 
Second, within each decision region (a particular group of input surfaces as per our model definition), we assume that the ``one-hot'' condensate encoded by the complementary hidden species is always the only steady-state with negligible composition of other species i.e., even far away from the decision boundaries which, by construction, are the coexistence plane for two ``one-hot'' condensate compositions. While this steady-state should naturally exhibit WTA behavior in the mean-field limit (see eq. \ref{eq:linear-score-wta}), this is unlikely in 3D liquids, where pockets of distinct phases may coexist within the same surface. Third, we assume the selective inclusion of hidden-output favorable links do not destabilize or change overall boundaries. Finally, since we require liquid-like condensates, the range of allowable $\chi_\text{pen}$ values are constrained. For increasingly complex decision boundaries that require many species, the entropic costs increase and may not be offset by the bounded value of  $\chi_{ij}$ values. This potential failure mode can, in principle, be alleviated by considering polymeric molecules, where entropy costs of demixing are significantly reduced by the  degree of polymerization.

\newpage
\section{Lattice liquid model}\label{sec:SI-Lattice}
\subsection*{Lattice setup and boundary conditions}

Simulations are performed on a three-dimensional cubic lattice of dimensions $24 \times 24 \times 24$, except where specified otherwise. We find that classifier performance saturates with bigger lattice sizes, saturating after a lattice width of $L\sim16$, motivating our choice above (Fig. \ref{fig:sup-lattice}A). Each lattice position $\mathbf{p}=(z,y,x)$ can be occupied by a single species from the set $\{0,1,\dots,N\}$. Species $0$ is treated as an inert solvent with zero chemical potential and inert interactions. The boundaries are \emph{walls} meaning that no interactions wrap around from one lattice face to its opposite face. Consequently, any site on a boundary has fewer neighbors than an interior site. 
To capture interactions out to $\sqrt{2}$ in Euclidean distance, each lattice site has up to 18 neighbors. Note that this is consistent with short-range interactions like the standard Lennard-Jones potential that typically only has an impact 1-2 molecule lengths away. Specifically, if $\mathbf{p}_1$ and $\mathbf{p}_2$ differ by at most 1 in up to two of their three coordinates, then $\mathbf{p}_2$ is in the neighborhood of $\mathbf{p}_1$. Positions outside the lattice bounds are ignored.

\subsection*{Free energy model}

Let $\epsilon_{ij}$ denote the pairwise interaction parameter between species $i$ and $j$, and let $\gamma_i$ denote the chemical potential of species $i$. We work at inverse temperature $\beta = 1/(k_{B}T)$. For a given configuration $\sigma$, the total interaction energy is computed by summing over all lattice sites. A configuration \(\sigma\) induces a count \(n_i\) for each molecule type \(i\). Defining \(\delta_{a,b}\) as the Kronecker delta, which is 1 if \(a = b\) and 0 otherwise, the count \(n_i\) is given by
\begin{equation}
n_i \;=\; \sum_{\mathbf{p}}\;\delta_{\sigma(\mathbf{p}),\,i},
\end{equation}
where the sum is taken over all lattice sites \(\mathbf{p}\) in the system. To calculate the system energy, for each site $\mathbf{p}$ with species $i = \sigma(\mathbf{p})$ and each neighbor $\mathbf{q}$ with species $j = \sigma(\mathbf{q})$, we add $\beta\,\epsilon_{ij}$. To avoid double counting, we include a factor of $\tfrac12$ in the total. The free energy, including chemical potentials, for a particular configuration may be written as:
\begin{equation}
   \beta H(\sigma) \;=\; 
    \tfrac12\,\beta \sum_{\mathbf{p}} \sum_{\mathbf{q}\in \eta(\mathbf{p})} \epsilon_{\sigma(\mathbf{p}),\,\sigma(\mathbf{q})}
    \;+\;\beta \sum_{i=1}^{N} \gamma_{i} \, n_{i},
\end{equation}
where $\eta(\mathbf{p})$ denotes the neighborhood of site $\mathbf{p}$, and species~$0$ (solvent) has $\mu_{0} = 0$ by definition.

\subsection*{Parameter mapping from Model A}

The interaction energies and chemical potentials used in these lattice Monte Carlo simulations are derived from Model A. Specifically, a mapping is applied to convert the Model A parameters (denoted $\chi,\vec{\mu}_\text{res}$) to lattice-gas (LG) parameters ($\epsilon_{ij},\gamma_i$). First, to convert from the mean field description (at $\beta=1$) to our lattice gas formulation, the pairwise interaction coefficients are scaled by a factor of \(\tfrac{1}{z}\), where the coordination number $z$ indicates the number of neighbors for interior lattice sites. Thus we have
\begin{equation}
\epsilon_{ij} \;=\; \frac{1}{18}\,\chi_{ij}
\end{equation}
when we incorporate the assumptions that effective solute-solvent interactions are negligible (by setting $\epsilon_{i0}=\epsilon_{00}=0$) and that solute monomers are non-self-interacting (by setting $\epsilon_{ii}=0$), where $0$ indexes the solvent.  These assumptions are consistent with $\chi_{i0}=0$, as seen more clearly in a later subsection. We set the solvent potential also to be $\gamma_0=0$, and under these assumptions

\begin{equation}
    \gamma_i = -\mu_{\text{res},i} \quad\forall i\in (N_{in}+1, N)
\end{equation}

\subsection*{Canonical vs.\ Grand Canonical moves}

We implement two fundamental move types in each MC step:

(1) \emph{Canonical (NVT) moves}, which exchange species between two lattice sites to conserve particle counts;

(2) \emph{Grand Canonical (μVT) moves}, which insert or remove species at a single site, exchanging with an implicit infinite reservoir.

We treat the ``input'' species (for example, species~1 and~2) as \emph{clamped}, meaning they cannot exchange position within the lattice or identity with the reservoir. For every move proposition, if an original site holds an \emph{input} species, the replacement probability is zero (no replacement allowed), keeping the counts and positions of input species fixed. By contrast, all other species (including the solvent) can freely exchange within the remaining sites of the lattice and with the reservoir. These species are handled using both canonical and grand canonical moves. This mixed ensemble preserves the total amount and positions of each input species while allowing all other components to exchange with an infinite reservoir. 

\subsubsection*{Canonical ($NVT$) moves}

Starting from a selected collection of positions (described in more detail in GPU-Accelerated Implementation and Masking):
\begin{itemize}
  \item Pair up any two sites $(\mathbf{p_1},\mathbf{p_2})$ (global swaps).
  \item Propose swapping the species at $\mathbf{p_1}$ and $\mathbf{p_2}$.  
  \item Compute the change $\Delta H$ in interaction energy $H(\sigma)$ induced by swapping the two species, and accept with the Metropolis-Hastings probability
  \begin{equation}
    P_{\mathrm{accept}} = \min\{1,\exp[-\beta\,\Delta H]\}.
  \end{equation}
  \item If accepted and neither position contains a clamped species, exchange the species; otherwise leave them unchanged.
\end{itemize}

\subsubsection*{Grand Canonical ($\mu VT$) moves}

Starting from a selected collection of positions: 
\begin{itemize} %
  \item For a position $\mathbf{p}$ with current species $i$, propose one of the free (unclamped) species as new species $j$.  
  \item Compute the combined energy change $\Delta H$ that includes both interaction energies and the chemical‐potential difference, then accept with probability
  \begin{equation}
    P_{\mathrm{accept}} \;=\;
    \min\!\Bigl\{\,1,\;\exp\bigl[-\beta\,\Delta H\bigr]\Bigr\}.
  \end{equation}
  \item If accepted and the original species is unclamped, update the site to species $j$; otherwise leave it unchanged.
\end{itemize}

\subsection*{Initialization and equilibrium}

The lattice is initialized with only \emph{input} species and solvent. Input species are assigned randomly to lattice sites according to a total sum of a fraction $\phi_i$ for each input species~$i$. All remaining sites are filled with the inert solvent ($i=0$).

Once initialized, the system is evolved via repeated MC moves (either NVT or $\mu$VT with equal probability) until the total free energy and the species counts remain stable over a sufficiently long period (on the order of $10^5$ accepted moves). Asymptotic behavior was verified to be independent of sampling protocol (i.e. frequency of NVT vs $\mu$VT swaps or lattice size), as expected for equilibrium.
We record 1000 equally spaced lattice configurations over the MC protocol and use the last 100 frames to estimate average species counts as the near-equilibrium state for  analysis. Each simulation condition is repeated in triplicate, i.e. with 3 different random seeds but identical parameters, for averaging.

\subsection*{GPU-accelerated implementation and masking}

All Monte Carlo sweeps are implemented in JAX and executed through the Accelerated Linear Algebra (XLA) compiler, combining just-in-time compilation (JIT), batched parallelism (via \texttt{vmap}), and functional key splitting for pseudorandom number generation (PRNG). To avoid race conditions involving calculation of the energy within the neighborhood of each site, we parallelize each step by considering lattice positions spaced modulo four and synchronize all accepted moves each step. Note that each step updates the entire lattice and each move within a step proposes an independent exchange (of position or identity).

\subsubsection*{PRNG and data‐flow structure}

At the start of each full step, a master PRNG key is split into subkeys to select the candidate positions of the 'reference grid', pick per-site swap directions or replacement species, and draw Metropolis acceptance variates. All arrays of positions, energy calculations, or acceptance evaluations are computed under a single JIT-decorated function.

\subsubsection*{Single‐mask strategy with offsets}

We generate a 'reference grid' by partitioning the cubic lattice into discrete modulo \emph{4×4×4 blocks} of sites:
\begin{equation}
    \mathrm{ref\_grid} \;=\; \bigl\{\,4\,(k,l,m)\mid 0\le l<L/4,\;0\le l<L/4,\;0\le l<L/4\bigr\},
\end{equation}
so that no two reference points share an edge. From this 'reference grid' we can generate a shared 'offset grid' by applying a global offset to one of the 64 possible offset positions within each 4³ cube.

\subsubsection*{Generating positions at each MC Step}

At each step, we randomly select one of the 64 offset grids to parallelize the MC moves. During a $\mu$VT step, we propose replacements at each site in the offset grid (excluding input species). During an NVT step, one of 26 neighbor vectors (unit step in any x, y, and/or z) is independently applied to each offset grid point. From these shifted points, we then draw a random permutation, split the points into equal halves, and pair them, such that each point appears in at most one pair. This process yields non-overlapping global swap proposals across the whole lattice.

Acceptance or rejection is then computed according to the Metropolis-Hastings criterion described above, independently and in parallel for all proposed swaps, and all accepted moves are synchronized across the lattice. This parallelized procedure ensures that every lattice site has an opportunity to update while respecting the non-periodic boundary conditions and the mixed canonical-grand-canonical setup.

\subsubsection*{Correctness versus efficiency}

By using one unified, randomized mask with per‐site offsets and directions, our procedure guarantees:
\begin{itemize}
  \item \emph{Correctness:} By construction, any two candidate sites generated from different reference points are at least two lattice steps apart, so edges cannot be shared, and updates commute exactly. No two sites ever race to read or write the same neighbor; counts cannot drift or desynchronize. Boundary sites (hard walls) simply have fewer neighbor offsets.
  \item \emph{Efficiency:} The entire nested‐scan loop over all MC steps is traced once into a single XLA computation—there are no host‐side Python loops or repeated JIT invocations. All random draws, vectorized grid updates, and other per-site operations are executed in one fused GPU kernel via \texttt{vmap}, giving efficient parallel throughput and minimizing host/device synchronization and overhead.
\end{itemize}

Together, this approach delivers robust statistical correctness (no hidden synchronization bugs or particle number errors) and optimized performance on modern hardware. 

\subsection*{Deriving the mean-field model from the lattice liquid formulation} \label{subsec:mfe_lattice}

In this section, we establish a correspondence  between the lattice model and the mean-field model discussed in the paper. Briefly, the lattice model defines an energy for each lattice configuration (or microstate). In the mean-field limit, we consider sets of configurations that share the same species counts (or macrostates). In what follows, we show that eq. 5 arises from the macrostate energies in the bulk limit, under certain assumptions.  Furthermore, we relate the lattice model parameters $\epsilon_{ij}$ and $\gamma_i$ to the mean field parameters $\chi_{ij}$ and $\mu_{\text{res},i}$.  Except at the very end of this section, we make no assumptions constraining the possible values of $\epsilon_{ij}$ and $\gamma_i$.  This is a standard treatment, included here with consistent terminology only as a convenience for the reader.

The lattice is a set $R$ of positions, with $\|R\|=S$, such that the total volume of $R$ is $S \nu$, where $\nu$ is the volume per position.  For an $L \times L \times L$ cubic lattice, $S=L^3$. We use $\eta(\mathbf{p})$ to denote the neighbors of position $\mathbf{p}$, with $||\eta(\mathbf{p})||=z$ being the effective valence of each particle.  For a system with $N$ distinct solute species and 1 solvent species, a microstate configuration is $\sigma$, where $\sigma(\mathbf{p}) \in \{0, \ldots, N\}$ is the species of the particle at position $\mathbf{p}$ and $0$ indexes the solvent.

The energy of the lattice when in microstate $\sigma$ is
\begin{align}
    H(\sigma) & = \frac{1}{2} \sum_{\mathbf{p} \in R}\ \sum_{\mathbf{q} \in \eta(\mathbf{p})} \epsilon_{\sigma(\mathbf{p}),\sigma(\mathbf{q})} + \sum_{\mathbf{p} \in R} \gamma_{\sigma(\mathbf{p})} \\
    & = \sum_{i=0}^N \sum_{j=i}^N n_{ij} \epsilon_{ij} + \sum_{i=0}^N n_i \gamma_i
       \label{eq:Hq}
\end{align}
where $\epsilon_{ij}=\epsilon_{ji}$ is the microscopic nearest-neighbor contact energy between species $i$ and species $j$, and $\gamma_i$ relates to the reservoir chemical potential of species $i$.  The number of $i$ particles (a Delta function sum over all lattice sites)  and $i:j$ interfaces are, respectively,
\begin{align}
    n_i &= \sum_{\mathbf{p} \in R} \delta_{i,\sigma(\mathbf{p})} \\
    n_{ij} &= \frac{1}{1+\delta_{ij}} \sum_{\mathbf{p} \in R} \ \sum_{\mathbf{q} \in \eta(\mathbf{p})} \delta_{i,\sigma(\mathbf{p})}\  \delta_{j,\sigma(\mathbf{q})} \ \ .
\end{align}

Note that the factor of $1/(1+\delta_{ij})$ in the second sum ensures interfaces are not double counted for each pair of \textit{positions} when $i=j$. To compute the sums in $H(\sigma)$ symmetrically, we rewrite eq. (\ref{eq:Hq}) as
\begin{align}
    H(\sigma) &= \frac{1}{2} \sum_{i=0}^N \sum_{j=0}^N n_{ij} \epsilon_{ij} (1+\delta_{ij}) + \sum_{i=0}^N n_i \gamma_i \\
    &= S \left( \frac{1}{2} \sum_{i \neq j} \frac{n_{ij}}{S} \epsilon_{ij} + \sum_i \frac{n_{ii}}{S} \epsilon_{ii} + \sum_i\frac{n_i}{S} \gamma_i \right) \ \ .
\end{align}
The sum above has a prefactor of 1/2 for the total pair contact energies since, unlike eq. \ref{eq:Hq}, the sum double counts over all pairs of distinct species $i\neq j$. 

Now consider a macrostate $M_{\vec n}$ consisting of all microstates whose counts of species $i$ are $n_i$.  As the Monte Carlo 
sampling satisfies detailed balance with respect to $H$ and the state space is fully connected, at equilibrium the probabilities of microstates and macrostates will obey the Boltzmann distribution:
\begin{align}
    P(\sigma) &= \frac{1}{Z} e^{-H(\sigma)/k_\text{B} T} \qquad &\text{where} \qquad & Z = \sum_{\sigma} e^{-H(\sigma)/k_\text{B} T} \\
    P(M_{\vec n}) &= \sum_{\sigma \in M_{\vec n}} P(\sigma) = \frac{1}{Z} e^{-G(M_{\vec n})/k_\text{B} T} \qquad &\text{where} \qquad & G(M_{\vec n}) = -k_\text{B} T \ln \left(\sum_{\sigma \in M_{\vec n}} e^{-H(\sigma)/k_\text{B} T}\right) \ \ .
\end{align}
We define $\phi_i = \frac{n_i}{S}$ to be the volume fraction of species $i$ and note that for well-mixed states, $\frac{n_{ij}}{S} \approx z \frac{n_i}{S} \frac{n_j}{S} = z \phi_i \phi_j$ when $i \neq j$, and otherwise $\frac{n_{ii}}{S} \approx \frac{z}{2} \phi_i^2$ . Such states $\sigma$ all have similar energy
\begin{align}
    H(\sigma) &\approx S \left( \frac{z}{2} \sum_{i \neq j} \phi_i \phi_j \epsilon_{ij} 
    + \frac{z}{2} \sum_i \phi_i^2 \epsilon_{ii} + \sum_i \phi_i \gamma_i \right) \\
    &= S \left( \frac{z}{2} \sum_{i,j} \phi_i \phi_j \left(\epsilon_{ij} - \frac{\epsilon_{ii}+\epsilon_{jj}}{2}\right) 
    + \frac{z}{2} \sum_i \phi_i \epsilon_{ii} + \sum_i \phi_i \gamma_i \right) \ \ .
\end{align}

In the mean-field limit, we assume that these well-mixed states dominate the free energy, and that the number of such states is approximately $\| M_{\vec n}\|$, which we can estimate using Stirling's approximation that $\ln n! \approx n \ln n/e$, so
\begin{align}
    \ln \| M_{\vec n}\| &= \ln {S \choose \vec n}\\
    &= \ln \frac{S!}{\prod_{i=0}^N n_i !} \\
    & \approx S \ln S/e - \sum_i n_i \ln n_i/e \\
    & = S \left( \ln S/e - \sum_i \phi_i \ln S \phi_i / e \right)\\
    &= -S \sum_{i=0}^N \phi_i \ln  \phi_i \ \ .
\end{align}

The free energy for this macrostate of the lattice is therefore
\begin{align}
    G(M_{\vec n}) &\approx -k_\text{B} T \ln \| M_{\vec n}\| e^{-H(\sigma)/k_\text{B} T} \\
    &\approx H(\sigma) + k_\text{B} T S \sum_i \phi_i \ln \phi_i 
\end{align}
and the (non-dimensionalized) free energy of the lattice per unit volume is
\begin{align}
    \Omega_\text{surface} &\equiv \beta\nu \frac{G(M_{\vec{n}})}{\nu S}\\
    &\approx \beta\frac{H(\sigma)}{S} + \sum_{i=0}^N \phi_i \ln \phi_i \\
    &=\beta \left( \frac{z}{2} \sum_{i=0}^N\sum_{j=0}^N \phi_i \phi_j \left(\epsilon_{ij} - \frac{\epsilon_{ii}+\epsilon_{jj}}{2}\right) + \frac{z}{2} \sum_{i=0}^N \phi_i \epsilon_{ii} + \sum_{i=0}^N \phi_i \gamma_i \right) + \sum_{i=0}^N \phi_i \ln \phi_i\\
    &= \frac{1}{2}\sum_{i=0}^N\sum_{j=0}^N\phi_i\chi_{ij}\phi_j+\sum_{i=0}^N \phi_i \ln \phi_i-\beta \sum_{i=0}^N \phi_i \mu'_{\text{res},i}\\
    &=\beta\nu f(\vec{\phi},\chi)-\beta\vec{\mu}'_{\text{res}}\cdot \vec{\phi}
\end{align}
where $\beta = 1/k_\text{B} T$ and $\chi_{ij} = \beta z (\epsilon_{ij} - \frac{1}{2}(\epsilon_{ii}+\epsilon_{jj}))$ and $\mu'_{\text{res},i} = -(\gamma_i+\frac{z}{2}\epsilon_{ii})$. We used the fact that $\phi_0=1-\phi_T$ is the solvent volume fraction and $\chi_{i0}=0$ by construction to equate the first term with eq. 6. Finally, since the input species are non-exchanging, and recalling that $\mu'_{\text{res},i}$ is the reservoir chemical potential of species $i$, with $\vec{\mu}_\text{res}'=0\circ\vec{\mu}_\text{res}^\text{(in)}\circ\vec{\mu}_\text{res}$ defined in SI Note \ref{sec:SI-Dynamics}, we have that
\begin{equation}
    \vec{\mu}'_\text{res}\cdot\vec{\phi}=\vec{\mu}_\text{res}\cdot\vec{\phi}_\text{oh}+\text{const.}
\end{equation}
and so, up to a constant,
\begin{equation}
    \Omega_\text{surface}=\beta\nu f(\vec{\phi},\chi)-\beta\vec{\mu}_\text{res}\cdot\vec{\phi}_\text{oh}
\end{equation}
is in agreement with eq. 5. Note that because elsewhere in this work we assume an inert solvent and non-self-interacting solute monomers, such that $\chi_{i0}=0$, lattice simulations are run with $\epsilon_{ii}=0=\epsilon_{0i}$, so $\epsilon_{ij} = \frac{\chi_{ij}}{\beta z}$ for $z=18$ and $\gamma_i = -\mu'_{\text{res},i}$. 

\newpage
\section{Analyses}\label{sec:SI-Analyses}
\subsection*{Phase number and composition calculation}

The steady-state compositions of the $n_\text{set}$ surfaces from the mean-field dynamics are gathered into a matrix  $B =n_\text{set}\times(N_\text{out} + N_\text{h})$, and given the large number of surfaces, we generically assume $N_\text{out} + N_\text{h} \ll n_\text{set}$. Subsequently, the matrix is normalized (mean-centered and standard-deviation set to $1$) and the covariance matrix's eigenvalues (i.e. eigenvalues of $\frac{B^TB}{N_\text{out} + N_\text{h}}$) is computed. If the normalized matrix was populated purely with i.i.d values from $N(0,\sigma=1)$, the Marchenko-Pastur distribution \citep{marcenko_distribution_1967} guarantees that the eigenvalues would be smaller than $\lambda= \left(1+\sqrt{\frac{N_\text{out} + N_\text{h}}{n_\text{set}}}\right)^2$. Thus, eigenvalues larger than this are unlikely to arise from compositions sampled randomly around a typical composition (i.e. of a particular phase) and when no eigenvalues are significant, we assume that there is only 1 typical phase composition. Note that this is an approximation since the MP distribution does not generically guarantee that eigenvalues from ``signal'' cannot be less than the above $\lambda$, and only that the eigenmodes from ``noise'' cannot be larger---so the number of phases we estimate may be lower than actually present. With that caveat, we use the number of significant modes (larger than above threshold) to estimate number of phases as $n_\text{phases}=n(\text{eig}>\lambda)+1$. With this estimate, we employ a hierarchical clustering method to group the $n_\text{set}$ surfaces into $n_\text{phases}$ compositions. The average composition of each phase is computed as the mean composition of all the surfaces clustered into the same phase and reported in Fig. 4.

\subsection*{Phase detection and composition analysis in lattice}

To identify distinct phases and their compositions from equilibrium lattice configurations, we employ a composition-based clustering algorithm as described below. Conceptually, the process includes smoothing particle numbers into normalized compositions and performing unsupervised clustering to identify bulk phase compositions and volume fractions. In Fig. \ref{fig:sup-lattice-phases}, this methodology is used to analyze that phase composition for the AND decision boundary with 2 hidden species at each surface.

Before clustering, we convert each discrete lattice snapshot into a matrix P (of size $N_{loc}\times N$ where $N_{loc}$ is the number of \textit{bulk} lattice positions as defined below and $N$ is the number of species including solutes and the solvent) that describes composition at each bulk location as follows.

\begin{enumerate}
    \item For each species $i$, we compute a smoothed density field $\rho_i(\mathbf{r})$ by applying a Gaussian filter with width $\sigma = 2.0$ lattice units to the binary indicator field (1 where species $i$ is present, 0 elsewhere). The Gaussian convolution is implemented using \texttt{scipy.ndimage.gaussian\_filter} specifically to handle non-periodic boundary conditions by extending edge values outward, appropriate for the closed boundary simulations.

    \item To focus on phase identity rather than local density variations, we normalize the smoothed densities at each lattice position $\mathbf{r}$ to obtain composition fields:
\begin{equation}
c_i(\mathbf{r}) = \frac{\rho_i(\mathbf{r})}{\sum_j \rho_j(\mathbf{r}) + \epsilon}
\end{equation}
where $\epsilon = 10^{-10}$ prevents division by zero and $j$ runs from $0$ to $N$ . This normalization removes the magnitude axis from the feature space, ensuring that regions with identical composition but different total densities are recognized as the same phase, helping to identify small pockets of distinct phases within a larger bulk phase.

\item We identify bulk (low-gradient) regions suitable for phase analysis by computing the gradient magnitude of the composition field and applying a percentile-based threshold. For each species, we compute $\nabla c_i$ using central differences and define the total gradient magnitude as $g(\mathbf{r}) = \max_i \|\nabla c_i(\mathbf{r})\|^2$. Voxels are retained in the bulk mask in the bulk mask if $g(\mathbf{r}) < g_P$, where $g_P$ is the Pth percentile of gradient values. In this analysis, we retain voxels below the 85th percentile of gradient values. Additionally, voxels with total density below 0.05 are excluded to avoid spurious phase detection in nearly empty regions. The included positions make up the $N_{loc}$ that go into the matrix P.

\end{enumerate}

From the matrix P, we next perform unsupervised clustering to identify the number, composition, and volume fraction of different coexisting phases as below. Note that this unsupervised clustering requires choices of hyperparameters that were found after benchmarking on lattices with known phase counts and compositions.

\begin{enumerate}
    \item The number of phases $K$ is determined via Bayesian Information Criterion (BIC) model selection. We fit Gaussian Mixture Models (GMM) to the bulk voxel composition vectors for $K = 1, 2, \ldots, 20$ using scikit-learn's \texttt{GaussianMixture}. For each $K$, the BIC is computed as $\text{BIC}(K) = -2 \ln(L) + k \ln(n)$, where $L$ is the maximum log-likelihood, $k$ is the number of model parameters, and $n$ is the number of data points. The optimal $K$ is identified using the elbow method: we compute the second derivative of the BIC curve and select the $K$ where this quantity is maximal, indicating the point where additional clusters provide diminishing improvement in model fit.

    \item With $K$ determined, we cluster the bulk voxel composition vectors using GMM with the selected number of components. Each voxel is assigned to the cluster with highest posterior probability. The composition of each phase $\alpha$ is computed as the mean composition of all bulk voxels assigned to that cluster, normalized to sum to unity.

    \item To prevent over-clustering due to noise or minor compositional fluctuations, we compute pairwise cosine similarities between all phase compositions: $\text{sim}(\alpha, \beta) = \mathbf{c}_\alpha \cdot \mathbf{c}_\beta / (\|\mathbf{c}_\alpha\| \|\mathbf{c}_\beta\|)$. Phase pairs with similarity exceeding 0.85 are merged, with the merged phase composition computed as the weighted average of the constituent phases. Merging continues iteratively until no pair exceeds the threshold.

\end{enumerate}

\newpage
\section{Random fluids}\label{sec:SI-Random}
\subsection*{Fluids with random collection of interactions}
We explore whether fluids with a random interaction network (as reported in Fig. 5) can be trained to classify distinct decision boundaries by simply tuning concentrations. For this, we first initialize a system with 2 inputs, 2 outputs, and a large ensemble ($N_\text{h}=30$ in Fig. 5B) of hidden species. In a given trajectory, the relevant pairwise interaction ($\chi_{ij}$) are directly sampled as follows: first a random variable $x$ is sampled uniformly from $[-1,1]$ and transformed to obtain $\chi_{ij}=\chi_\text{m}\tanh(x)$, where $\chi_\text{m}=15$ is chosen to set a maximum strength of interactions  $\chi_\text{max}\approx 12$.  This resulting transformed distribution is not perfectly uniform and is biased a bit  towards higher values of $\chi$. The output-output interactions are set to favor demixing as above. With this initial \textit{frozen} interaction matrix, we perform training as described above, except over 3000 epochs, to minimize the loss by only changing the reservoir potential $\vec{\mu}_\text{res}$. Since the interaction matrix is sampled randomly and frozen, we repeat this training across 30 replicates and for distinct decision boundaries. The results of these tests are presented in the manuscript.

\newpage
\section{Sharp edges of the model}\label{sec:SI-Sharp}

\subsection*{Reservoir}
A central assumption of the model is that the trained reservoir potential ($\vec{\mu}_\text{res}$) will be maintained by the cellular milieu, likely through out-of-equilibrium mechanisms. Note that this assumption does  not directly posit any further requirements of such a reservoir. That is, it could exist as a single or multiple coexisting phases, and either be dense or dilute---as long as the reservoir potential remains constant ($\vec{\mu}_\text{res}$) and unaffected by the exchange with surfaces. While not explicitly modeled in our study, we briefly discuss potential considerations in \textit{designing} biological/physical models of reservoirs. 

\textbf{Biological reservoirs}: The cellular millieu typically contains the same molecular repertoire but is generically coupled to various active processes. For example, molecules are routinely created and destroyed through active reactions, and cytoplasm/nucleoplasm resident molecules like chaperones and disaggregases \citep{bard_chaperone_2024,narayanan_first_2019} contribute to partial solubilization of the reservoir. Thus, explicit models of the chemical potential remain challenging to describe. 

\textbf{Physically realizable reservoirs:} In physical or synthetic systems, particularly those at equilibrium, one pertinent question relates to properties of the reservoir. In particular, what are its corresponding composition and stability? This requires a \textit{specific} model of the reservoir. For example, one could allow for the \textit{same} mean-field like treatment of the entropy/interactions for the reservoir as was used for the surface, except it could exist at a different, larger volume $V_\text{res}$. If we further assume that the reservoir is input-free---comprised of only hidden, output, and solvent species---one can invert the $\vec{0}$ input surface composition to get a reservoir composition from the Model A dynamics. 

We discuss next how this inferred composition is guaranteed to  be thermodynamically stable i.e., outside of the spinodal, and as such, will not spontaneously phase separate. This emerges because the criteria for the thermodynamic stability of the surface is $\frac{d^2\Omega_\text{surface}}{d\phi_i \phi_j} = \frac{\delta_{ij}}{\phi_i} + \frac{1}{\phi_T} + \chi_{ij}$ is positive semi-definite ($i,j \, \forall \text{ non-input species}$). This is guaranteed by construction, since the gradient descent procedure employed in the mean-field model finds local minima of $\Omega_\text{surface}$ that must satisfy this constraint. Importantly,  the input-associated terms and linear reservoir terms do not explicitly show up in this Hessian. The above term is identical to the Hessian of the free energy that describes a box of finite volume comprising non-input only species at the identified steady-state composition. Physically, this can be interpreted as the stability of non-input species in a canonical ensemble, or in other words, following the $\beta \nu f$ like-term that we describe in eq. 6 only for the pertinent species. The lack of input-related terms, despite their contribution in the free-energy, stems from their constraints in the model. Since inputs are both clamped in space and position, $\phi_{\text{in},i}$ is not a free parameter that is capable of fluctuations. Thus, their interactions with non-input species can be re-interpreted as an (linear) ``internal'' chemical potential coupling, i.e., $ (\chi_{ik}\phi_{in,i})\phi_{k} \equiv \mu_{ik}^{int} \phi_k \,\,\  \forall k \in (N_\text{h},N_\text{out}),\ \forall i \in {N_\text{in}}$. An important caveat to note is that this Hessian does not guarantee stability of a mixture where inputs \textit{also} contain translational entropy i.e. the ability to move in space. Although their counts are fixed, input species can now undergo spatial fluctuations, and thus can change the stability of the surface. Evaluating the stability of the whole surface requires determining: $\frac{d^2\Omega_\text{surface}}{d\phi_i d\phi_j}, \,\,\, i,j \, \forall (N_\text{in},N_\text{out},N_\text{h})$ - inherently not possible directly in the spatially unresolved mean-field model described here but could be studied by incorporating spatial gradient terms (as in eq. \ref{eq:LGH}) leading to a Cahn-Hilliard type formulation or through lattice models. Note that this stability would also depend on input-associated parameters like input-input interactions, that are not directly learned or modified in our model. 
Preliminary investigations of lattice simulations with mean-field parameters, but with inputs no longer immobilized, suggest a loss in classifier performance as well as stronger intra-surface demixing. In such cases, since inputs strongly prefer distinct outputs and are still constrained to remain in the box, they demix to form pockets of coexisting phases with distinct outputs and compositions. These suggest the possibility of novel, or only partially overlapping, class of (microscopic) solutions may be discovered in a model where inputs are free to move within the surface but still incapable of exchanging with the reservoir---bearing  resemblance to the model explored for MNIST classification by \citep{chalk_learning_2024}.

However, generally such an effective composition requires the multiple assumptions stated above. More generally, it may be experimentally advantageous to directly specify a desired reservoir composition ($\vec{\phi}_{\text{res}}^{*}$)---for instance, an equimolar, dilute reservoir. One could incorporate this constraint by suitably modifying our formulation to instead require that as the molecular interaction parameters $\chi$ evolve in the optimization procedure, the reservoir potential is implicitly derived as $\vec{\mu}_\text{res}=\vec{\mu}(\vec{\phi}_\text{res}^*,\chi)$ from the mean-field model. This will need to include an additional constraint that the Hessian matrix of the free-energy $H_{ij}=\frac{d\mu_{\text{res},i}}{d\phi_{j}}$ be positive semi-definite at $\vec{\phi}_\text{res}^*$ \citep{gibbs_equilibrium_1879,sear_instabilities_2003}. 
However, in both methods outlined here, there is still no guarantee that the reservoir composition is stable to fluctuation-driven nucleation.

\subsection*{Stability and properties of surface phases}
In the model formulation, the surface is treated in the well-mixed mean-field limit. Thus, we don't explicitly consider whether the surface itself can demix \textit{within} the volume that it occupies. In this section, we discuss the assumptions that underlie this model and where they may break down. 

\textbf{Biological motivation for mean-field treatment:} We begin with the context presented in the paper i.e., DNA-bound transcription factors (TFs) as input species on genetic loci and mobile species (polymerases, cofactors etc.) that exchange with the nucleoplasm. DNA-bound TFs (inputs) are treated as fixed in position and space within our framework. This is motivated by the fact that the time-scales of free diffusion and exchange from the nucleoplasm of mobile molecules is significantly faster than for DNA-bound TFs. For example, the diffusion coefficients of chromatin, and thus molecules stably bound to it, are typically 2-3 orders of magnitude slower than those of nucleoplasmic proteins. We ignore any internal organization of the inputs within the surface that may emerge from the 3D DNA conformation and treat it as uniform, i.e., well-mixed. Thus the \textit{mobile} species (hidden and outputs) in our model framework effectively live in a mean-field environment created by the well-mixed inputs.  More generally, there may exist other active mechanisms like, for example, chromatin associated remodeler proteins that stir DNA, that may further contribute to keeping the input species well-mixed.

\textbf{Stability of a surface}: With the above assumption that inputs are effectively randomly well-mixed in the surface, the composition of the exchanging species (as queried by the Model A dynamics) is found as a minimum of the effective free-energy of the surface. This means the surface will not spontaneously phase separate but may still form multiple phases from nucleation. As described in the next section, we generally find that the explicit 3D lattice model shows a single phase in most regions except for the region adjacent to the decision boundary---see SI Notes \ref{sec:SI-Lattice} and \ref{sec:SI-Analyses} as well as Fig. \ref{fig:sup-lattice-phases}.

\textbf{Remarks from the lattice liquid model}: In the lattice liquid (see SI Note \ref{sec:SI-Lattice}), for each trajectory both the overall composition of the input species as well as their positions are held fixed to mimic immobile, non-exchanging TFs on short timescales. Note that the initial positions of the inputs are randomly assigned in the lattice. With this implementation, we find that parameters trained from the mean-field model  successfully \textit{translate} to 3D lattice fluids as measured by the classifier performance. This supports the idea that under the assumption of immobile, localized input species, the lattice model generally predicts a major, single phase within the surface. When closer to the decision boundary, we see that the lattice models deviate from the mean-field predictions (Fig. 7B). At these points, we empirically find that multiple phases can form \textit{within} the surface that are enriched in the two distinct outputs. 

\textbf{Input-free surfaces}:  Biologically, the no-input surface is explicitly considered as a finite volume DNA loci that has \textit{no} binding sites for any of the input molecules. Thus, the ``output function'' of a (0,0) surface is ascribed by condensing the appropriate output (`green' in AND, `pink' in XOR, and so on). More generally, a surface absent of input species is nonetheless described by a fixed volume $V$ that can freely exchange with the reservoir. 
As discussed above, in a (non-biological) physical reservoir that is not actively disaggregated and is constrained to a finite volume (i.e. a canonical ensemble), we expect the same condensed phase to emerge in the reservoir as in the (0,0) surface volume. More generally, if an input-free surface or ``low-input'' surface is desired to not recruit any outputs, a suitably altered loss will aid in discovery of parameters that achieve this. Fig \ref{fig:sup-alternates}A represents an AND gate where the system is explicitly trained (with an alternate loss) to not recruit any output in regions that are not high in either inputs, and thus, near the (0,0) region as well. Thus, the composition of the input-free surface would not recruit any outputs.

\newpage

\onecolumn
\newpage

\renewcommand{\thefigure}{S\arabic{figure}}
\setcounter{figure}{0}  %

\renewcommand{\thetable}{S\arabic{table}}
\setcounter{table}{0}  %

\captionsetup*{format=largeformat}
\newcommand{\thirdtitle}{%
    \begin{flushleft}
        {\fontsize{14}{16}\bfseries \color{titlecolor} Supplementary Figures: Combinatorial decision-making driven by multicomponent surface condensates}\\[1em]
        \vspace*{0.2cm}
    \end{flushleft}
}
\thirdtitle

\begin{figure*}[h]
    \centering
    \includegraphics[width=\linewidth]{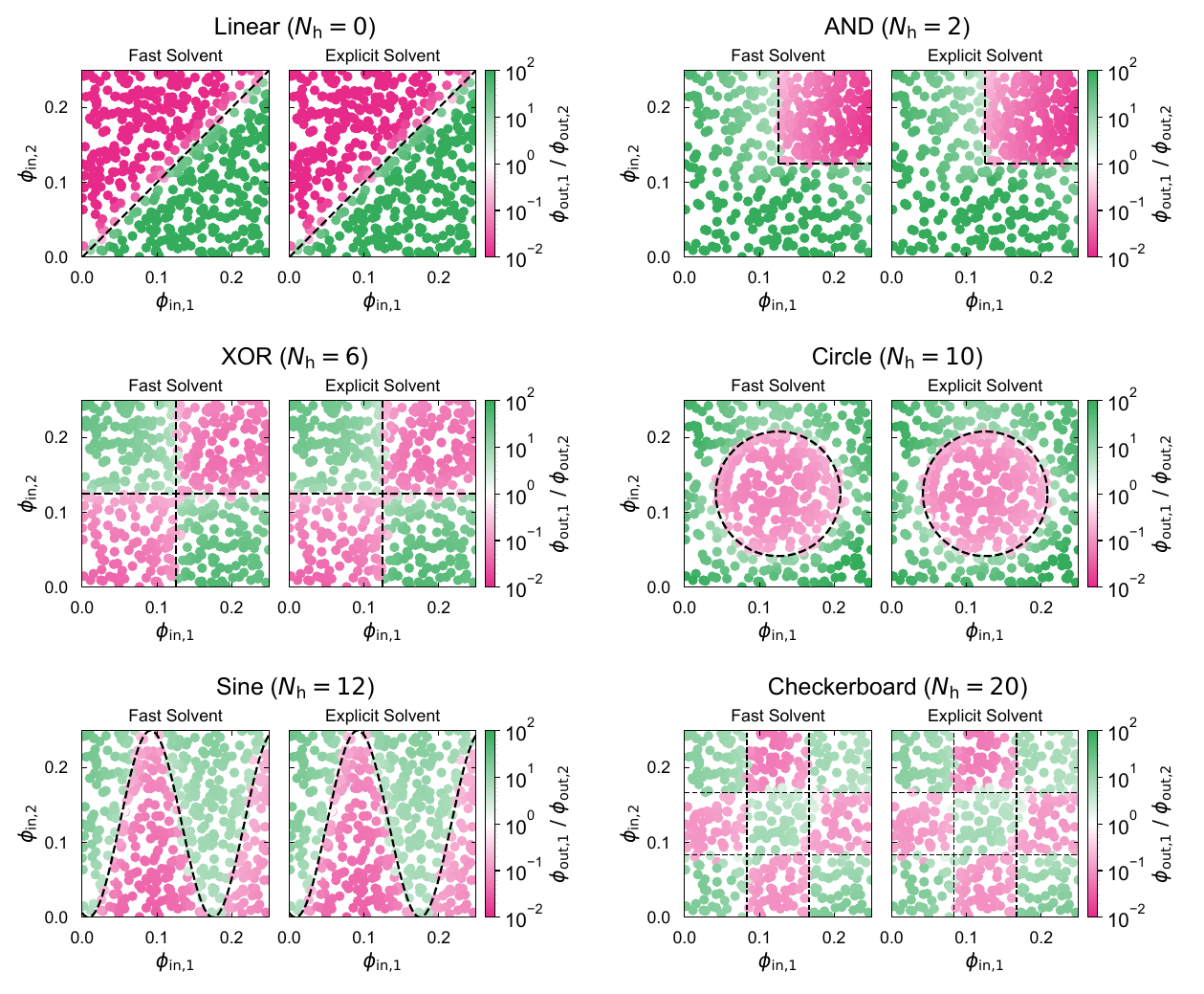}
    \caption{In training and testing the classifiers in the manuscript, we assume that the solvent has fast dynamics and can therefore be treated implicitly according to the mass constraint of the system. However, the steady states of surfaces are largely insensitive to the choice of solvent dynamics, as shown above. For each of the decision boundaries tested in Figs 1-3 (reproduced here for ease of comparison as the “fast solvent” panels), we produce the same plot using dynamics in which the solvent is treated explicitly in the dynamics and is given the same mobility as the solutes (presented as the “explicit solvent” panels). The result is a decision boundary that looks nearly identical for all cases.}
    \label{fig:sup-dynamics}
\end{figure*}

\clearpage
\begin{figure*}[h]
    \centering
    \includegraphics[width=\linewidth]{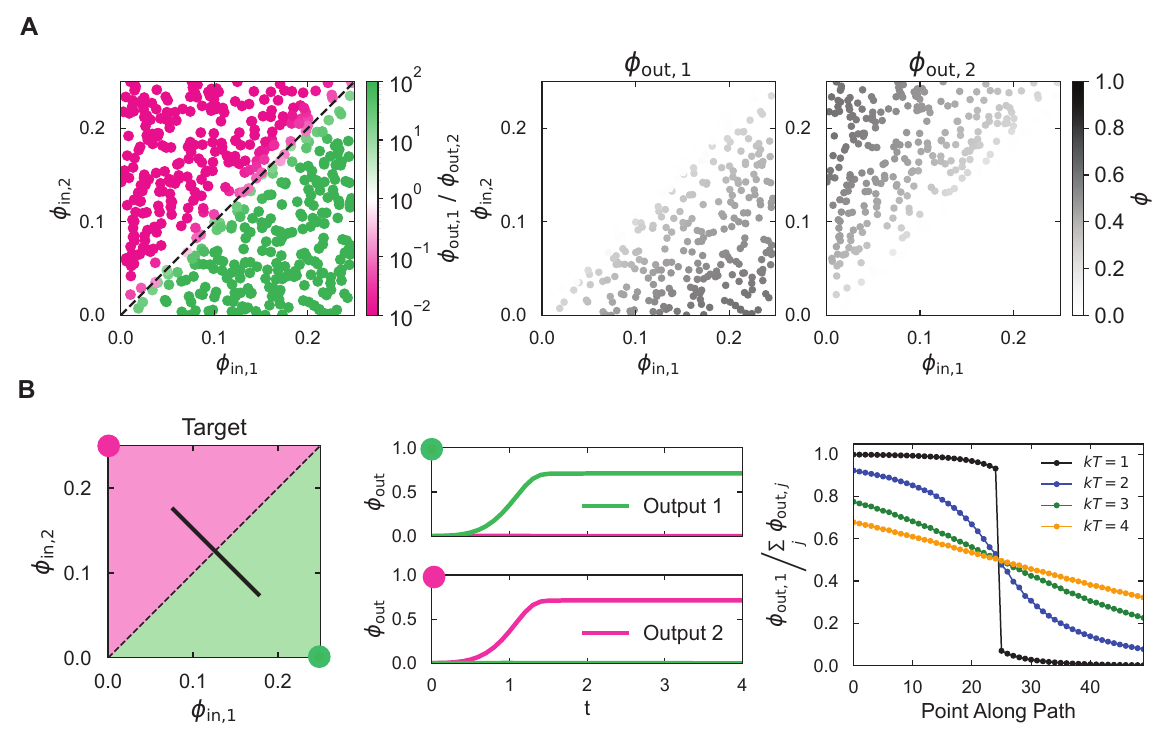}
    \caption{\textbf{(A)} The absolute concentrations of the two output species, with the test prediction in the Fig. 2A reproduced on the left for comparison. \textbf{(B)} The middle panel shows the dynamics of the mean-field composition at two points (labelled by the green and pink dots) far away from the decision boundary. As an extension of Fig. 2B, the right panel depicts how the mean-field composition changes across the decision boundary at multiple temperatures.}
    \label{fig:sup-linear}
\end{figure*}

\clearpage
\begin{figure*}[h]
    \centering
    \includegraphics[width=\linewidth]{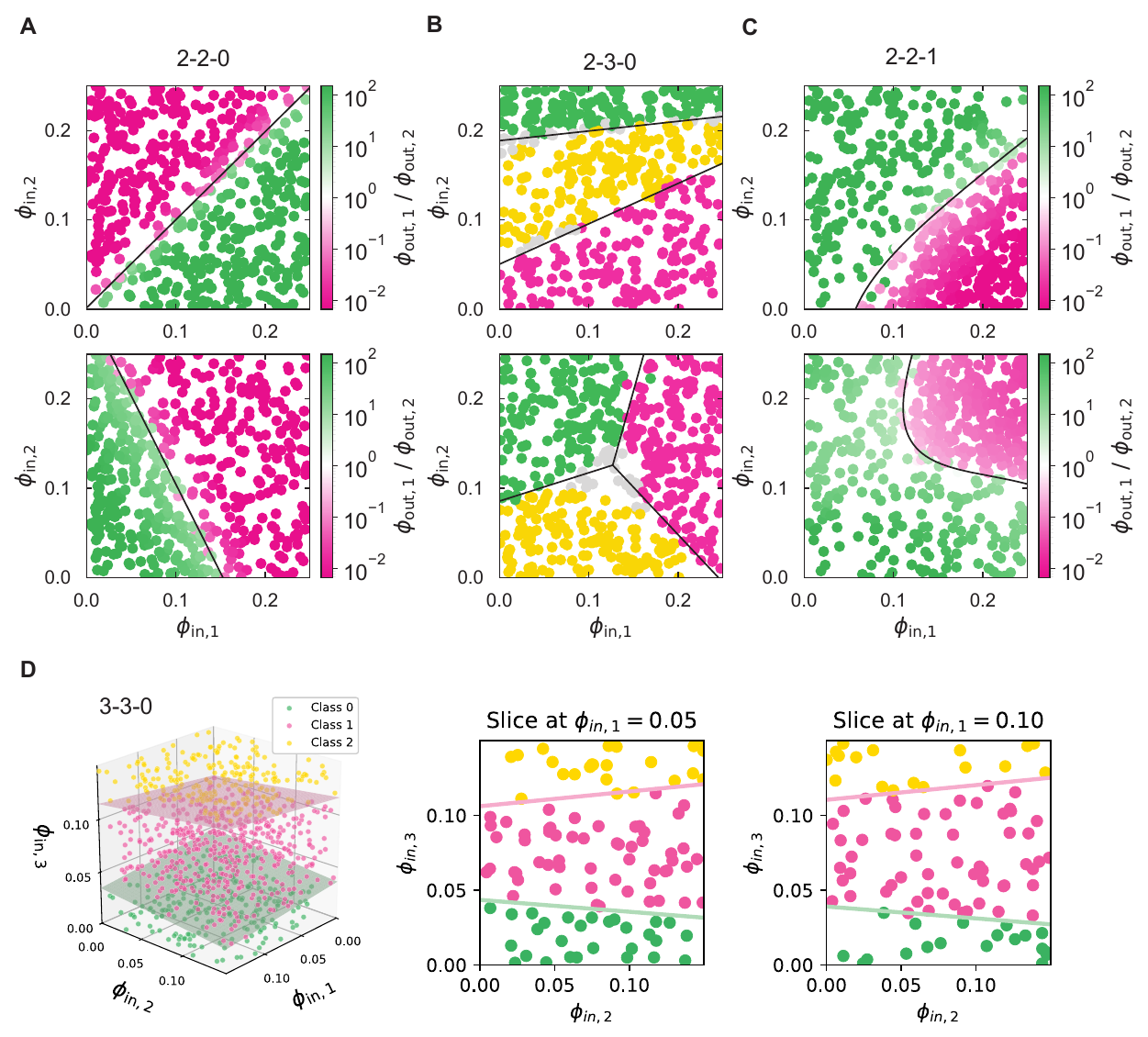}
    \caption{\textbf{(A)} Two theoretical boundary solutions to systems with only 2 input and 2 output species (computed using eq. \ref{eq:2-2-linear}). Only linear boundaries are possible. \textbf{(B)} Two theoretical boundary solutions to systems with only 2 input and 3 output species (computed using eq. \ref{eq:m-n-linear}); gray points indicate surfaces for which none of the output species were recruited by a significant amount above others. Away from intersection points between more than two classes, the decision boundaries remain linear. \textbf{(C)} Two theoretical boundary solutions to systems with only 2 input, 2 output and 1 hidden species (computed using eq. \ref{eq:2-2-1-implicit}), enabling the construction of nonlinear boundaries. \textbf{(D)} An example of a 3-input, 3-output, 0-hidden system, with classes separated by 2D planes. The 2D slices compare two different planes of constant concentration of species 1; as before, the solid lines are predictions of the decision boundary from theory, and each point is false-colored by its true class. }
    \label{fig:sup-theory}
\end{figure*}

\clearpage
\begin{figure*}[h]
    \centering
    \includegraphics[width=\linewidth]{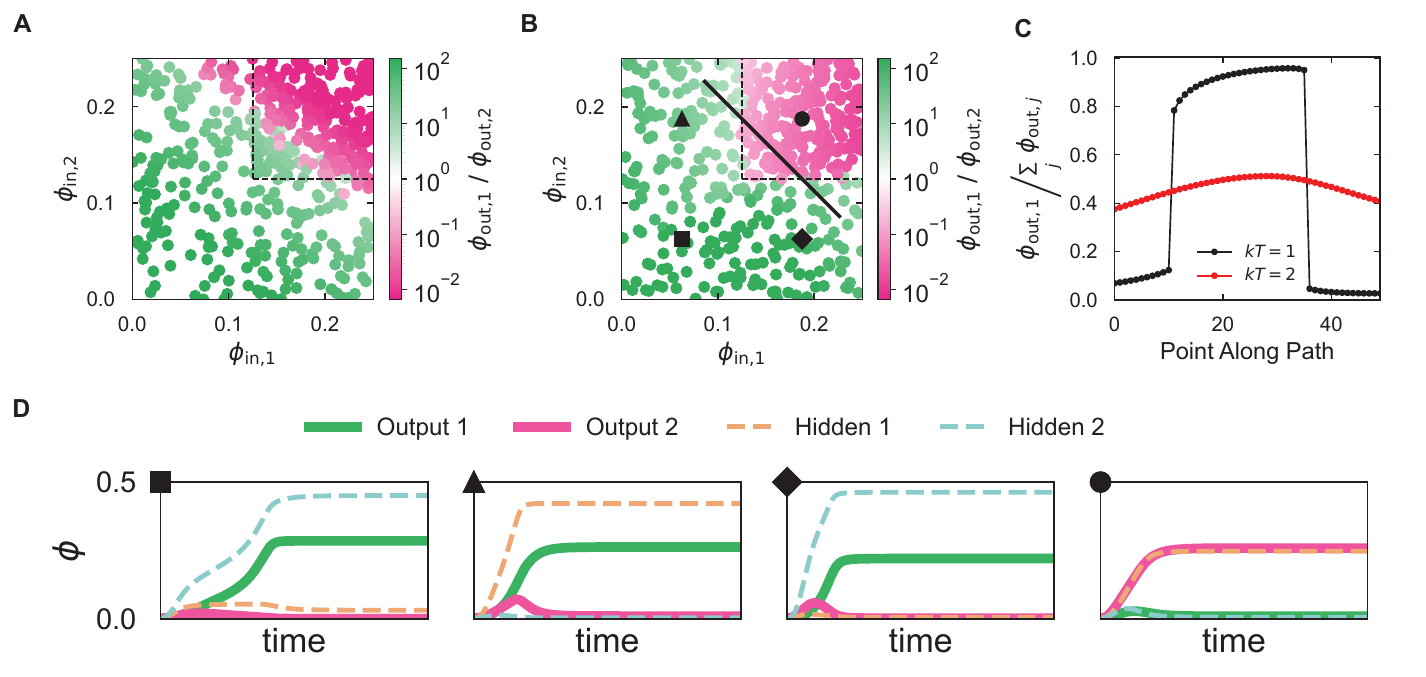}
    \caption{\textbf{(A)} The optimal solution to an AND-like upper quadrant decision boundary with 0 hidden species, which is a best-fit linear cut of the boundary. \textbf{(B)} The 0-hidden solution in (A) can be compared with the solution using 2 hidden species, reproduced from Fig. 3C for convenience. \textbf{(C)} Moving along the solid black line and across the decision boundary in (B), we find that the system undergoes an abrupt transition in the recruited output species that is destroyed at higher temperatures. \textbf{(D)} The mean-field dynamics for the AND-like upper quadrant solution from a point in each of the quadrants of the input space.}
    \label{fig:sup-and}
\end{figure*}

\clearpage
\begin{figure*}[h]
    \centering
    \includegraphics[width=\linewidth]{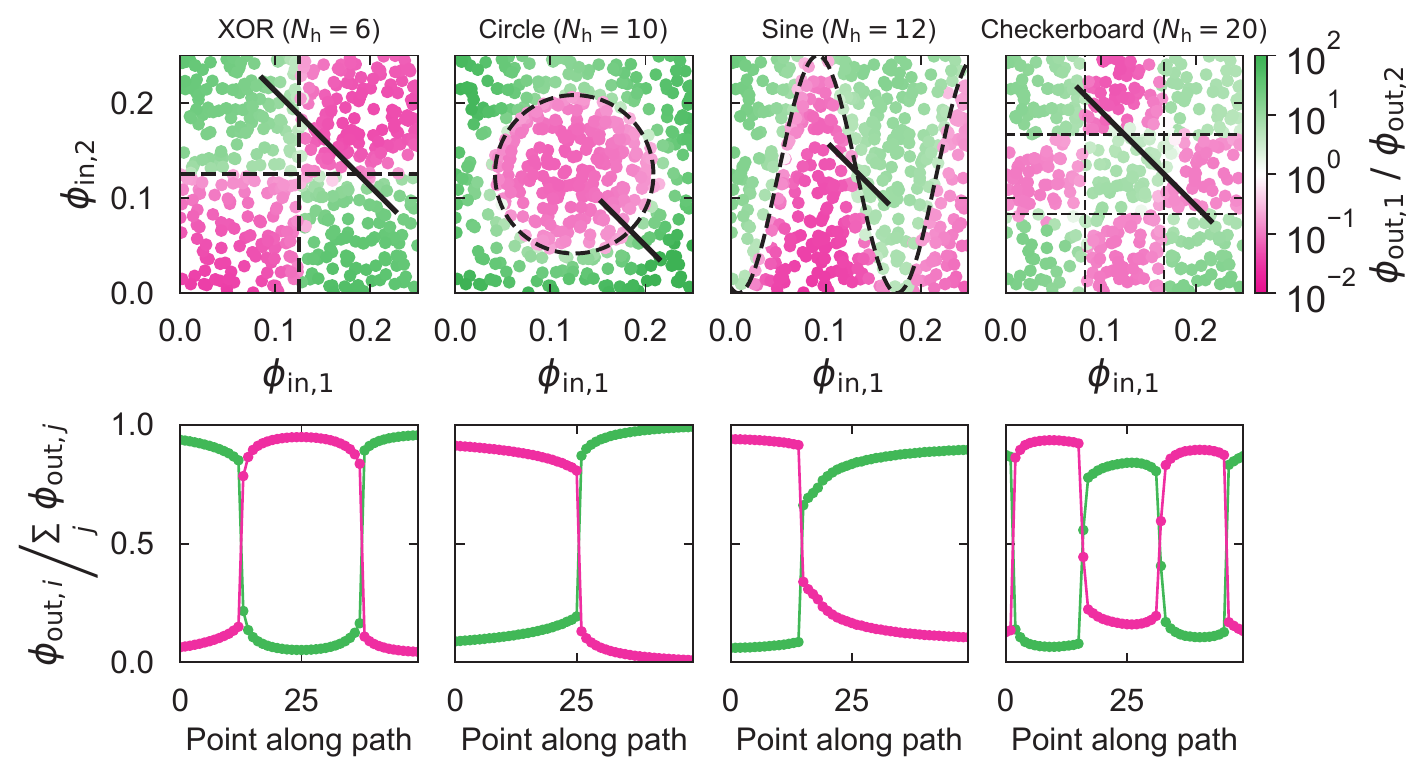}
    \caption{The analogous simulations for the remaining 4 decision boundaries studied in Fig 3D. All show abrupt transitions in the order parameter across a decision boundary.}
    \label{fig:sup-other_boundaries}
\end{figure*}

\begin{figure*}
    \centering
    \includegraphics[width=\linewidth]{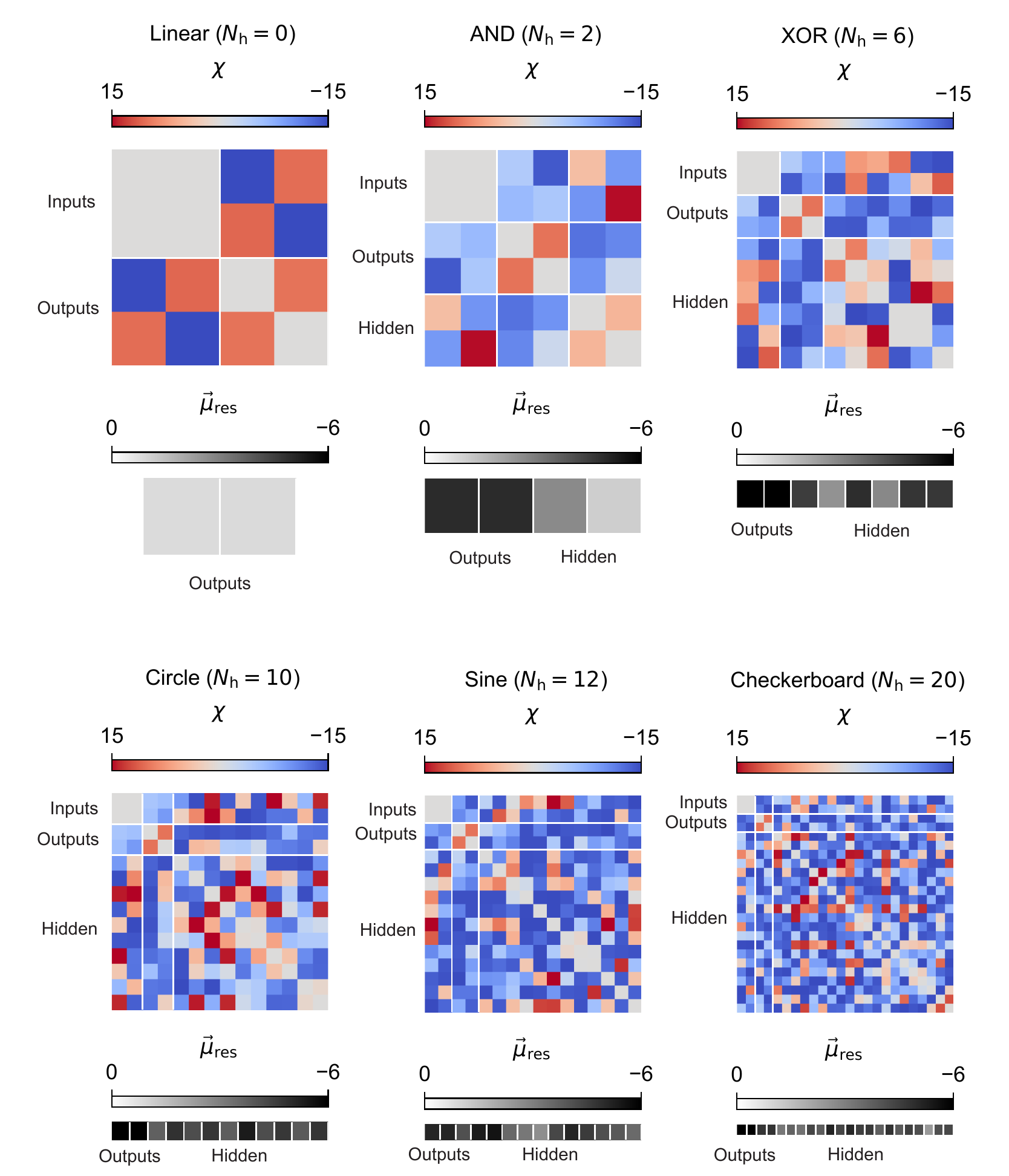}
    \caption{The parameters for the trained networks that were used on the test data shown in Figs. 2 and 3.}
    \label{fig:sup-parameters}
\end{figure*}

\clearpage
\begin{figure*}[h]
    \centering
    \includegraphics[width=0.7\linewidth]{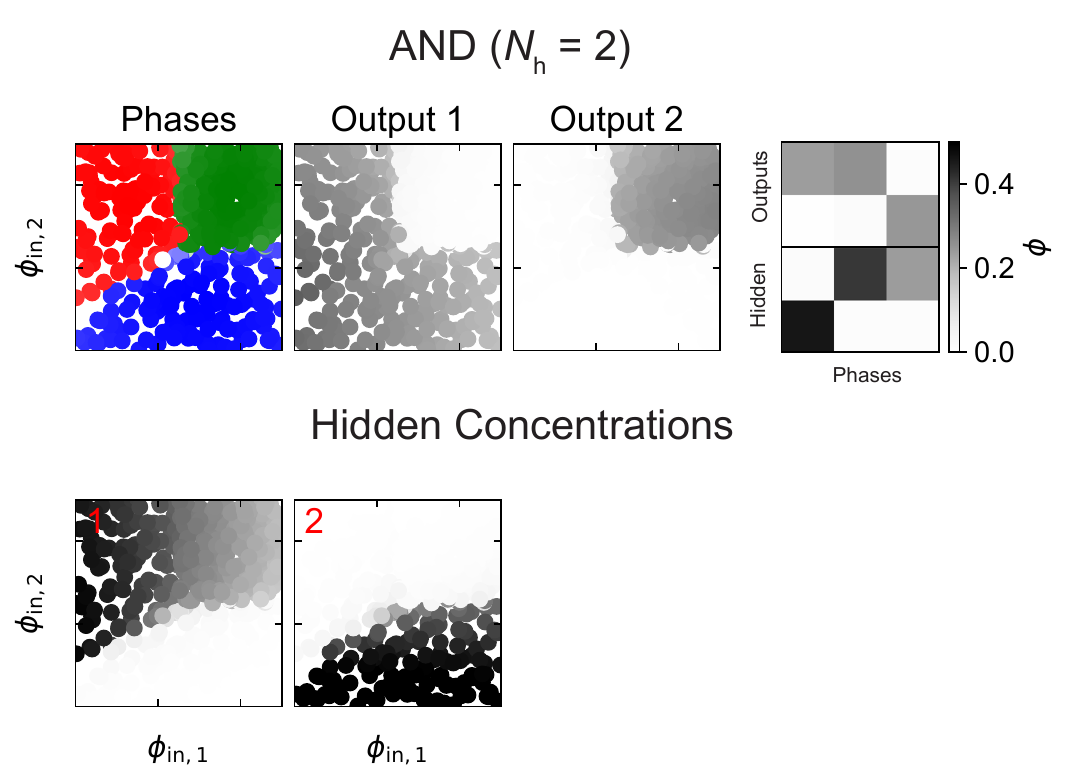}
    \caption{The encrypted phases in the AND classifier and the volume fraction of each hidden species in the input space. The phase decomposition aligns with the decision boundary shown in Fig. 3C. Following the convention in Fig. 4, the color of a point in the phase panel indicates the phase that the surface exhibits, while the color intensity of the point indicates the cosine similarity between the surface's phase vector and the mean concentration vector of all surfaces within the phase.}
    \label{fig:phases_and}
\end{figure*}

\clearpage
\begin{figure*}[h]
    \centering
    \includegraphics[width=0.7\linewidth]{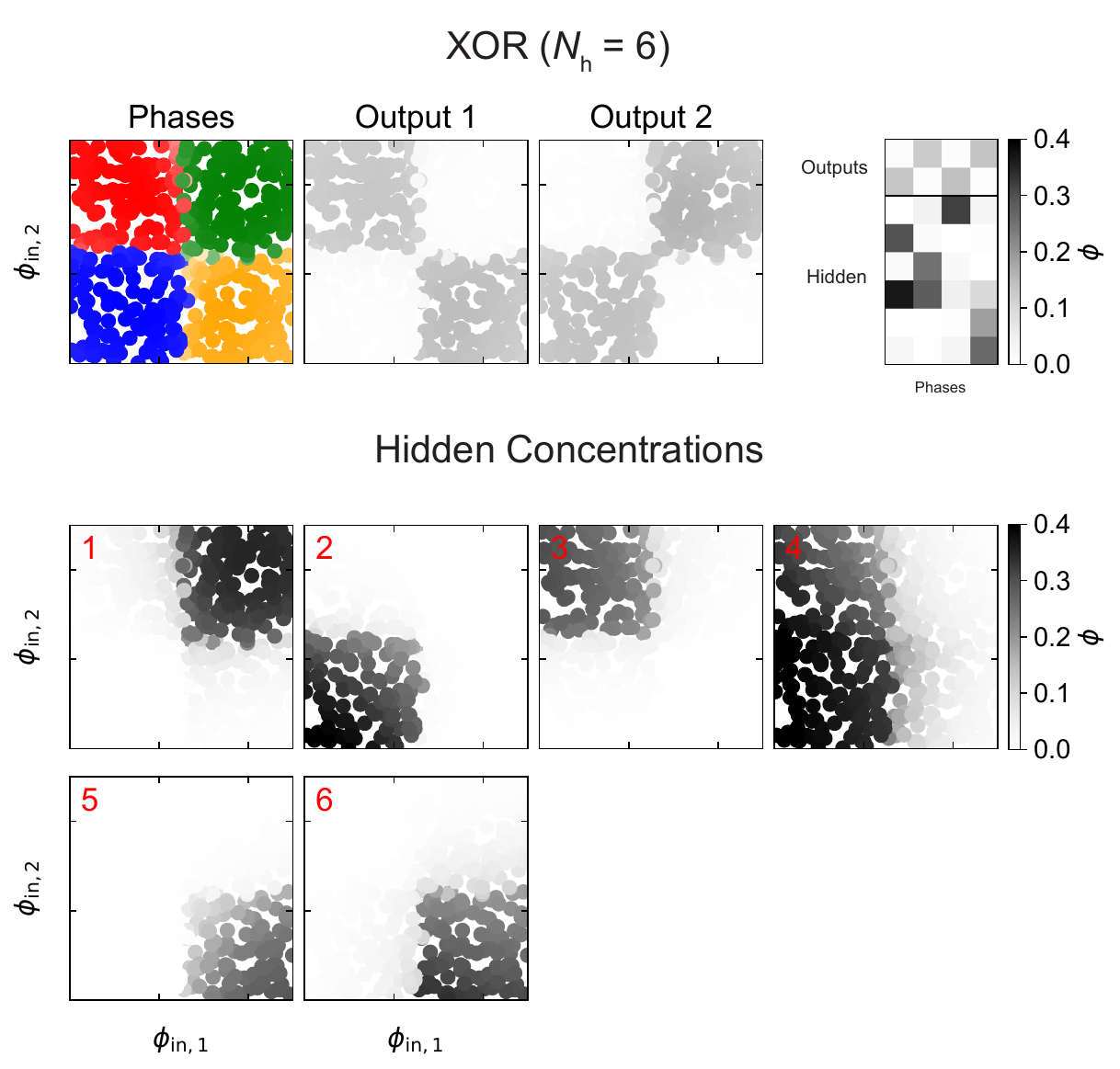}
    \caption{The encrypted phases in the XOR classifier and the volume fraction of each hidden species in the input space. The phase decomposition aligns with the decision boundary shown in Fig. 3D. Following the convention in Fig. 4, the color of a point in the phase panel indicates the phase that the surface exhibits, while the color intensity of the point indicates the cosine similarity between the surface's phase vector and the mean concentration vector of all surfaces within the phase.}
    \label{fig:phases_xor}
\end{figure*}

\clearpage
\begin{figure*}[h]
    \centering
    \includegraphics[width=0.7\linewidth]{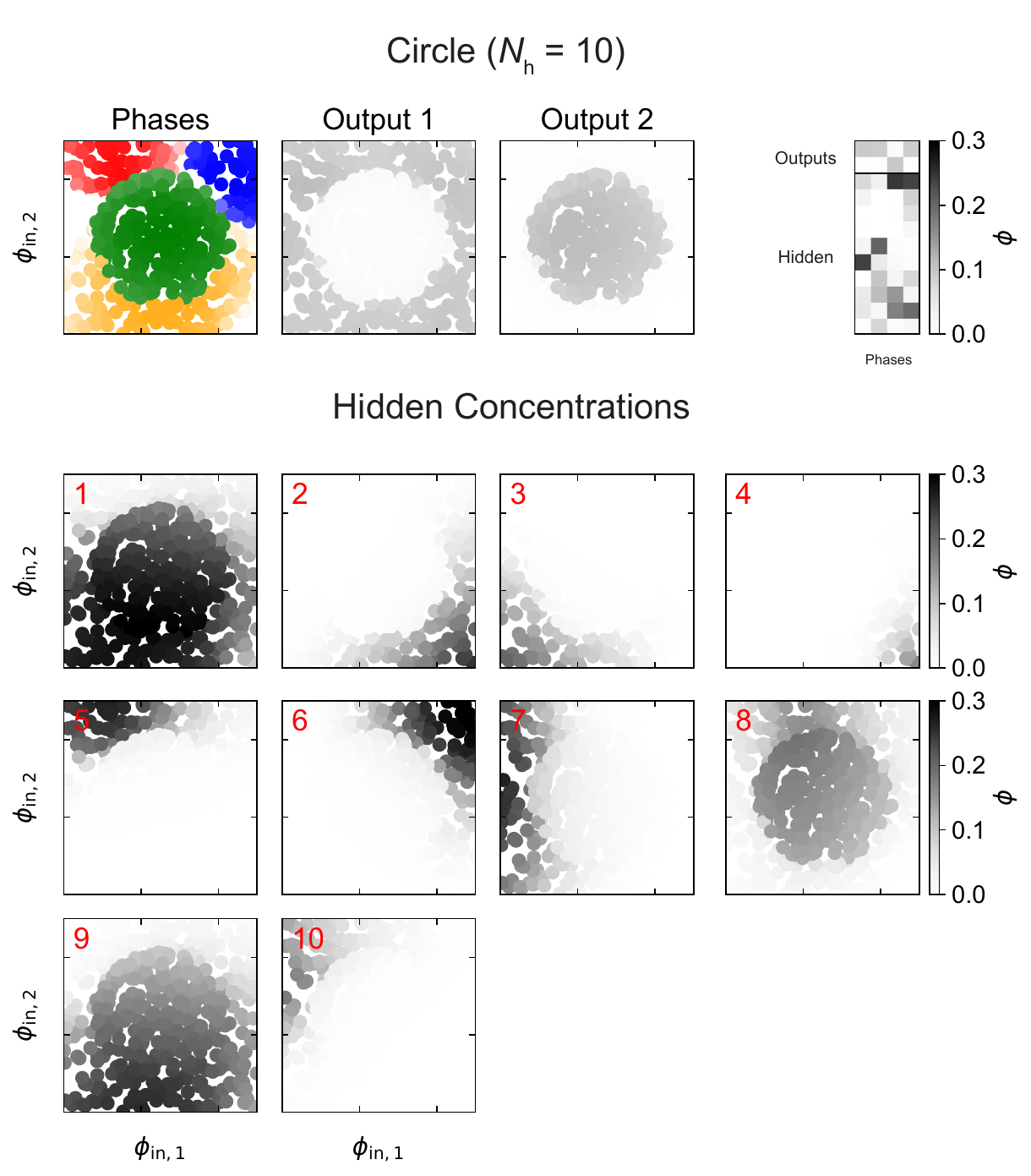}
    \caption{The encrypted phases in the Circle classifier and the volume fraction of each hidden species in the input space. The phase decomposition aligns with the decision boundary shown in Fig. 3D. Following the convention in Fig. 4, the color of a point in the phase panel indicates the phase that the surface exhibits, while the color intensity of the point indicates the cosine similarity between the surface's phase vector and the mean concentration vector of all surfaces within the phase. Note that in this case, there is a mismatch in the phase decomposition and the concentration of constituent species in the yellow phase. This disagreement likely results from our choice of clustering algorithm---the Marchenko-Pastur distribution does not guarantee that there are no significant signals below its cutoff threshold, and can therefore underestimate the number of phases in the input space.}
    \label{fig:phases_circle}
\end{figure*}

\clearpage
\begin{figure*}[h]
    \centering
    \includegraphics[width=0.7\linewidth]{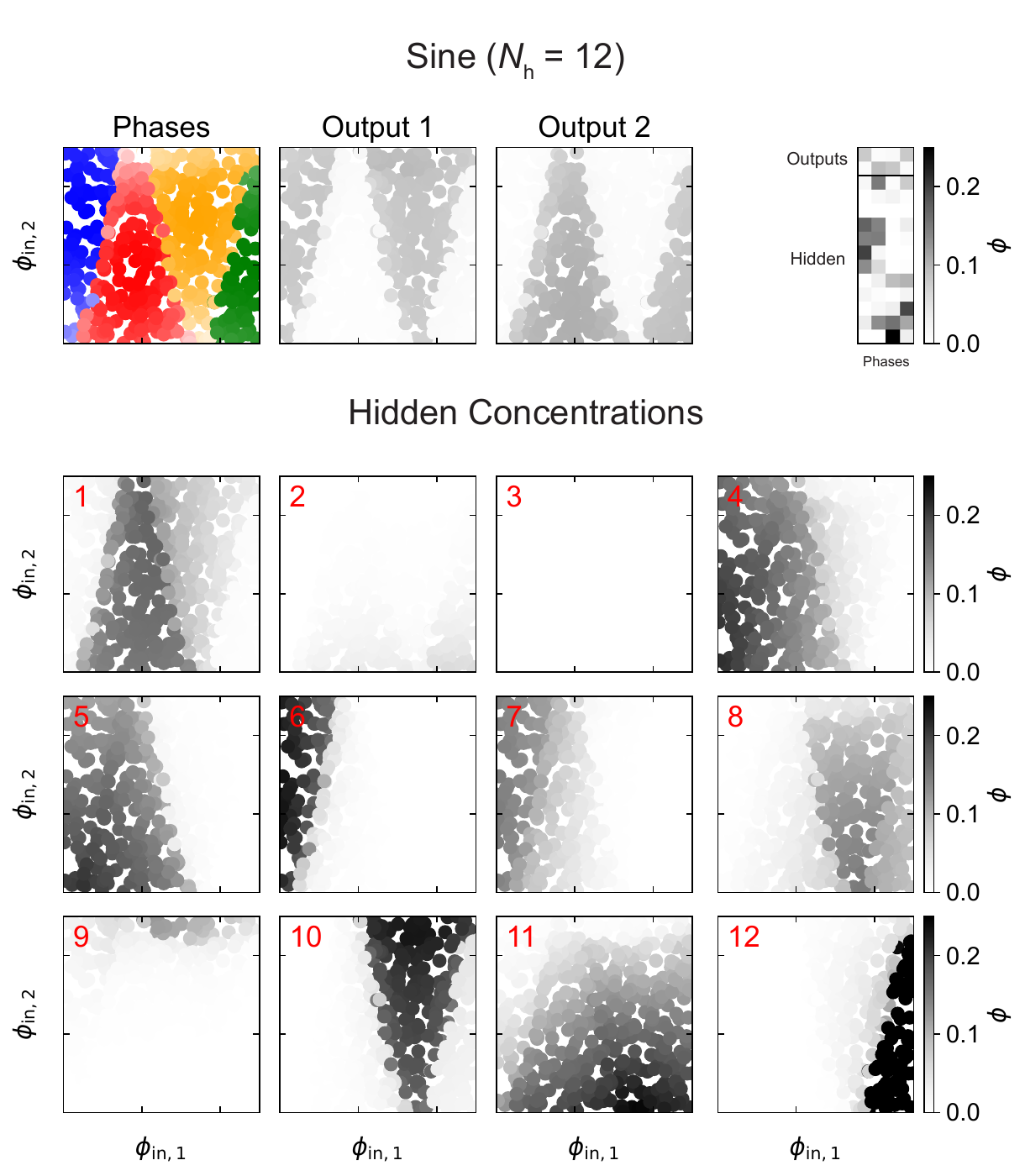}
    \caption{The encrypted phases in the Sine classifier and the volume fraction of each hidden species in the input space. The phase decomposition aligns with the decision boundary shown in Fig. 3D. Following the convention in Fig. 4, the color of a point in the phase panel indicates the phase that the surface exhibits, while the color intensity of the point indicates the cosine similarity between the surface's phase vector and the mean concentration vector of all surfaces within the phase.}
    \label{fig:phases_sine}
\end{figure*}

\clearpage
\begin{figure*}[h]
    \centering
    \includegraphics[width=0.7\linewidth]{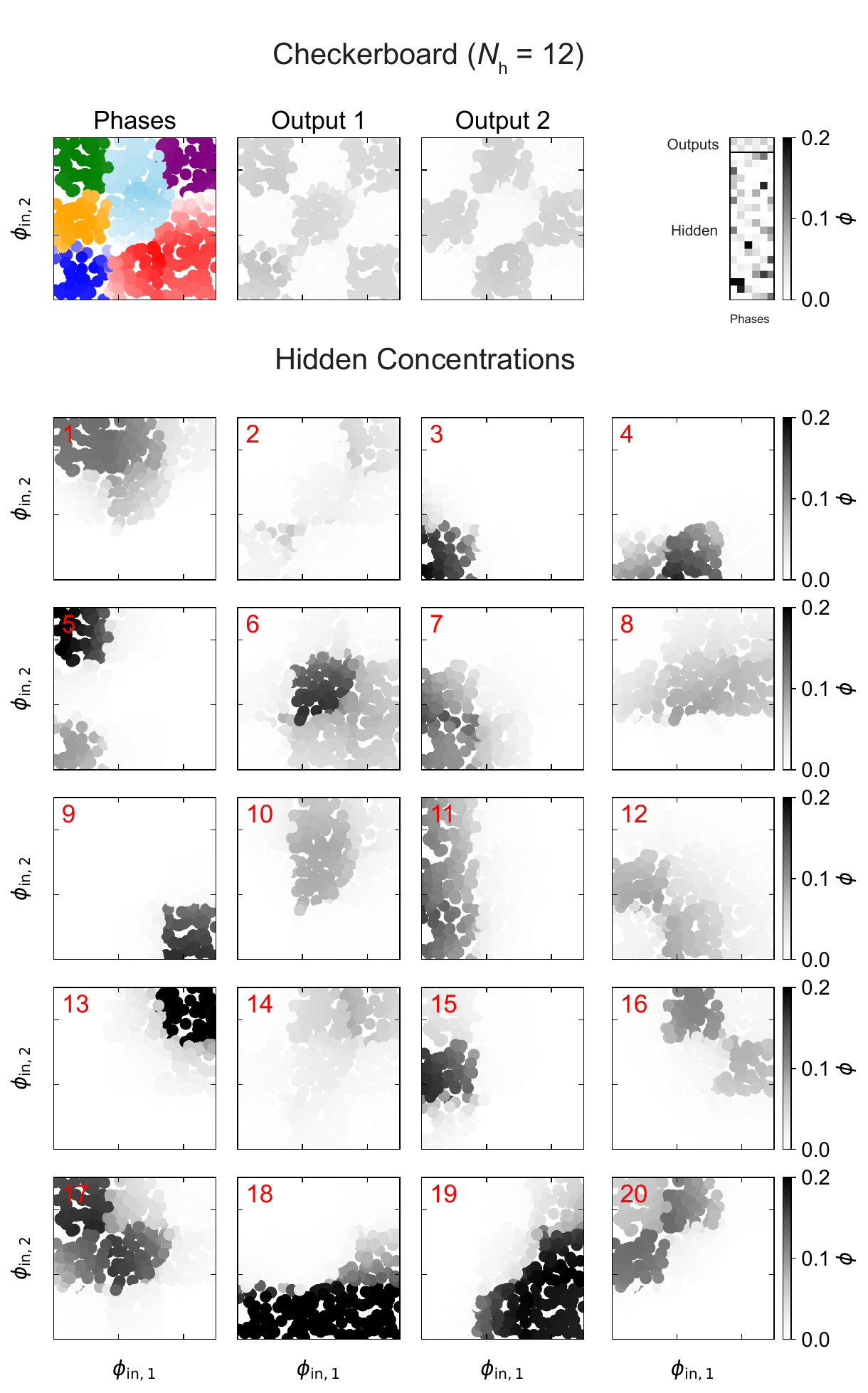}
    \caption{The encrypted phases in Checkerboard classifier and the volume fraction of each hidden species in the input space. The input space has distinct phase compositions in its combination of hidden species. Following the convention in Fig. 4, the color of a point in the phase panel indicates the phase that the surface exhibits, while the color intensity of the point indicates the cosine similarity between the surface's phase vector and the mean concentration vector of all surfaces within the phase. The discrepancy between the phase decomposition and the decision boundary in Fig. 3D likely results from our choice of clustering algorithm---the Marchenko-Pastur distribution does not guarantee that there are no significant signals below its cutoff threshold, and can therefore underestimate the number of phases in the input space.}
    \label{fig:phases_checkerboard}
\end{figure*}

\clearpage
\begin{figure*}[h]
    \centering
    \includegraphics[width=\linewidth]{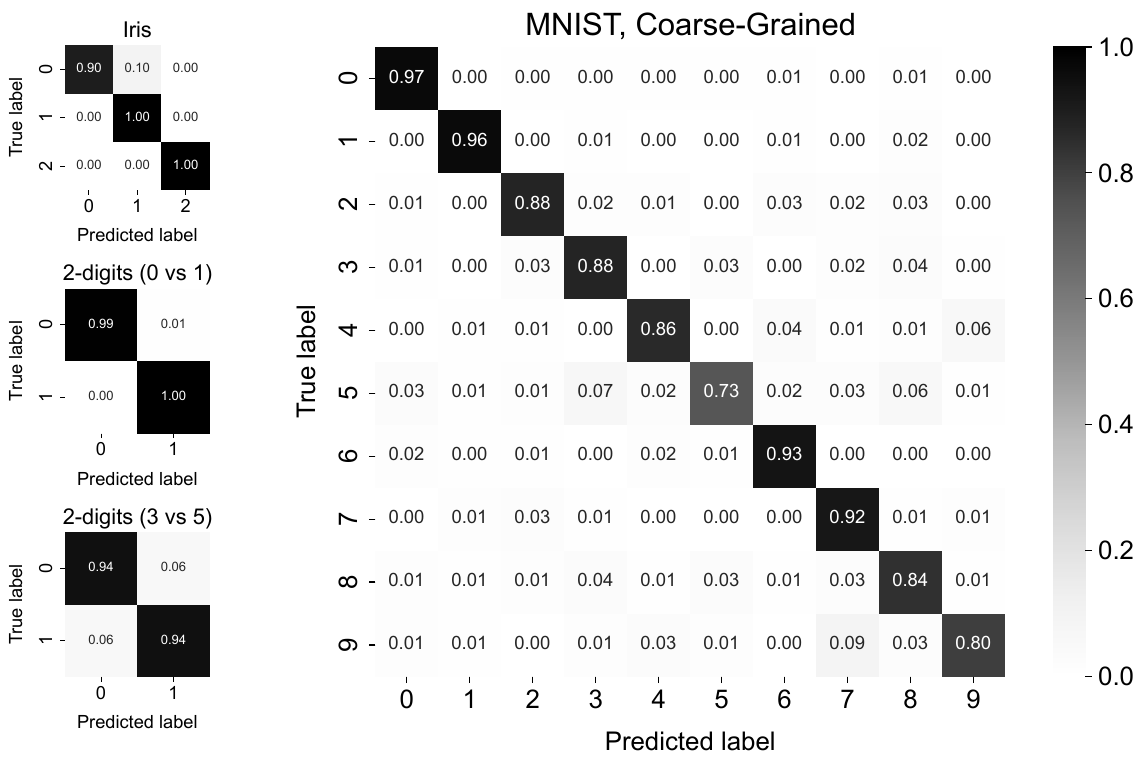}
    \caption{Confusion matrices for traditional classification problems, where rows are the true class and columns are the predicted class. The confusion matrices show the proportion of correct predictions (diagonal entries) vs. incorrect predictions (off-diagonal entries) made by the molecular network for each of these cases. Left column, top to bottom: the Iris dataset (0 hidden species---classes 0, 1, and 2 denote setosa, versicolor and virginica, respectively), 2-digit 0 vs 1 MNIST (2 hidden species), 2-digit 3 vs 5 MNIST (10 hidden species), and full 10-digit MNIST (20 hidden species). Here, the success criterion is made more lenient as in Fig. 6. In the case of 10-digit MNIST, this success criterion yields an average correct percentage of $87\%$, the worst-performing digit being $5$ at $73\%$ and the best-performing digit being 0 at $97\%$.}
    \label{fig:sup-mnist}
\end{figure*}

\clearpage
\begin{figure*}[h]
    \centering
    \includegraphics[width=0.85\linewidth]{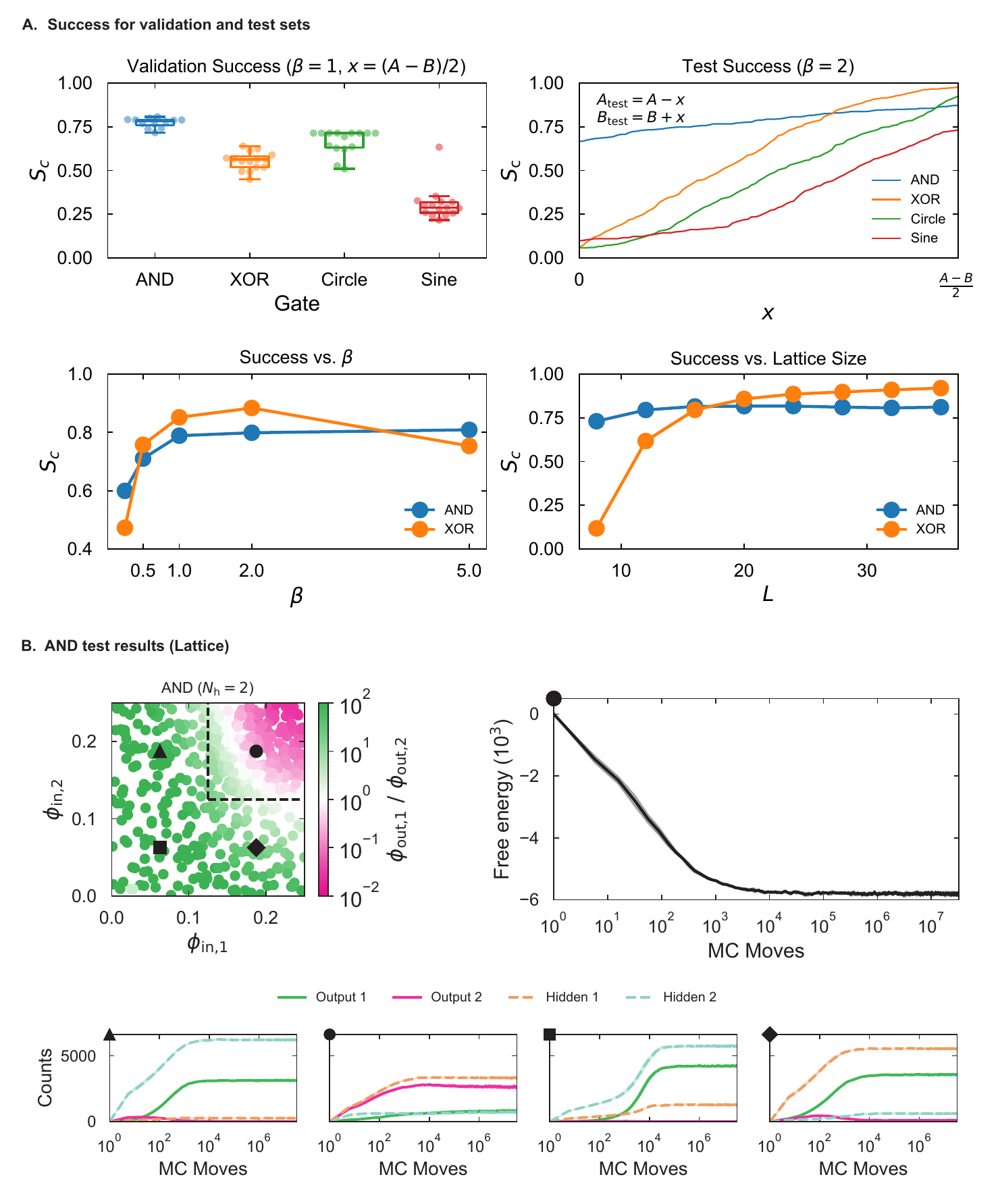}
    \caption{\textbf{(A)} The success score $S_c$ over the validation set for 15 separate mean-field solutions for AND, XOR, Circle, and Sine decision boundaries evaluated using the lattice liquid at $\beta=1$ (top left). In plotting the validation success, we used the maximally lenient (or asymptotic) values of $A_\text{test}=B_\text{test}=(A+B)/2$, which allows for a clearer assessment of boundary correctness in blurry regions. The classification success on the test set improves with increasing threshold leniency $x$, to above $\sim75\%$ (with several of the boundaries reaching $\sim90\%$) when $x$ is maximally lenient (top right); this test success uses $\beta=2$, based on the temperature scan performed on the AND and XOR boundaries (bottom left), where we found that using lower temperatures (or, higher $\beta$) generally improved performance after translating the mean-field parameters to lattice parameters. We also confirm that the lattice size $L$ does not meaningfully alter $S_c$ for $L\gtrsim24$. \textbf{(B)} The test results of the AND decision boundary at $\beta=2$. The plot from Fig. 7 is reproduced, with the test point nearest to the center of that quadrant for which the dynamics of the corresponding surface are shown. For the point in the upper-right quadrant, we also show the free energy of the surface as the simulation progresses, confirming that the surface reaches a free energy minimum.}
    \label{fig:sup-lattice}
\end{figure*}

\clearpage
\begin{figure*}[h]
    \centering
    \includegraphics[width=0.9\linewidth]{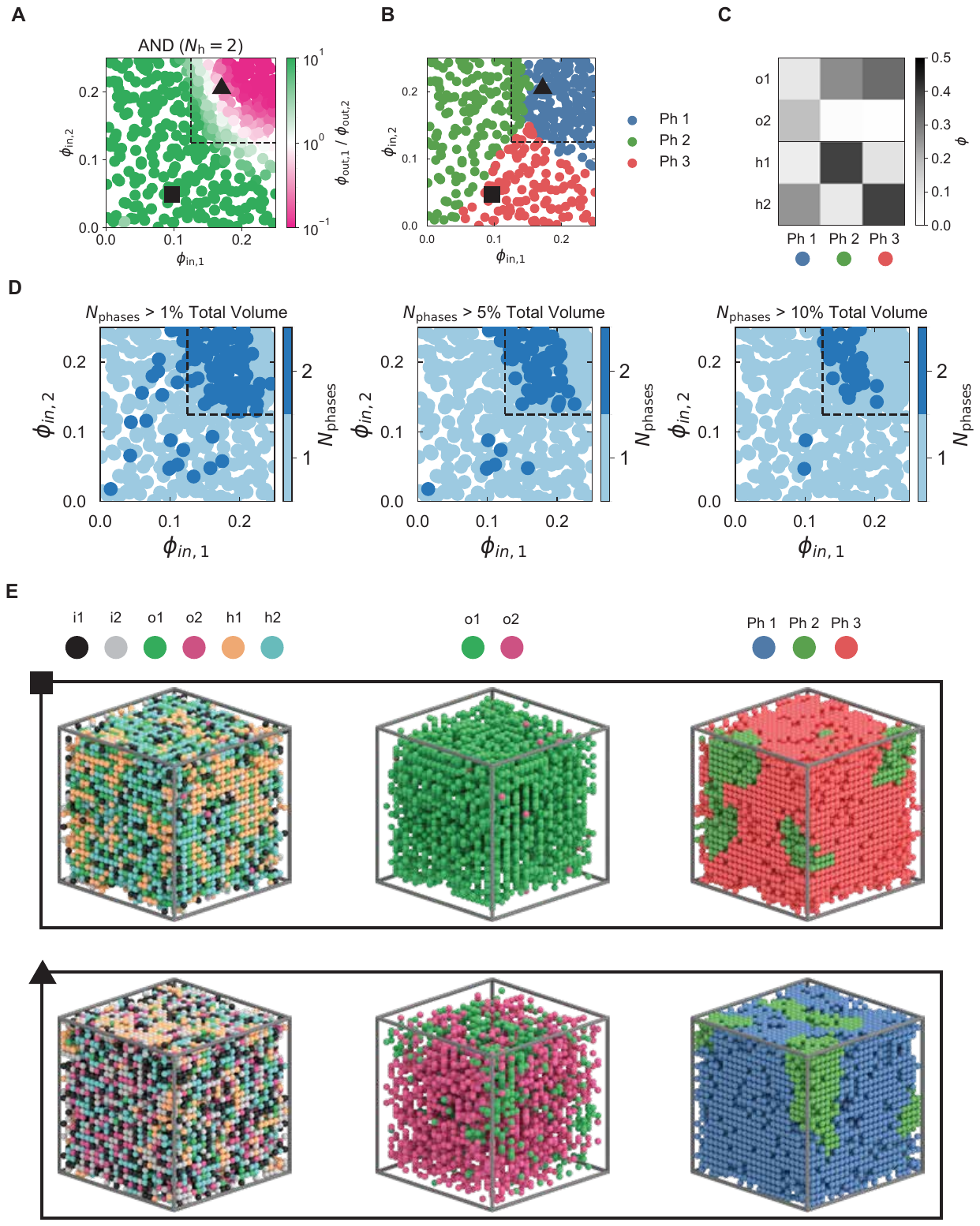}
    \caption{The structure within the lattice for the molecular network trained to solve the AND boundary. \textbf{(A)} The trained decision boundary for Fig. 7, reproduced for clarity. \textbf{(B)} The phase decomposition of the input space. Using the methodology presented in SI Note 5, three distinct phases are predicted, consistent with the phase analysis in the mean-field limit (see Fig. \ref{fig:phases_and}). \textbf{(C)} The concentration vectors defining the 3 discovered phases. \textbf{(D)} Near the phase boundaries, the number of phases within the simulation box grows from 1 to 2. \textbf{(E)} This change indicates phase separation within the box: at the points indicated in panel (A), which are near phase boundaries, the surface separates into phases 2 and 3 (square) and phases 1 and 2 (triangle).}
    \label{fig:sup-lattice-phases}
\end{figure*}

\clearpage
\begin{figure*}[h]
    \centering
    \includegraphics[width=0.8\linewidth]{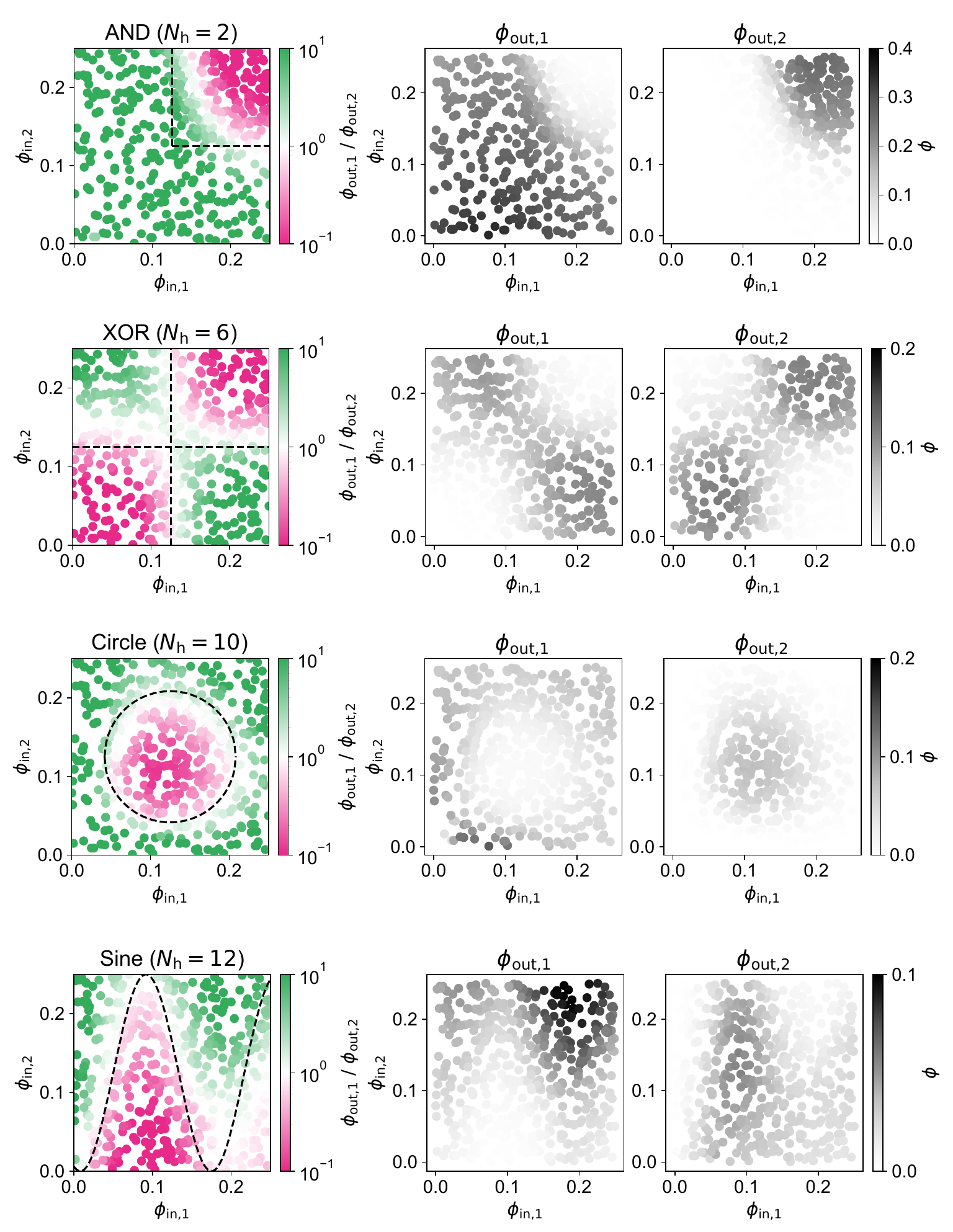}
    \caption{In the left column are the test predictions from Fig. 7 reproduced with a truncated colorbar ranging from $0.1$ to $10$ for greater visual clarity. Next to each test prediction are the absolute concentrations of the two output species across the input space.}
    \label{fig:sup-lattice-concentrations}
\end{figure*}

\clearpage
\begin{figure*}[h]
    \centering
    \includegraphics[width=0.9\linewidth]{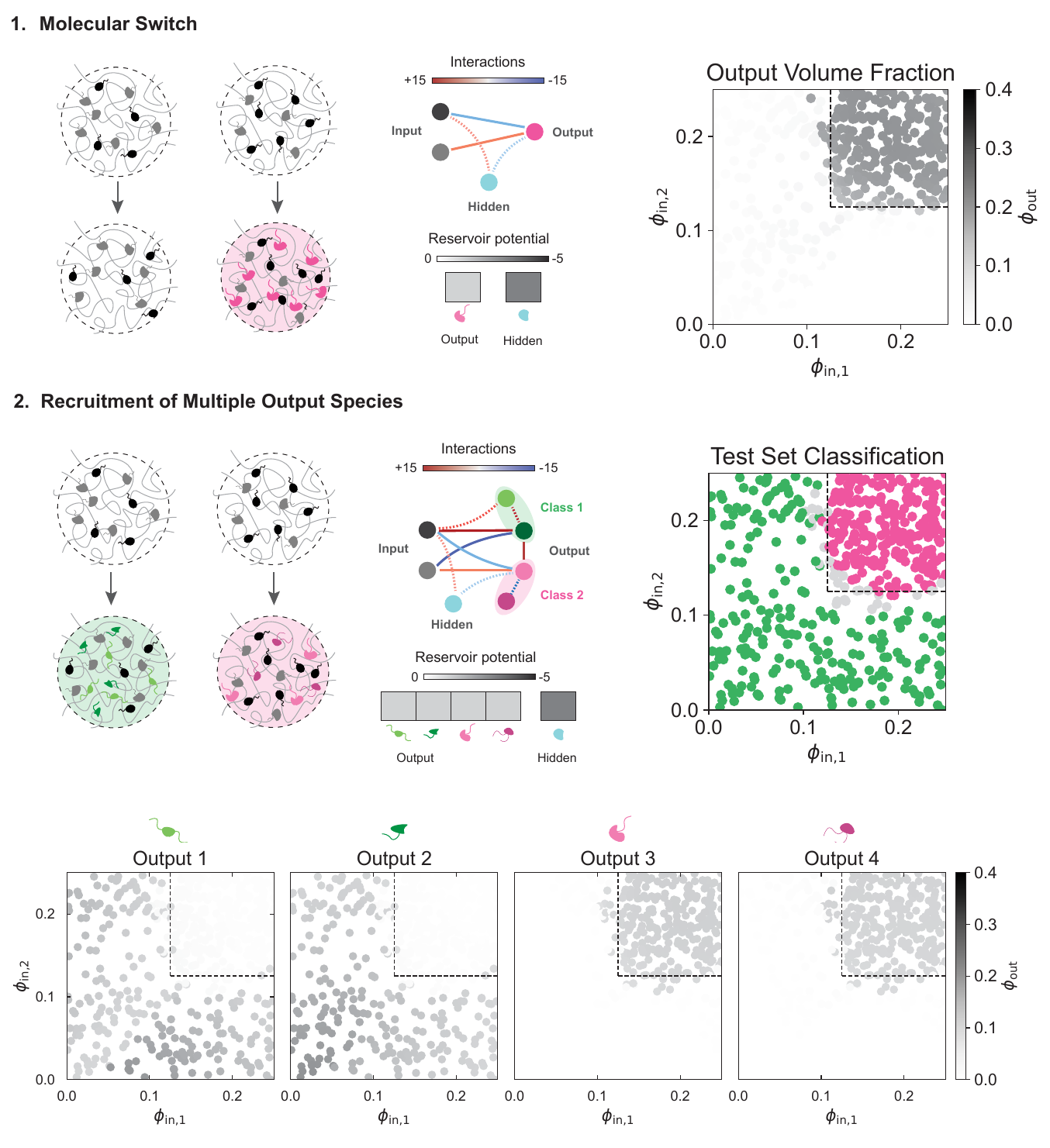}
    \caption{Examples of additional possibilities for the proposed system. Rather than recruit one output or another, it may be desirable to recruit, for instance, no outputs vs one output (the ``molecular switch''---see the top row), or multiple distinct outputs per class (bottom two rows). Many other variations are possible.}
    \label{fig:sup-alternates}
\end{figure*}

\clearpage
\begin{figure}
    \centering
    \includegraphics[width=\linewidth]{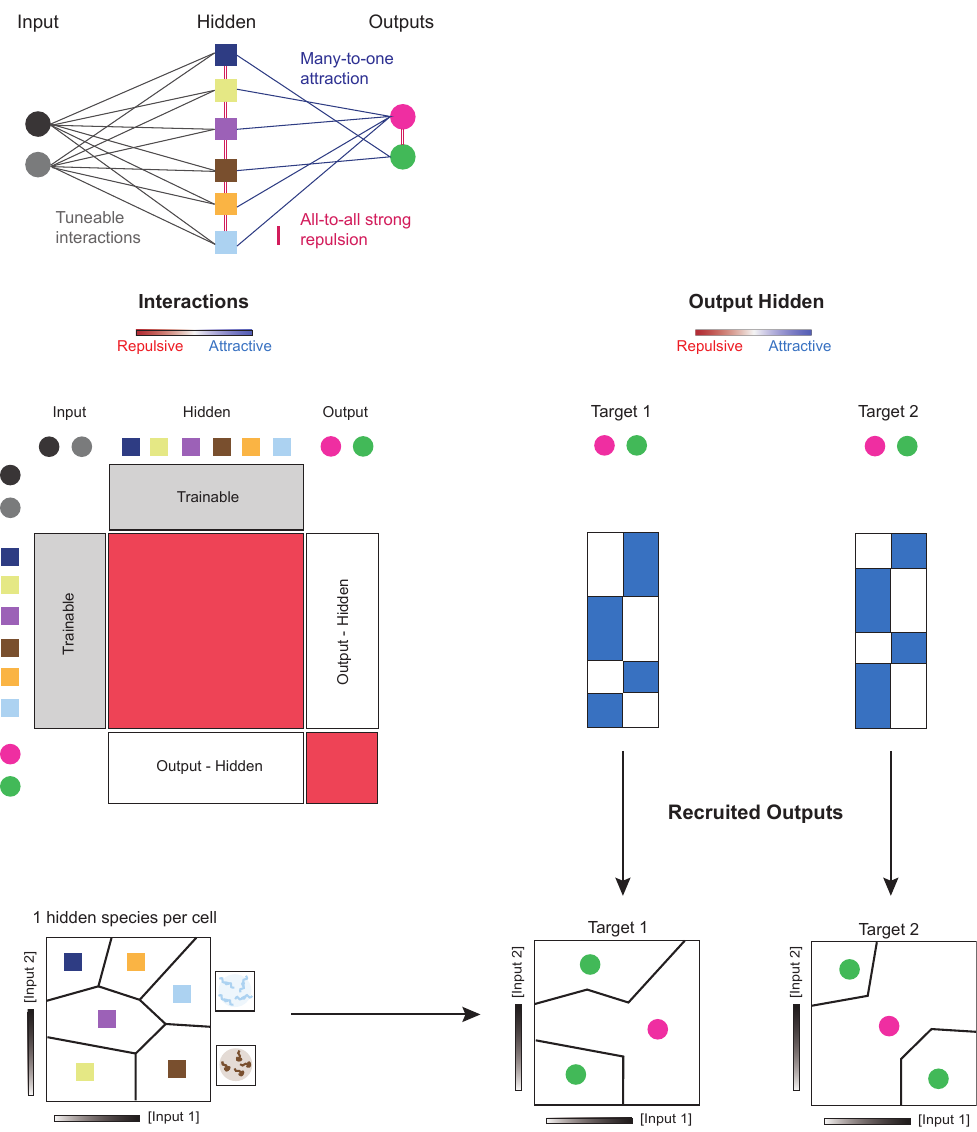}
    \caption{A schematic illustrating the proposed route towards proving that multiphase systems can perform universal approximation. The input-hidden interactions are tuned while the hidden-hidden and output-output interactions are taken to be repulsive. After training, the input space is partitioned into regions in which condensates are enriched in a single hidden species. Subsequently, the output-hidden interactions can be chosen such that each cell of the input space partition recruits the target output species, in line with the decision boundary being approximated. Increasing the number of hidden species could allow for a finer partitioning of the input space, leading to a better approximation of the target decision boundary.}
    \label{fig:sup-universal-approx}
\end{figure}

\end{document}